\documentclass[12pt]{article}

\usepackage{amsmath,amssymb,amsfonts,bbm}
\usepackage{epsfig,cancel,cite}

\numberwithin{equation}{section}

\newlength{\xtrawidth}
\newlength{\xtraheight}
\newcommand{\beq}{\begin{equation}}
\newcommand{\eeq}{\end{equation}}
\newcommand{\ba}{\begin{array}}
\newcommand{\ea}{\end{array}}
\newcommand{\bea}{\begin{eqnarray}}
\newcommand{\eea}{\end{eqnarray}}
\newcommand{\bean}{\begin{eqnarray*}}
\newcommand{\eean}{\end{eqnarray*}}
\newcommand{\eref}[1]{(\ref{#1})}

\newcommand{\comment}[1]{}

\newcommand{\IP}{\mathbb{P}}

\newcommand{\IZ}{\mathbb{Z}}
\newcommand{\cO}{{\cal O}}

\newcommand{\cF}{{\cal F}}

\newcommand{\cL}{{\cal L}}

\def\fnote#1#2{\begingroup\def\thefootnote{#1}\footnote{#2}
     \addtocounter{footnote}{-1}\endgroup}

\begin{document}

\title{{\LARGE
Exploring Positive Monad Bundles And \\A New Heterotic Standard Model
}}
\author{
Lara B. Anderson${}^{1}$,
James Gray${}^{2}$,
Yang-Hui He${}^{2,3,4}$,
Andre Lukas${}^{2}$
}
\date{}
\maketitle
\begin{center}
{\small
${}^1${\it Department of Physics, University of
      Pennsylvania,\\ 209 South 33rd Street, Philadelphia, PA 19104-6395, U.S.A.}\\[0.2cm]
${}^2${\it Rudolf Peierls Centre for Theoretical Physics, Oxford
  University,\\
$~~~~~$ 1 Keble Road, Oxford, OX1 3NP, U.K.}\\[0.2cm]
${}^3${\it Merton College, Oxford, OX1 4JD, U.K.}\\[0.2cm]
${}^4${\it Department of Mathematics, City University London,\\
Northampton Square, London EC1V 0HB, U.K.}
\fnote{}{andlara@physics.upenn.edu}
\fnote{}{yang-hui.he@merton.ox.ac.uk} 
\fnote{}{j.gray1@physics.ox.ac.uk}
\fnote{}{lukas@physics.ox.ac.uk}
}
\end{center}

\abstract{A complete analysis of all heterotic Calabi-Yau compactifications based on 
  positive two-term monad bundles over favourable complete intersection
  Calabi-Yau threefolds is performed. We show that the original data set of about 7000
  models contains 91 standard-like models which we describe in detail. A closer analysis
  of Wilson-line breaking for these models reveals that none of them gives rise to precisely
  the matter field content of the standard model. We conclude that the entire set of positive two-term
  monads on complete intersection Calabi-Yau manifolds is ruled out on phenomenological grounds.
  We also take a first step in analyzing the larger class of non-positive monads. In particular, we construct
  a supersymmetric heterotic standard model within this class. This model has the standard model gauge group and an additional $U(1)_{B-L}$ symmetry, precisely three families of quarks and leptons, one pair of Higgs doublets and no anti-families or exotics of any kind.}

\newpage

\tableofcontents

%
%
\section{Introduction}
For many years, one of the canonical approaches to string
phenomenology has been the compactification of heterotic string or
M-theory on smooth Calabi-Yau three-folds \cite{Candelas:1985en,Witten:1985bz,GSW}. The main aim of
these constructions is to find four-dimensional theories which are as
close as possible to the minimal supersymmetric standard model (MSSM)
and, ultimately, to construct a fully realistic standard model from
string theory. In the M-theory limit, these models have an underlying five-dimensional
brane-world structure~\cite{Lukas:1998yy}. Building such models is normally achieved in two
steps. Firstly, one obtains GUT models with gauge symmetries $E_6$,
$SO(10)$ or $SU(5)$ and, secondly, these unified gauge groups are
broken down to the standard model group by adding Wilson lines in the
internal space. Initially, heterotic Calabi-Yau models were based on
the standard embedding where the internal gauge bundle is chosen to be
the spin connection of the Calabi-Yau manifold~\cite{GSW}. This gives
rise to models with an $E_6$ grand unified group. In recent years, the
focus has moved to a wider class of models based on general
holomorphic vector bundles on the internal space which can lead to any
of the three GUT groups mentioned above
\cite{Anderson:2007nc,Anderson:2008uw,Anderson:2008ex,Donagi:2004ia,Donagi:2004qk,Braun:2005ux,Braun:2005nv,Bouchard:2005ag,Donagi:2004su,Blumenhagen:2006wj,Bouchard:2006dn,Bak:2008ey}. One
common feature of the majority of the work that has been published in
this field, is that small numbers of models tend to be considered at a
time. Typically, a single, very carefully constructed, gauge bundle
over a single manifold is given, which leads to a phenomenologically
attractive four-dimensional theory.

The present paper is the latest work in a programme
\cite{Anderson:2007nc,Anderson:2008uw,Anderson:2008ex,Anderson:2009ge,He:2009wi}
which is developing techniques to perform more comprehensive scans of
heterotic Calabi-Yau compactifications. The goal of this programme is to consider
complete classes of models at a time, rather than restricting to
single specific cases, and identify interesting models by successively imposing
physical constraints. In effect, this is the equivalent for smooth
Calabi-Yau compactifications of the comprehensive scans of orbifold
compactifications which have been carried out in recent years
\cite{Lebedev:2008un,Lebedev:2007hv,Lebedev:2006tr,Lebedev:2006kn,Kobayashi:2004ud,Kobayashi:2004ya,Buchmuller:2005jr,Buchmuller:2006ik,Forste:2004ie,Kobayashi:2006wq}. In this article we will present the final results of a scan over a complete class of heterotic Calabi-Yau compactifications. We discuss all positive two-term monad bundles over all favourable
complete intersection Calabi-Yau manifolds (CICY) \cite{Candelas:1987kf,Candelas:1987du,Green:1987cr,He:1990pg,Gagnon:1994ek,Candelas:2008wb,hubsch}. This is a relatively small initial data set, consisting of some 7118 consistent, supersymmetric models, which has been constructed in Ref.~\cite{Anderson:2008uw}. Models of this kind have been considered in the physics literature the 1980's, with specific cases being analysed in a number of papers \cite{Distler:1987ee,Kachru:1995em,Douglas:2004yv,Blumenhagen:2006wj}.

In this work we perform a comprehensive analysis of the physical properties of this class of compactifications. Of the
initial 7118 models we show that only 91 pass a certain test which is necessary, but not sufficient, in order to obtain three generations of matter fields. Among these 91 models are
87 $E_6$ based models, 3 $SO(10)$ based models, and a single $SU(5)$
example. It turns out that, of the 87 $E_6$ models, most cannot be
broken to the standard model, and all those that remain suffer
from a fatal lack of doublet triplet splitting. Further, none of the three $SO(10)$ models
can give rise to three generation models, despite passing the necessary but not sufficient
check mentioned above. Finally, the single $SU(5)$ models leads to an interesting spectrum with
three families in $\bar{\bf 5}+{\bf 10}$ and one pair of Higgs multiplets in ${\bf 5}+\bar{\bf 5}$.
However, it turns out that the $SU(5)$ group cannot be broken to the standard model group with the available Wilson lines. Thus, we can conclude that the entire set of positive two-term monad bundles on favourable CICYs is ruled out on phenomenological grounds.

We note that this class of models has been put forward as a possible setting for string phenomenology since the early days of string theory. For this reason, we consider our results, albeit negative, to be of some relevance. 
Our analysis of the positive monads has led to two further insights. We have shown that the condition of positivity, although conductive to the stability of the vector bundle, is in fact not necessary for Calabi-Yau manifolds $X$  with $h^{1,1}(X)>1$. This means that semi-positive or even ``slightly negative'' monads can be stable~\cite{stabpaper,Anderson:2009nt,Anderson:2009sw,Anderson:2008ex} and as a consequence, that positive monads are likely to be a small sub-set of all stable monad bundles. In addition, in our work so far we have developed many of the tools necessary for a systematic analysis of this more general class of monads.

In this paper, we will take a first, preliminary step towards analyzing general monad bundles by showing that such a scan is a worthwhile enterprise which will lead to phenomenologically attractive models. Concretely, we will construct a new heterotic standard model based on a semi-positive monad bundle and the bi-cubic CICY. This model is supersymmetric and anomaly-free and its four-dimensional gauge group is the standard model group and an additional $U(1)_{B-L}$ symmetry which stabilises the proton. Its matter field content consists precisely of three families of quark and leptons (including three right-handed neutrinos) and one pair of Higgs doublets. There are no anti-families or exotic matter fields of any kind. A systematic scan of monad bundles in order to find all models with similar properties is already underway~\cite{zeropaper}.

The rest of the paper is structured as follows. In the next section we
introduce the favourable complete intersection Calabi-Yau manifolds
and positive monad bundles over them. In section \ref{equimon} we
describe how one can decide whether these bundles admit an equivariant
structure under a given discrete symmetry acting on the base
three-fold. This is a crucial step required for breaking the GUT group
down to the standard model group. In section \ref{results} we state the
results of our scan over the positive monads. Finally, in section
\ref{the_model}, we introduce a new heterotic standard model. We
conclude in section \ref{conc}. Various technical considerations and
lists of positive monad data are provided in the appendices. In particular, as an interesting case study in the use of non-Abelian Wilson lines, we construct a model on the tetra-quadric CICY, based on the quaternionic group $\mathbb{H}$. 

\section{The data set: positive monads over favourable
  CICYs}\label{setup}

In this section, we shall review the data set which will be considered
in the rest of the paper. This is the set of positive monads over favourable
complete intersection Calabi-Yau in products of projective
spaces. This data set has been analysed in quite some detail, within
the context of this programme of research, in previous papers
\cite{Anderson:2007nc,Anderson:2008uw,Anderson:2008ex}. We will start
with a brief description of the manifolds themselves, before moving on
to discuss the class of bundles which we will consider.

\subsection{The Calabi-Yau: favourable CICYs}

Complete intersection Calabi-Yau manifolds in projective spaces (CICYs)
\cite{He:1990pg,Gagnon:1994ek,Green:1987cr,Candelas:1987du,Candelas:1987kf} are defined as the common zero locus of
homogeneous polynomials in an ambient space ${\cal
  A}=\mathbb{P}^{n_1}\times\dots\times\mathbb{P}^{n_m}$. We denote the
(canonically normalised) K\"ahler form of each projective space by
$J_r$. Line bundles on ${\cal A}$ are then written as ${\cal O}_{\cal
  A}(k^1,\ldots ,k^m)={\cal
  O}_{\mathbb{P}^{n_1}}(k^1)\times\ldots\times{\cal
  O}_{\mathbb{P}^{n_m}}(k^m)$, where ${\cal
  O}_{\mathbb{P}^{n_r}}(1)$ is the line bundle associated to the
divisor which is Poincar\'e dual to $J_r$.

To define a three-fold as a complete intersection within such an
ambient space we need $K=\sum_{r=1}^mn_r-3$ polynomials. We denote the
multi-degrees of these polynomials by ${\bf q}_i=(q^1_i,\ldots
,q^m_i)$, where $q^r_i$ is the degree of the $i^{\rm th}$ polynomial
in the coordinates of the $r^{\rm th}$ projective space. A customary
way to encode this information is by a {\em configuration matrix},
\beq\label{cy-config} \left[\ba{c|cccc}
  \IP^{n_1} & q_{1}^{1} & q_{2}^{1} & \ldots & q_{K}^{1} \\
  \IP^{n_2} & q_{1}^{2} & q_{2}^{2} & \ldots & q_{K}^{2} \\
  \vdots & \vdots & \vdots & \ddots & \vdots \\
  \IP^{n_m} & q_{1}^{m} & q_{2}^{m} & \ldots & q_{K}^{m} \\
  \ea\right]_{m \times K}\; .  \eeq The normal bundle of a complete
intersection $X$, defined by a configuration matrix \eqref{cy-config},
over an ambient space ${\cal A}$, is given by,
\begin{equation}
{\cal N}|_{X}=\bigoplus_{i=1}^KO_{\cal A}({\bf q}_i)|_X\; . \label{N}
\end{equation}
For such a three-fold to be a Calabi-Yau manifold its first Chern
class must vanish. This translates into the conditions, \beq
\sum_{j=1}^K q^{r}_{j} = n_r + 1 \qquad \forall \; r=1, \ldots, m\;
. \label{c10} \eeq 

The CICYs can be classified, essentially by finding all configuration
matrices subject to the constraints~\eqref{c10}. This has been done
some time
ago~\cite{He:1990pg,Gagnon:1994ek,Green:1987cr,Candelas:1987du,Candelas:1987kf}
and results in a list of 7890 manifolds. The corresponding data set
has recently been revived in Ref.~\cite{Anderson:2008uw}, and our work
will be based on this new version.

In fact, from these 7890 manifolds, we will only consider those for
which the second cohomology descends entirely from the ambient space
${\cal A}$, that is, those for which $h^{1,1}(X)=m$. There are 4515
such manifolds and we will refer to them as {\em favourable}
CICYs. This restriction is adopted for technical reasons. For
favourable CICYs, the second cohomology is spanned by the ambient
space K\"ahler forms pulled back to $X$, which we will, by abuse of
notation, also denote by $J_r$. The K\"ahler cone of the favourable
CICYs can then simply be described by $\{J=t^rJ_r\, |\, t^r\geq
0\}$. Further, if we introduce a basis of harmonic four-forms
$\{\nu^r\}$ dual to $J_r$, the effective classes $W\in
H^2(X,\mathbb{Z})$ correspond to positive integer linear combinations
of the $\nu^r$. This fact considerably simplifies the task of checking
the heterotic anomaly cancellation condition. Another advantage of the
favourable CICYs becomes apparent when we consider line bundles, which
will be the main building blocks of our monad vector bundles. In
general, line bundles are classified by their first Chern
class. Hence, on a favourable CICY we can label line bundles by $m$
integers ${\bf k}=(k^1,\ldots, k^m)$ and denote them by ${\cal
  O}_X({\bf k})$ such that $c_1({\cal O}_X({\bf
  k}))=k^rJ_r$. Moreover, labelled in this way, the line bundles on
$X$ are the restrictions of their ambient space counterparts, that is
${\cal O}_X({\bf k})={\cal O}_{\cal A}({\bf k})|_X$. The positive line
bundles are those ${\cal O}_X({\bf k})$ with all $k^r>0$. For positive
line bundles, Kodaira vanishing implies that $h^i(X,{\cal O}_X({\bf
  k}))=0$ for all $i>0$, that is, the zeroth cohomology is the only
non-trivial one. This fact helps considerably in cohomology
calculations.

\subsection{The bundles: positive monads}\label{thebuns}

We now move on to review the bundle construction we shall use
\cite{monadbook,
  maruyama,beilinson,HM,Anderson:2007nc,Anderson:2008uw}. On the
favourable CICYs, described in the previous subsection, we would like
to construct holomorphic vector bundles $V$ with structure group
${\cal G}=SU(n)$, where $n=3,4,5$, which break the ``observable'' $E_8$ gauge
group in ten dimensions to the grand-unified groups $E_6$, $SO(10)$ or
$SU(5)$, respectively. Such bundles $V$ can be obtained from the monad
construction~\cite{monadbook, maruyama,beilinson,HM}, that is from
short exact sequences of the form\footnote{More generally, a monad
  bundle is defined as the middle homology of a sequence of the form
  $0\to A \stackrel{m_1}{\longrightarrow} B \to C\to 0$. This sequence
  is exact at $A$ and $C$, and $\textnormal{Im}(m_1)$ is a sub-bundle
  of $B$ \cite{monadbook}. In this paper we restrict ourselves, as is
  often done in the physics literature, to the case where
  $\textnormal{Im}(m_1)$~vanishes. We thus recover the description
  \eqref{V}.},
\begin{equation} \label{V}
0 \to V \to B \stackrel{f}{\longrightarrow} C \to 0\ ,
\end{equation}
where
\begin{equation}
B = \bigoplus_{i=1}^{r_B} \cO_X({\bf b}_i) \ , \quad
C = \bigoplus_{a=1}^{r_C} \cO_X({\bf c}_a)
\label{BC}
\end{equation}
are sums of line bundles. The map $f$ is an element of ${\rm
  Hom}(B,C)\simeq\Gamma(X,B^*\otimes C)=\bigoplus_{i,a}\Gamma(X,{\cal
  O}_X({\bf c}_a-{\bf b}_i))$ and we can think of it as a $r_C\times
r_B$ matrix of sections, with the entry $(a,i)$ corresponding to an
element of $\Gamma(X,{\cal O}_X({\bf c}_a-{\bf b}_i))$.  In practice,
this means $f$ is given by a matrix of homogeneous polynomials with
multi-degrees ${\bf c}_a-{\bf b}_i$. From the exactness of the sequence, the bundle $V$ is isomorphic to ${\rm Ker}(f)$. We shall ask for all of the
entries of $f$ to be non-trivial and thus for all $\Gamma(X,{\cal
  O}_X({\bf c}_a-{\bf b}_i))$ to be non-vanishing. To guarantee this
we require that~\footnote{In this case, $C^*\otimes B$ is globally generated, so that, due to a theorem by Fulton and Lazarfeld~\cite{lazarsfeld}, $V$ is indeed a vector bundle rather than merely a sheaf. It may well be possible to relax the condition~\eqref{fex} and still obtain a vector bundle $V$. However, this requires a detailed case by case analysis which we will not consider in the present paper.}
\begin{equation}
  c^r_a\geq b^r_i \;\; \forall \;\; a, i, r\; . \label{fex}
\end{equation}
In order to obtain bundles $V$ with rank $3$, $4$ or $5$ we need that
\begin{equation}
 {\rm rk}(V)=r_B-r_C\stackrel{!}{=}3,4,5\; . \label{rk}
\end{equation}
Further, for the structure group to be $SU(n)$ (rather than $U(n)$) the first Chern class $c_1(V)=c_1^r(V)J_r$ of $V$ must vanish:
\begin{equation}
 c_1^r(V) = \sum_{i=1}^{r_B} b^r_i - \sum_{a=1}^{r_C} c^r_a\stackrel{!}{=}0\; \quad \forall \; r \label{c1}~.
\end{equation} 
The heterotic anomaly cancellation condition imposes a constraint on
the second Chern class $c_2(V)=c_{2r}(V)\nu^r$ if we wish to preserve
supersymmetry in the four dimensional theory. The two-cycle dual to
$c_2(TX)-c_2(V)$ must be an effective class in $H_2(X,\mathbb{Z})$,
where $c_2(TX)=c_{2r}(TX)\nu^r$ is the second Chern class of the
tangent bundle of $X$. Written out in terms of components, this
condition becomes \cite{Anderson:2008uw}, in our cases,
\begin{equation}
c_{2r}(V) = \frac12  d_{rst} \left(\sum_{a=1}^{r_C} c^s_a c^t_a- \sum_{i=1}^{r_B} b^s_i b^t_i \right)\stackrel{!}{\leq}c_{2r}(TX)\; \quad \forall \; r, \label{anomaly}
\end{equation}
where $d_{rst}=\int_XJ_r\wedge J_s\wedge J_t$ are the triple
intersection numbers of $X$. It is also useful to provide the
expression for the third Chern class,
\begin{equation}\label{c3}
  c_3(V) = \frac13 d_{rst} \left(\sum_{i=1}^{r_B} b^r_i b^s_i b^t_i - \sum_{a=1}^{r_C} c^r_ac^s_a c^t_a \right)\;,
\end{equation}
since the chiral asymmetry of the model (the net number of families) is given by the index ${\rm ind} (V)=\frac{1}{2}\int_{X}c_3(V)$. 

Another important requirement on our models is that they preserve
supersymmetry. In particular, the gauge fields must preserve
supersymmetry and, translated into mathematical terminology, this
means that the bundles $V$ must be poly-stable~\cite{duy}. Stability
is a condition which is typically difficult to prove and, for monad
bundles on CICYs, has been studied in detail in Refs.~\cite{maria,
  Anderson:2007nc,Anderson:2008ex,stabpaper}. These papers provide an
explicit algorithm for checking stability and in this way many monad
bundles have been shown to be stable. Still, proving stability remains
a complicated task which is best performed after filtering out
physically uninteresting models. This is the general attitude we will
follow in this paper. Indeed, for the new heterotic standard model
presented in Section \ref{the_model}, we explicitly verify
slope-stability of the bundle.

Apart from the constraints on monad bundles described above, there is
one more condition that we would like to impose. In this paper, we
will focus on {\em positive} monad bundles, that is monad bundles
defined by Eqs.~\eqref{V} and \eqref{BC} with
\begin{equation}
 b^r_i>0\; ,\;\; c_a^r>0\; \mbox{ for all } r,i,a\; . \label{positivity}
\end{equation}
Unlike the previous constraints, positivity is by no means necessary
either from a mathematical or a physical point of view. However,
positive monads bundles are the ones which have been traditionally
studied in the
literature~\cite{Distler:1987ee,Kachru:1995em,Douglas:2004yv,Blumenhagen:2006wj,Anderson:2007nc}
and for this reason it is of interest to provide a comprehensive study
of their physical properties. They also offer a number of considerable
technical advantages. The building blocks of positive monads are
positive line bundles to which Kodaira vanishing applies, as discussed
above. This dramatically simplifies cohomology calculations. In
particular, one can show that positive monads only lead to families
but not to anti-families~\cite{Anderson:2008uw}. There is also a
helpful connection between positivity and stability. This can be made
explicit for cyclic CICYs, that is CICYs with $h^{1,1}(X)=1$. For
cyclic CICYs it can be shown~\cite{Anderson:2007nc} that the stable
monad bundles are precisely the positive ones. For non-cyclic CICYs
($h^{1,1}(X)>1$) the connection is less clear, but it seems probable
that all positive monads on such CICYs are stable. On the other hand,
it is clear from the examples in
Refs.~\cite{Anderson:2009nt,Anderson:2009sw} that non-positive monads
on non-cyclic CICYs can still be stable. We will come back to this
important observation towards the end of the paper.

In Ref.~\cite{Anderson:2008uw}, it was shown that the set of positive
monads bundles defined by~\eqref{V}, \eqref{BC}, satisfying the
conditions~\eqref{fex}, \eqref{rk}, \eqref{c1}, \eqref{anomaly} and
\eqref{positivity} on favourable CICYs, is finite. A complete
classification of all such bundles was given. It was found that
positive monad bundles exist on only 36 of the 4515 favourable
CICYs. The total number of bundles is 7118, of which 5680 have
structure group $SU(3)$, 1334 structure group $SU(4)$ and 104
structure group $SU(5)$. One of the main purposes of the present paper is to
analyze this class of positive monad bundles in detail and to extract
the physically interesting cases.\\

What do we require to carry out such an analysis of physical
properties? So far, our models only provide grand-unified theories
(GUTs) with gauge groups $E_6$, $SO(10)$ or $SU(5)$. For realistic
low-energy models we need to break these groups down to the standard
model group (possibly with additional $U(1)$ factors). This is done,
following the standard heterotic model building route, by quotienting
the geometric construction by a discrete symmetry and introducing
Wilson lines. We need to find, among our 7000 models, those which
allow for a discrete symmetry, $G$, freely-acting on $X$, which is
also respected by the bundle $V$. In more technical terms this means
that $V$ must admit a $G$-equivariant structure so that it descends to
a bundle $\hat{V}$ on the quotient $\hat{X}=X/G$. In addition, the
matter field content of the ``downstairs'' model, specified by
$\hat{X}$ and $\hat{V}$, must be that of the (supersymmetric) standard
model. We will carry all of this out in great detail in the following
sections.

\section{Equivariance and monad bundles} \label{equimon}

In this section we shall describe the quotienting process by which the
positive monad constructions over the favourable CICYs, as described
in the previous section, can be used to produce standard model like
theories.  It should be intuitively clear that checking for the
existence of discrete symmetries of $X$ and $V$, and analyzing the
downstairs field content, is not a straightforward task and would be
extremely laborious to carry out for all of our 7118 models. What we
need, therefore, are simple but necessary conditions, that we require
our bundles to satisfy, which can be used to cut down the number of
models which we need to consider. The physically promising ones can
then be subjected to a more detailed analysis. In this section we will
achieve this by first imposing the physical constraint that there
exists a freely-acting discrete symmetry of the Calabi-Yau manifold
which, when divided out, leads to three families of matter.  As we
will see, this simple constraint already provides a substantial
reduction in the number of models. Once we have reduced the number of
cases that we need to consider, we will then proceed to a more
detailed analysis of the remaining models.

\subsection{An initial physical constraint: obtaining three
  families}\label{initconst}

For a given CICY, what discrete symmetries are available to us in our
efforts to obtain a standard like model? An obvious necessary
condition for the existence of a freely-acting discrete symmetry $G$
of $X$ is that the Euler number $\chi (X)$ be divisible by the group
order $|G|$. This condition can be considerably refined by using the
indices discussed in Ref.~\cite{Candelas:1987du}. In addition to the
Euler number, the Euler characteristics $\chi({\cal N}^k\otimes TX^l)$
and Hirzebruch signatures $\sigma ({\cal N}^k\otimes TX^l)$ of the
``twisted'' bundles ${\cal N}^k\otimes TX^l$ (where we recall from
Eq.~\eqref{N} that ${\cal N}$ is the normal bundle of $X$) must be
divisible by the group order $|G|$ for all integers $k,l\geq 0$. It
was shown in Ref.~\cite{Candelas:1987du}, that is it sufficient to
consider the cases $(k,l)=(0,1),(1,0),(2,0),(3,0)$ for the Euler
characteristic and $(k,l)=(1,1)$ for the Hirzebruch signature without
loosing information. We have computed these indices for all of the
relevant CICYs, using the equations provided in
Ref.~\cite{Candelas:1987du}, and their common divisors in any one case
provides us with a list, $S(X)$, which must include the orders
of all freely-acting symmetry groups for $X$. It turns out that this
list is quite restrictive and in many cases provides precisely the
orders of the actual symmetries available. 

The Euler characteristic of the upstairs bundle $V$ and the downstairs
bundle $\hat{V}$ are related by $\chi (\hat{V})=\chi (V)/|G|$. Hence,
given that the Euler characteristic determines the number of families
present in the model, only bundles $V$ satisfying
\begin{equation}
  \chi (V)\in 3\, S(X)\;  \label{3fam}
\end{equation}
can lead to cases with three families. We have scanned all $7118$
bundles and find that only $91$, on five different CICYs, pass this
``three-family'' criterion. More specifically, these are:
\begin{itemize}
\item Five bundles on the quintic, $X=[\mathbb{P}^4\,|\,5\,]$, one each with
  structure groups $SU(5)$ and $SU(4)$ and the other three with
  structure group $SU(3)$. All but one require a group of order
  $|G|=25$ which can indeed be realized by the well-known
  freely-acting $\mathbb{Z}_5\times\mathbb{Z}_5$ symmetry of the
  quintic. A single bundle with structure group $SU(3)$ requires a
  group of order $5$, realized by either of the ${\mathbb{Z}_{5}}$'s
  mentioned above (see Table~\ref{tablequintic}).
\item Three bundles on $X=[\mathbb{P}^5\,|\,3\, 3\,]$, two with structure
  group $SU(4)$ and requiring symmetry orders $|G|=18$ and $|G|=12$
  respectively, and a third with structure group $SU(3)$, requiring
  $|G|=9$. Only the last case can be realized by a freely-acting
  symmetry, namely $G=\mathbb{Z}_3\times\mathbb{Z}_3$ (see Table~\ref{table33}).
\item One bundle with structure group $SU(3)$ on
  $X=[\mathbb{P}^7\,|2\,2\,2\,2\,]$ with required symmetry order
  $|G|=16$. This CICY has two freely-acting symmetries of order $32$,
  namely $\mathbb{Z}_8\times \mathbb{Z}_4$ and $\mathbb{H}\times\mathbb{Z}_4$,
  where $\mathbb{H}$ is the quaternionic group. Any subgroup $G$ of
  order $16$ of one of these two groups can be used (see Table~\ref{table2222}).
\item One $SU(3)$ bundle on the bi-cubic,
\begin{equation}
 X=\left[\begin{array}{c}\mathbb{P}^2\\\mathbb{P}^2\end{array}\right.\left|\begin{array}{c}3\\3\end{array}\right]\; ,
\end{equation} 
requiring a symmetry order $|G|=9$, which can be realized by a
freely-acting symmetry $G=\mathbb{Z}_3\times\mathbb{Z}_3$ (see Table~\ref{tablebicubic}).
\item 81 bundles with structure group $SU(3)$ on the tetra-quadric,
\begin{equation}
  X=\left[\begin{array}{c}\mathbb{P}^1\\\mathbb{P}^1\\\mathbb{P}^1\\\mathbb{P}^1\end{array}\right.\left|\begin{array}{c}2\\2\\2\\2\end{array}\right]\; ,
\end{equation}
all with required symmetry order $16$. This symmetry order is realized
by the freely-acting, non-Abelian, symmetry
$G=\mathbb{H}\times\mathbb{Z}_2$ (see Table~\ref{tabletetraquadric}).
\end{itemize}
While the criterion~\eqref{3fam} is necessary for a realistic model,
it is by no means sufficient. In particular, it is not yet clear
whether the above $91$ bundles $V$ indeed descend to bundles $\hat{V}$
on the quotient manifold, that is, if they admit an equivariant
structure under the corresponding symmetry groups $G$. In the rest of
this section we shall discuss how this can be decided. In the next
section we will then apply this knowledge to exhaustively study the
list given above.

\subsection{Equivariant Structures} \label{equivariance}

We wish to consider Calabi-Yau three-folds $X$ with a fixed point free
discrete group action, $G$. With the goal in mind of creating
three-generation heterotic models (via the use of Wilson lines), we
are interested in constructing new smooth three-folds $X/G$ with
$\pi_{1}(X/G)\neq 0$. That is, we will construct a multi-degree cover,
$q: X \rightarrow X/G$, of degree equal to the order of $|G|$. We now need
to discuss how to deal with the gauge bundle in such a quotient construction~\cite{mumford,knop,Donagi:2003tb}.

More precisely, we wish to find out if a bundle $V \stackrel{\pi}{\rightarrow} X$
descends to a bundle $\hat{V}$ on $X/G$, in the sense that $V\cong q^*\hat{V}$. For a bundle to descend
to the quotient space it is necessary that the automorphisms $G$
of $X$ ``lift" to automorphisms of the bundle $V$ over $X$. That is,
for each $g \in G$, there must exist a bundle morphism, that is a map $\phi_g: V
\rightarrow V$ which commutes with the projection $\pi : V\rightarrow X$
(i.e. $\phi_g \circ \pi = \pi \circ \phi_g$) and covers the action $g:X\rightarrow X$ on the base.
Such a lifting of the group action is called an {\em invariant structure} on $V$.
All this can be expressed by saying that the diagram
\begin{equation}
  \begin{array}{lllll}
  &V&\stackrel{\phi_g}{\longrightarrow}&V&\\
  \pi &\downarrow&&\downarrow&\pi\\
  &X&\stackrel{g}{\longrightarrow}&X&
 \end{array}
 \label{invalt}
\end{equation} 
commutes for all $g \in G$. Invariance
alone, however, is not enough for the bundle to descend to $X/G$. We
must further require that the $\phi_g$ satisfy a so-called {\em co-cycle
condition}, namely that for all $g,h \in G$, \beq\label{cocycle} \phi_g
\circ \phi_h =\phi_{gh} \;. \eeq

An invariant structure on $V$ with morphisms $\phi_g$ which satisfy the cocycle
condition is called an {\it equivariant structure} on $V$. If $V$ allows for such a set of morphisms it is said to admit an equivariant structure and, in this case, it descends to a bundle $\hat{V}$ on $X/G$. Moreover, the set of vector bundles on $\hat{X}$ is in one-to-one correspondence with the set of equivariant vector bundles on $X$. 

Direct sums, tensor products, and dualizations of equivariant bundles
are equivariant by the obvious induced representations \cite{FH}. As
we shall now see, if a bundle, $V$, is defined by a short exact
sequence of equivariant bundles, as in \eref{V}, then there is an
induced equivariant structure on $V$ as well.
\subsubsection{Equivariant structures on monad bundles}\label{glob_gen}
The simplest way to ensure that a monad bundle, $V$, admits an
equivariant structure is to build equivariant structures on the terms
$B$ and $C$ in the monad, \eref{V}, and then induce an equivariant
structure on $V$. Let $B$ and $C$ be sums of line bundles over $X$,
and \beq 0\rightarrow V\stackrel{\imath}{\rightarrow}
B\stackrel{f}{\rightarrow}C\rightarrow 0\; , \label{monad} \eeq where
$\imath$ is the injection, $f: B \rightarrow C$ is a bundle morphism
covering the identity on $X$ and $V\cong {\rm Ker}(f)$. Assume that
$B$ and $C$ admit equivariant structures under $G$ with associated
isomorphisms $\phi_{B,g}$ and $\phi_{C,g}$\footnote{In fact, it can be
  shown that if $B$ admits an equivariant structure, $C$ admits an
  invariant structure and the diagram \eqref{moninv2} commutes in its
  right-hand block, then it follows that $C$ admits an equivariant
  structure as well.}. Then, we have the following diagrams, one for
each $g \in G$, built from two exact sequences, which are written over
$X$ and $g(X) \approx X$, respectively.
\begin{equation}
 \begin{array}{ccccccccc}
  0&\rightarrow&V&\stackrel{\imath}{\longrightarrow}&B&\stackrel{f}{\longrightarrow}&C&\rightarrow&0\\
    &                    &\quad  & \quad    & \quad \downarrow\phi_{B,g}&& \quad\downarrow\phi_{C,g}&&\\
  0&\rightarrow&V&\stackrel{\imath}{\longrightarrow}&B&\stackrel{f}{\longrightarrow}&C&\rightarrow&0
\end{array}\label{moninv2}
\end{equation}  
If we ask that the right hand sides of these diagrams commute, so that
\beq\label{intertwining}\phi_{C,g}\circ f = f\circ \phi_{B,g} \;, \eeq
then we can construct bundle morphisms $\phi_{V,g}$ by setting
$\phi_{V,g}=\imath^{-1} \circ \phi_{B,g} \circ \imath$ where the
inverse map $\imath^{-1}$ is understood to be defined only over the
image of $V$ in $B$.

Given that $\phi_{B,g}$ and $\phi_{C,g}$ are bundle isomorphisms, the
Snake Lemma \cite{AG,AG2},
\begin{equation} \nonumber 0\rightarrow{\rm
    Ker}(\phi_{V,g})\rightarrow{\rm Ker}(\phi_{B,g})\rightarrow{\rm
    Ker}(\phi_{C,g})\rightarrow{\rm Coker}(\phi_{V,g})\rightarrow{\rm
    Coker}(\phi_{B,g})\rightarrow{\rm Coker}(\phi_{C,g})\rightarrow 0
  \;,
\end{equation}
then shows that the $\phi_{V,g}$ are bundle isomorphisms and, hence,
that $V$ is invariant. Furthermore, if the intertwining condition
\eref{intertwining} is satisfied, and $B$ and $C$ are equivariant
(that is, if $\phi_{B,g}$ and $\phi_{C,g}$ are isomorphisms satisfying the
cocycle condition \eref{cocycle}), then we can show that, in fact, $V$
is equivariant.  \beq
\phi_{V,g}\circ \phi_{V,h}=\imath^{-1} \circ \phi_{B,g} \circ \imath\circ\imath^{-1} \circ \phi_{B,h} \circ \imath \\=\imath^{-1} \circ \phi_{B,g} \circ \phi_{B,h} \circ \imath \\
=\imath^{-1} \circ \phi_{B,gh}\circ \imath \\
=\phi_{V,gh}~. \eeq

This can be summarized in the following\\[0.3cm]
{\bf Lemma :} Let $B$ and $C$ be $G$-equivariant bundles over $X$. The
bundle $V$ over $X$ is defined by the short exact
sequence~\eqref{monad}. If $f\circ\phi_{B,g}=\phi_{C,g}\circ f \;\forall\; g$, that
is, if the right-hand side of the diagrams~\eqref{moninv2} commute,
then $V$ is $G$-equivariant. In this case, the bundle isomorphisms
$\phi_{V,g}:V\rightarrow V$ covering $g$ on $X$ are given by
$\phi_{V,g}=\imath^{-1}\circ\phi_{B,g}\circ\imath$ and satisfy
$\phi_{V,g} \circ \phi_{V,h}=\phi_{V,gh}$.\\[0.3cm]

\subsubsection{Action on sections and globally generated line bundles}\label{sections}
To actually construct the necessary isomorphisms $\phi_{B,g}$ and
$\phi_{C,g}$, on the sums of line bundles $B$ and $C$, we will find it
useful to observe that, for the positive monads we consider in this
work, $B$ and $C$ are composed of line bundles generated by their
global sections. Using this fact, we find that we can build the
equivariant morphisms on $B$ and $C$ by constructing explicit actions
on the spaces of their global sections, $\Gamma(X,B)$ and $\Gamma(X,C)$.\\

We begin by describing how an equivariant structure on a bundle $U$ on $X$ 
induces actions on the section $s: X\rightarrow U$. We
have the following diagrams, one for each $g \in G$:
\begin{equation}
  \begin{array}{lllll}
  &U&\stackrel{\phi_g}{\longrightarrow}&U&\\
  s &\uparrow&&\uparrow&s'\\
  &X&\stackrel{g}{\longrightarrow}&X&
 \end{array} \;.
 \label{invalts}
\end{equation}
Demanding commutativity of these diagrams implies the existence of
maps $s\rightarrow s'$ between sections, which cover the action of $G$
on the base. Such maps, $\Phi_g:\Gamma (X,U)\rightarrow \Gamma(X,U)$
are evidently given by
\begin{equation}
 s'=\Phi_g(s)=\phi_g\circ s\circ g^{-1}\; . \label{sPhi}
\end{equation} 
Using the fact that the $\phi_g$ define an equivariant structure, and
thus satisfy \eqref{cocycle}, we have that $ \Phi_h \circ \Phi_g (s) = \phi_h \circ \phi_g \circ g^{-1}\circ h^{-1} = \phi_{hg} \circ s \circ (hg)^{-1} = \phi_{hg}(s)$ and hence,
\begin{equation}
\Phi_h\circ \Phi_g=\Phi_{hg}\; . \label{eqPhi}
\end{equation}
That is, written as an action on a basis of sections, the maps
$\Phi_g$ form a representation on $\Gamma(X, U)$ of the discrete group G.

We can now apply this discussion to $B$ and $C$ provided that they
both admit an equivariant structure.  This leads to representations
$\Phi_B:G\rightarrow\Gamma(X,B)$ and $\Phi_C:G\rightarrow\Gamma(X,C)$
of $G$ on the spaces of sections and the intertwining condition
\eref{intertwining}, can be re-written as
\beq \label{section_intertwining} \Phi_{C,g}\circ\tilde{f} =
\tilde{f}\circ \Phi_{B,g} \eeq where, $\tilde{f}:\Gamma(X,B)
\rightarrow \Gamma(X,C)$ is a polynomial map between the sections of
$B$ and $C$ (induced from the bundle morphism $f$ of \eref{monad}). It
is worth noting that choosing a monad map $\tilde{f}$ which satisfies
the intertwining condition \eref{section_intertwining} is equivalent
to choosing a section $f \in \Gamma(X, B^* \otimes C)$ that is invariant under the group action on
$\Gamma(X,B^*\otimes C)$ induced by the equivariant structures of $B$
and $C$ \cite{FH}.

For any equivariant vector bundle, the above discussion can be used to
determine the action of the group on the space of sections. However,
for globally generated bundles \cite{AG,AG2}, it is possible to reverse
the logic above. That is, given the section-wise mappings, $\Phi_g$,
it is possible to construct the full bundle morphisms $\phi_g$ and
hence an equivariant structure on $V$. This provides us with a practical and systematic method of constructing equivariant structures for globally generated bundles and we will apply this method to the bundles $B$ and $C$.
The first step involves choosing a suitable basis on the spaces $\Gamma(X,B)$ and $\Gamma(X,C)$ of sections, which is typically given by sets of vectors with homogeneous polynomial (or even monomial) entries. From Eq.~\eqref{sPhi} we should then carry out a $g$-actions on this basis, that is $s\rightarrow s\circ g^{-1}$, and combine them with morphisms $\phi_g$. If we can find suitable $\phi_g$ such that the combined linear transformations form a representation of $G$, then we have succeeded in constructing an equivariant structure.

To see how this works explicitly, let us discuss a simple toy example constructed from line bundles on $\mathbb{P}^1$. We consider a symmetry $G=\mathbb{Z}_2^{(1)}\times\mathbb{Z}_2^{(2)}$ of $\mathbb{P}^1$, with generators $g_1$ and $g_2$ acting on the homogeneous coordinates $[x_0,x_1]$, (with indices defined mod $2$) by
\begin{equation}
 g_1: x_k\rightarrow x_{k+1}\; ,\quad g_2:x_k\rightarrow (-1)^kx_k\; . \label{simpgen}
\end{equation}
(The fact that this symmetry is not freely acting is irrelevant for the purpose of illustrating our method.)
First, we consider the line bundle $L={\cal O}_{\mathbb{P}^1}(1)$. The space of sections for this line bundle is represented by linear polynomials in the homogeneous coordinates, so $\Gamma(\mathbb{P}^1,L)$ has a basis $\{x_0,x_1\}$. From Eq.~\eqref{sPhi}, the $g$-action of the two generators~\eqref{simpgen} in this basis is described by the matrices
\begin{equation}
 g_1^{(1)}=\left(\begin{array}{cc}0&1\\1&0\end{array}\right)\; ,\quad
 g_2^{(1)}=\left(\begin{array}{rr}1&0\\0&-1\end{array}\right)\; . \label{gaction}
\end{equation} 
Each of these matrices generates $\mathbb{Z}_2$, so $L$ has an equivariant structure under $\mathbb{Z}_2^{(1)}$ and $\mathbb{Z}_2^{(2)}$ (choosing the bundle morphism to be the identity). However, the two matrices do no commute, so they do not, by themselves, represent $\mathbb{Z}_2^{(1)}\times\mathbb{Z}_2^{(2)}$. Can this be fixed by combining the $g$-action with a suitably chosen bundle morphism? We have ${\rm Hom}(L,L)\cong \Gamma(\mathbb{P}^1,{\cal O}_{\mathbb{P}^1})\cong\mathbb{C}$ so the bundle morphisms are parametrized by a single complex number. It acts on sections by simple multiplication. This means, we are free to modify the matrices~\eqref{simpgen} by multiplying them with a complex number each but whichever numbers we choose, the matrices will still be non-commuting. Hence, a $\mathbb{Z}_2^{(1)}\times\mathbb{Z}_2^{(2)}$ equivariant structure on $L$ does not exist.

Next, we consider $L^{\oplus 2}={\cal O}_{\mathbb{P}^1}(1)\oplus{\cal O}_{\mathbb{P}^1}(1)$. The sections $\Gamma(\mathbb{P}^1,L^{\oplus 2})$ of this bundle can be described by two-dimensional vectors with linear polynomial entries, so the space is four-dimensional with basis
\begin{equation}
\left\{ \left(\begin{array}{r}x_0\\0\end{array}\right)\; ,\;
 \left(\begin{array}{r}x_1\\0\end{array}\right)\; ,\;
  \left(\begin{array}{r}0\\x_0\end{array}\right)\; ,\;
   \left(\begin{array}{r}0\\x_1\end{array}\right)\;\right\} .\label{l2basis}
\end{equation}   
The available bundle morphisms are ${\rm Hom}(L^{\oplus 2},L^{\oplus 2})\cong\Gamma(\mathbb{P}^1,{\cal O}_{\mathbb{P}^1})^{\oplus 4}\cong\mathbb{C}^{\oplus 4}$ and are explicitly given by complex $2\times 2$ matrices which act linearly on the two-dimensional polynomial vectors (while the matrices~\eqref{gaction} act ``within each component''of these vectors). The freedom of having arbitrary two-dimensional matrices available can now be used to ``fix'' the non-commutativity of $g_1^{(1)}$ and $g_2^{(1)}$.
Relative to the basis~\eqref{l2basis}, we can write down the following $4\times 4$ matrices
\begin{equation}
\Phi_{g_1}=g_1^{(1)}\otimes g_1^{(1)}\; ,\quad \Phi_{g_2}=g_2^{(1)}\otimes g_2^{(1)}\; , \label{l2equiv}
\end{equation}
where the first matrix in each tensor product corresponds to the bundle morphism and the second matrix is the $g$-actions~\eqref{gaction}. These matrices represent the action of an invariant structure on the sections. Moreover, $\Phi_{g_1}$ and $\Phi_{g_2}$ both square to one and commute and, hence, they define a representation of $\mathbb{Z}_2^{(1)}\times\mathbb{Z}_2^{(2)}$ on the sections $\Gamma(\mathbb{P}^1,L^{\oplus 2})$. This means that, unlike a single line bundle ${\cal O}_{\mathbb{P}^1}(1)$, the rank two bundle ${\cal O}_{\mathbb{P}^1}(1)\oplus{\cal O}_{\mathbb{P}^1}(1)$ does admit a $\mathbb{Z}_2^{(1)}\times\mathbb{Z}_2^{(2)}$ equivariant structure.

For a further, more elaborate example of an equivariant structure, see section \ref{the_model}.

\subsection{Spectra Downstairs: The group action on cohomology}\label{chars}

If a bundle, $V$, admits a $G$-equivariant structure, then there is a
natural action of the group $G$ on the cohomology groups $H^i(X,
V)$. Since the cohomology of $\hat{V}$ and its wedge powers encodes
the particle content of our low energy effective theory, we are
interested in determining $H^i(X/G,\hat{V})$ and its relationship to
$H^i(X, V)$. As has been discussed in detail in \cite{Donagi:2004su},
the cohomology of $\hat{V}$ on $X/G$ is precisely the $G$-invariant part
of the cohomology on $X$ (where invariance is relative to the group
action induced from the equivariant structure, as discussed for the
case of sections in section \ref{sections}). That is,
\beq\label{coho_descent} H^i(X/G, \hat{V})=H^{i}_{inv} (X, V) \;.\eeq

In the presence of Wilson lines, the physical spectrum changes still
further. We have considered vector bundles with structure group ${\cal G}=SU(n)$ for $n=3,4,5$, so that the commutant within $E_8$
will be a GUT symmetry, $E_6$, $SO(10)$, or $SU(5)$,
respectively. The GUT symmetry of this four-dimensional effective
theory can then be broken with Wilson lines to a group ${\cal H}$ which contains
the standard model gauge group. To analyse the particle content in the presence of the Wilson line we
should decompose the ${\bf 248}$ adjoint representation of $E_8$ under the sub-group
${\cal G}\times G\times{\cal H}\subset E_8$. Formally, this decomposition can be written as
\begin{equation}
 {\bf 248}\rightarrow\bigoplus_a ({\cal R}_a,R_a,{\cal S}_a)\; , \label{decomp}
\end{equation}
where a triple $({\cal R},R,{\cal S})$ denotes a representation of  ${\cal G}\times G\times{\cal H}$.
A Wilson line, $W$, is a flat bundle on $X/G$ induced via the
embedding of the discrete group $G$, which is the fundamental group of
our quotiented manifolds, into the visible sector gauge group,
$H$. The complete ``downstairs" bundle is now given by
\beq
 U=\hat{V}\oplus W \;,
\eeq
We denote by $U_a$ the bundle associated to $U$ in the representation $({\cal R}_a,R_a)$ and by $V_a$ the bundle associated to $V$ in the representation ${\cal R}_a$. The multiplets transforming in the representation ${\cal S}_a$ under the low-energy group ${\cal H}$ are given by the ``downstairs" cohomology $H^1(X/G,U_a)$.
For the purpose of calculating these cohomologies it is useful to relate them to ``upstairs" cohomologies.
The relevant relation is
\begin{equation}\label{wilson_inv}
 H^1(X/G,U_a)=(H^1(X,V_a)\otimes R_a)_{\rm inv}\; ,
\end{equation}
where the subscript ``inv" indicates the part which transforms as a
singlet under the discrete group $G$.  So, in practice, once the
upstairs cohomologies $H^1(X,V_a)$ have been found we need to
determine their representation content under $G$, tensor with the
$G$-representations $R_a$, and then extract the $G$-singlets.

Finding the $G$-representation content of $H^1(X,V_a)$ is most
conveniently done by the introduction of characters \cite{FH}.  Let us
denote by $R_{H^1(X,V_a)}$ the $G$-representation of $H^1(X,V_a)$ and
by $\chi_{H^1(X,V_a)}$ the associated character. These characters can
be computed from the equivariant structures on $\Gamma(X,B)$ and
$\Gamma(X,C)$ as will be shown in the next sub-section. Further, let
$R_p$ be a complete set of irreducible $G$-representations with
associated characters $\chi_p$. It is well-known that these
characters are orthonormal under the scalar product
\beq\label{chi_prod} (\chi_p, \chi_q)=\frac{1}{|G|}\sum_{g
  \in G}\chi_p(g)\overline{\chi}_q(g)\; .  \eeq
Parametrizing the representation content of $R_{H^1(X,V_a)}$ by
\begin{equation} \label{thellamabadger}
 R_{H^1(X,V_a)}=\bigoplus_p n_a^p R_p\; ,
\end{equation}
it is clear that the integers $n_a^p$ can be extracted from
\begin{equation}\label{char_sub}
 n^p_a=(\chi_p,\chi_{H^1(X,V_a)})\; .
\end{equation}

\subsubsection{The equivariant action on $H^1(X,V)$ and $H^1(X,\wedge^2 V)$}\label{h1v}

While the discussion of the previous subsection holds in general, it
is useful to consider explicitly the specific case of a positive monad
bundle, $V$, and the cohomology $H^1(X,V)$. Taking the long exact sequence in cohomology
associated to \eref{V} we obtain
\beq
 0 \rightarrow H^0(X,V) \rightarrow H^0(X,B)\stackrel{\tilde{f}}{\rightarrow} H^0(X,C) \rightarrow H^1(X,V)
\rightarrow H^1(X,B) \rightarrow \ldots
\eeq
For a positive monad, $H^{i}(X,B)=H^{i}(X,C)=0$ for $i>0$ by the Kodaira vanishing theorem, and
for a stable bundle $H^0(V)=0$. Thus, for a stable bundle, defined by
a positive monad, the only non-vanishing cohomology of $V$ is
\beq\label{h1}
  H^1(X,V)\cong\frac{\Gamma(X,C)}{\tilde{f}(\Gamma(X,B))} \;.
\eeq
The map $\tilde{f}$, is the induced map on cohomology associated to the
bundle map $f$ in \eref{V}. 

We can use the intertwining condition \eref{section_intertwining} to
relate the representations $\Phi_{B,g}, \Phi_{C,g}$, of the equivariant structure acting on $\Gamma(X,B)$
and $\Gamma(X,C)$. Since $\tilde{f}$ is injective, the condition \beq
\Phi_{C,g}\circ\tilde{f} =\tilde{f}\circ\Phi_{B,g} \eeq can be inverted on
$\textnormal{Im}(\tilde{f})$ to obtain \beq
\Phi_{C,g}|_{~\textnormal{Im}(\tilde{f})}=\tilde{f}\circ\Phi_{B,g}\circ\tilde{f}^{-1}|_{~\textnormal{Im}(\tilde{f})}~.
\eeq
This relation means that the restriction of $\Phi_{C,g}$ to $\textnormal{Im}(\tilde{f})$ is
equivalent (as a representation) to $\Phi_{B,g}$ \cite{FH}. Hence, we have
\begin{equation}\label{h1diff}
 \chi_{H^1(X,V)}=\chi_{\Gamma(X,C)}-\chi_{\Gamma(X,B)}\; .
\end{equation} 
The characters $\chi_{\Gamma(X,C)}$ and $\chi_{\Gamma(X,B)}$ can be computed from the explicit representation matrices $\Phi_{B,g}$ and $\Phi_{C,g}$.\\

In order to illustrate how to compute the representation content of a
space of sections, we shall return to the toy example which we
introduced at the end of sub-section~\ref{sections}. The group
$\mathbb{Z}_2\times\mathbb{Z}_2$ has four irreducible representations,
all one-dimensional, which we denote by $R_{(\pm 1,\pm 1)}$ with
characters $\chi_{(\pm 1,\pm 1)}$. Since we are dealing with an
Abelian group every element of $\mathbb{Z}_2\times\mathbb{Z}_2$ forms
its own conjugacy class, so that characters are specified by four
values. For the characters of the four irreducible representations we
have
\begin{equation}
\begin{array}{lllllll}
 \chi_{(1,1)}&=&(1,1,1,1)&,&\chi_{(1,-1)}&=&(1,1,-1,-1)\\
 \chi_{(-1,1)}&=&(1,-1,1,-1)&,&\chi_{(-1,-1)}&=&(1,-1,-1,1)
\end{array} 
\end{equation} 
By taking the traces of the matrices~\eqref{l2equiv} it is easily seen that the character for the $G$-representation on $\Gamma(\mathbb{P}^1,L^{\oplus 2})$ is given by
\begin{equation}
 \chi_{\Gamma(\mathbb{P}^1,L^{\oplus 2})}=(4,0,0,0)\; .
\end{equation}
From Eqs.~\eqref{chi_prod}--\eref{char_sub}, this means every irreducible representation of $\mathbb{Z}_2\times\mathbb{Z}_2$ is contained in $\Gamma(\mathbb{P}^1,L^{\oplus 2})$ precisely once.\\

To compute the number of Higgs multiplets we will also need to deal with equivariant cohomologies of $\wedge^2V$.
We start by writing down the exterior power sequence for $\wedge^2V$ associated to the monad sequence~\eqref{monad}. Splitting up this exterior power sequence by introducing a co-kernel $K$ we have
\bea\label{extp}
\nonumber 0 &\to& \wedge^2 V \to \wedge^2 B \to K \to 0 \\
0 &\to& K \to B \otimes C \to S^2 C \to 0 \ .
\eea
The associated long exact sequence contains
\bea 
\dots &\to& H^1(X,\wedge^2 B) \to H^1(X, K_3) \to H^2(X, \wedge^2 V) \to H^2(X, \wedge^2
B) \to \ldots\\
\label{secondso10}
\ldots &\to& H^0(X, B \otimes C) \stackrel{F}{\longrightarrow} H^0(X, S^2 C)
\to H^1(X, K_3) \to H^1(X, B \otimes C) \to \ldots \; . 
\eea 
For positive monads we have $H^1(X, \wedge^2 B) = H^2(X, \wedge^2 B)=0$ so that $H^1(X, K) \cong H^2(X, \wedge^2V)$ from the first long exact sequence above. Further, $H^1(X, B \otimes C) \cong 0$ for positive monads and combined with the second long exact sequence above this implies
\bea
 \label{thirdso10}
  H^1(X, \wedge^2 V) \cong\frac{\Gamma(X,S^2 C)}{\tilde{F}\left(\Gamma(X,B\otimes C)\right)}\; .
\eea
Here, $\tilde{F}$ is induced from the monad map $f$. It turns out that this map is typically not injective (in fact, in relevant examples the dimension of $\Gamma(X,B\otimes C)$ is larger than that of $\Gamma(X,S^2 C)$), but we still have the relation
\begin{equation}
 \chi_{H^1(X, \wedge^2 V)}=\chi_{\Gamma(X,S^2 C)}-\chi_{{\rm Im}(\tilde{F})}=\chi_{{\rm Coker}(\tilde{F})}\; ,
 \label{Higgschar}
\end{equation} 
between the various characters. It has been
shown~\cite{Anderson:2008uw} that ${\rm Coker}(\tilde{F})=0$ for
generic choices of the monad map $f$ and, hence, that the number of
Higgs multiplets vanishes generically. However, it can also be shown
that special choices for $f$ can lead to a non-vanishing number of
Higgs multiplets~\cite{Donagi:2004qk,Anderson:2007nc}. This is of
particular importance in the present context since the monad map is
restricted by the intertwining condition~\eqref{section_intertwining}
and is, hence, ``special" by construction. Our first task, therefore,
is to compute ${\rm Coker}(\tilde{F})$ for a map $\tilde{F}$ which is
induced from an monad map that obeys the intertwining condition, but
which is otherwise generic. If the result is non-zero, the appearance
of Higgs multiplets would be linked to the existence of an equivariant
structure of the monad and would, in this sense, be
automatic. Otherwise, one might want to specialise the monad map $f$
further, beyond what is dictated by the intertwining
condition~\eqref{section_intertwining}, until Higgs multiplets arise.
In either case, we then need to compute the associated character $
\chi_{H^1(X, \wedge^2 V)}$ from Eq.~\eqref{Higgschar} and determine
the number of surviving ``downstairs" Higgs multiplets following the
discussion of the previous sub-section.

\subsection{Further simple tests for equivariant structures}\label{simple_tests}

In the next section we shall discuss which of the bundles in the list
of Section \ref{initconst} admit equivariant structures. Before we do
this, however, it is worth observing that, now that we have an
understanding of equivariant structures, we can spot a few more simple
topological conditions which must be satisfied by a bundle $V$.

If $V$ is $G$-equivariant then it is isomorphic to the pull-back of a
bundle $\hat{V}$ on the quotient space $X/G$ (i.e. $V\approx
q^*(\hat{V})$ where $q: X \rightarrow X/G$). As a result of the simple
properties of Chern classes and pull-back maps, \beq
c_i(q^*(\hat{V}))=q^*(c_i(\hat{V}))~, \eeq we can make several
restrictions on the Chern classes of a bundle $V$ if it is to be the
pull-back of a bundle on $X/G$. Hence, we can rule out even more
bundles on the grounds that they admit no equivariant structure.

If we denote the generators of $H^2(X, \mathbb{Z})$ by $J_r$, where
$r=1\ldots h^{1,1}(X)$, and let $\hat{J}_{a}$ be the generators of
$H^2(X/G, \mathbb{Z})$, with $a=1,\ldots h^{1,1}(X/G)$, then we can
express the relationship between these sets of basis forms as
\beq\label{Js} q^*(\hat{J}_a)=K^{r}_{a} J_r \;, \eeq for some matrix
of integers $K^{r}_{a}$. Thus for a bundle $V=q^*(\hat{V})$, \beq
c_1(q^*(\hat{V}))^r J_r=c_1(q^*(\hat{V}))\\
=q^*(c_1(\hat{V})^a \hat{J}_a)=c_1(\hat{V})^a K^{r}_{a} J_r.  \eeq
Hence, the coefficients of the first Chern classes of $V$ and
$\hat{V}$ are related as follows. \beq c_1(q^*(\hat{V}))^r=c_1(\hat{V})^a
K^{r}_{a}~. \eeq

To illustrate this, let us consider the case of the quintic
$[\mathbb{P}^4\,|\,5]$, with a freely acting $\mathbb{Z}_5 \times
\mathbb{Z}_5$ symmetry. In this case, the Picard groups of both $X$
and $X/G$ are one-dimensional, and hence there is only a single
integer, $K^{1}_{1}$ in \eref{Js}, which is to be determined. We
denote the generator of $H^2(X, \mathbb{Z})$ by $J$ (the dual to the
divisor class $H$, the restriction of the hyperplane from
$\mathbb{P}^4$). It is straightforward to show that $5H$ is the
pullback of the generator of the second homology of $X/G$. Thus, if we
define $\hat{H}$ to be the basis of the divisor class on $X/G$, a simple analysis of the group action yields that

 \beq q^*(\cO(\hat{H}))=\cO(5H) \;. \eeq 
Hence,
$5J=c_1(q^*(\cO(\hat{H})))=q^*(c_1(\cO(\hat{H})))= q^*(\hat{J})$ and
it is clear $J$ is related to $\hat{J} \in H^2(X/G,\mathbb{Z})$ via
\beq q^*(\hat{J})=K^{1}_{1}J=5J \;. \eeq As a result, if a bundle, $V$, on
the quintic is to admit an $\mathbb{Z}_5 \times
\mathbb{Z}_5$-equivariant structure, it must be the case that
$c_1(V)=5m$ for some integer, $m$.

Using the relationship between $\hat{J}_a$ and $J_r$, we can derive
further conditions on the second and third Chern classes of
equivariant bundles. To begin, we note that the triple intersection numbers, $\hat{d}_{abc}$ of $X/G$ can be determined via
 \beq\label{triple}
 \int_X q^*(\hat{J}_a)\wedge q^*(\hat{J}_b) \wedge q^*(\hat{J}_c) =\int_X
q^*( \hat{J}_a\wedge \hat{J}_b\wedge \hat{J}_c) =|G| \int_{X/G} (\hat{J}_a\wedge \hat{J}_b\wedge \hat{J}_c) \;.\eeq
Expanding out both sides and using Eq.~\eref{Js} we find
\beq\label{intersec_down}
 K_{a}^{r}K_{b}^{s}K_{c}^{t}d_{rst}=|G| \int_{X/G} (\hat{J}_a\wedge \hat{J}_b\wedge \hat{J}_c)=|G|\hat{d}_{abc}
 \eeq
 where $d_{tsr}=\int_X J_t \wedge J_s \wedge J_r$ are the triple intersection numbers of $X$.
  
Next we note that the second Chern class must
satisfy
 \beq \int_X c_2(q^*(\hat{V})) \wedge q^*(\hat{J}_a) =\int_X
q^*((c_2(\hat{V}))\wedge \hat{J}_a) =|G| \int_{X/G} c_2(\hat{V})\wedge
\hat{J}_a  \;.
 \eeq 
 Therefore, expanding the integrands in a basis of
harmonic forms, and again using Eq.~\eref{Js}, we have 
\beq\label{c2cond}
c_2(q^*(\hat{V}))^{ts} d_{tsr} K^{r}_a\equiv c_2(q^*(\hat{V}))_r
K^{r}_a=|G|\int_{X/G} c_2(\hat{V})\wedge
\hat{J}_a~.
 \eeq 
From Eqs.~\eref{triple} and \eref{intersec_down} we can expand this further as
 \beq\label{c2cond2}
 c_2(q^*(\hat{V}))_rK^{r}_a=|G|c_2(\hat{V})^{cb}\hat{d}_{cba}=|G|c_{2}(\hat{V})_{a}=|G|m_a~,
 \eeq
where $m_a$ is an integer. This expression constrains the second Chern class of an equivariant bundle and is listed in Table \ref{constraints_table} for the specific manifolds and symmetries considered in this work. Furthermore, \eref{c2cond2} can be strengthened still further by requiring that both $c_2(q^*(\hat{V}))$ and $c_2(\hat{V})$ be integrally normalized. 

As an example, consider the quintic with $G=\mathbb{Z}_5 \times
\mathbb{Z}_5$. Using the fact that $K^{1}_{1}=5$, the condition \eref{c2cond2} becomes
$c_2(q^*(\hat{V}))_{1}=5m$ for some integer $m$. However, we must be careful and note that $J^2/5$ is an element of the integer cohomology, $H^4(X,\mathbb{Z})$, while using the intersection numbers, $\hat{d}$ from \eref{triple}, we see that $\hat{J}^{2}/{25} \in H^4(\hat{X}, \mathbb{Z})$. However, the pull-back does not preserve such normalization, that is, $q^*(\hat{J}^{2}/{25})=J^2$. Accounting for this difference, we have the stronger condition, $c_2(V)_{1}=25n$. That is, $c_2(V)$ must be divisible by $25$ if $V$ is to admit a $\mathbb{Z}_5 \times \mathbb{Z}_5$-equivariant structure. For a detailed discussion of integral cohomology and torsion in this context, see Ref. \cite{Braun:2007xh}.

Finally, for the third Chern class of the bundle, we have \beq \int_X
c_3(q^*(\hat{V}))=\int_X q^*(c_3(\hat{V}))=|G| \int_{X/G} c_3(\hat{V})
\eeq and thus we re-derive the well-known constraint that
$c_3(q^*(\hat{V}))=|G| n$ for some integer $n$. That is, the index of
the bundle must be divisible by the order of the group if it is to
have a $G$-equivariant structure \cite{GSW, Distler:1987ee}.

Rewriting these conditions, we find that the topological constraints
on $G$-equivariant vector bundles on $X$ are determined, in terms of a
given set of integers $K^{r}_a$, defined in \eref{Js}, to be \bea
c_1(V)^r=l^a K^{r}_{a} \\
c_2(V)_r K^{r}_a=|G| m_a\\
c_3(V)= |G| n \eea for some integers, $l^a, m_a$ and $n$. The
coefficients above are correct for the Chern classes as defined in
equations \eref{c1},\eref{anomaly} and \eref{c3}.

\section{The results}\label{results}
The positive monad bundles that pass the ``three-generation" test, \eref{3fam}, have been described in the previous section and are listed in Appendix \ref{posmonad}. Before we proceed to analyze
which of these bundles can give rise to physically relevant heterotic
theories, we should ask whether this list can be obviously reduced by
any other simple criteria.

From the results of Section \ref{simple_tests}, there are a series of simple checks to perform on the Chern classes of positive monad bundles $V$ in order to decide if they (or their constituent sums of line bundles $B$ and $C$) admit equivariant structures. The conditions for each of the manifolds listed in Section \ref{setup}, are given in the table below.
\begin{table}[h]
\begin{center}
\begin{tabular}{|l|l|l|l|l|l|l|}\hline
$X^{h^{1,1},h^{2,1}}$&$\tiny{\hat{X}^{h^{1,1},h^{2,1}}}$&$G$&$K^{r}_{a}$&$c_1(V)$&$c_2(V)$&$c_3(V)$\\\hline\hline

$\tiny{[\mathbb{P}^4\, |\, 5\, ]^{1,101}}$ &$\hat{X}^{1,21}$& $\tiny{\mathbb{Z}_5}$ & $\tiny{1}$ &$\tiny{l}$ & $\tiny{5m}$& $\tiny{5n}$\\\hline

$\tiny{[\mathbb{P}^4\, |\, 5\, ]^{1,101}}$ &$\hat{X}^{1,5}$& $\tiny{\mathbb{Z}_5 \times \mathbb{Z}_5}$ & $\tiny{5}$ & $\tiny{5l}$ & $\tiny{25m}$& $\tiny{25n}$\\\hline

$[\tiny{\mathbb{P}^5\, |\, 3\, 3\, }]^{1,73}$ &$\hat{X}^{1,9}$& $\mathbb{Z}_3 \times \mathbb{Z}_3$ & $\tiny{3}$ & $\tiny{3l}$ &$\tiny{9m}$&$\tiny{9n}$\\ \hline

$\tiny{[\mathbb{P}^7\,|\,2\, 2\, 2\, 2\,]^{1,65}}$ &$\hat{X}^{1,5}$& $\tiny{\mathbb{H} \times \mathbb{Z}_2}$ & $\tiny{2}$  &$2l$&$\tiny{8m}$ & $\tiny{16n}$\\\hline

$\tiny{[\mathbb{P}^7\,|\,2\, 2\, 2\, 2\,]^{1,65}}$ &$\hat{X}^{1,5}$& $\tiny{\mathbb{Z}_8 \times \mathbb{Z}_2}$ & $\tiny{2}$  &$2l$&$\tiny{8m}$ & $\tiny{16n}$ \\\hline

$\tiny{[\mathbb{P}^7\,|\,2\, 2\, 2\, 2\,]^{1,65}}$&$\hat{X}^{1,5}$ & $\tiny{\mathbb{Z}_4 \times \mathbb{Z}_4}$ & $\tiny{2}$  &$2l$  &$\tiny{8m}$ & $\tiny{16n}$\\\hline

$\tiny{\left[\begin{array}[c]{c}\mathbb{P}^2\\\mathbb{P}^2\end{array}
\left|\begin{array}[c]{ccc}3 \\3
\end{array}
\right.  \right]^{2,83}\;}$ &$\hat{X}^{2,11}$&$\tiny{\mathbb{Z}_3 \times \mathbb{Z}_3}$& $\tiny({\ba{cc}
3&0\\
1&1
\ea
)}$& $\tiny{(3l_1+l_2,l_2)}$&$\tiny{c_2(V)_r K^{r}_a=9m_a}$& $\tiny{9n}$ \\\hline

$\tiny{\left[\begin{array}{c}\mathbb{P}^1\\\mathbb{P}^1\\\mathbb{P}^1\\\mathbb{P}^1\end{array}\right.\left|\begin{array}{c}2\\2\\2\\2\end{array}\right]^{4,68}\;}$&$\hat{X}^{1,5}$& $\tiny{\mathbb{H} \times \mathbb{Z}_2}$ & $\tiny{(1,1,1,1)}$&$\tiny{l(1,1,1,1)}$&$\tiny{c_2(V)_r K^{r}_1=16m}$&$\tiny{16n}$\\\hline
\end{tabular}
\caption{Conditions on the Chern classes of $G$-equivariant vector bundles on $X$. Above, $\hat{X}=X/G$ is the quotient manifold and $m,n,l_i$ are integers.}
\label{constraints_table}
\end{center}
\end{table}
Using the results of Table \ref{constraints_table} we immediately
discover that the vast majority of the bundles listed in Appendix
\ref{posmonad} do not admit equivariant structures. For example, of
the data set of $E_6$ bundles listed in appendix \ref{posmonad}, $81$
of these models arise on the tetraquadric manifold in
$\mathbb{P}^1\times\mathbb{P}^1\times\mathbb{P}^1\times\mathbb{P}^1$. Applying
the constraint on the second Chern class of $V$ given in Table
\ref{constraints_table}, we find that none of these bundles can
descend to $X/{(\mathbb{H}\times \mathbb{Z}_2)}$. However, some of the
$81$ can admit equivariant structures for the quaternionic symmetry,
$\mathbb{H}$ alone (see appendix \ref{quat_eq} for an example). Since
we are interested in three generation models, we will not consider this
set further.

The single $SO(10)$ model resulting from the scan of Section
\ref{initconst} is also ruled out immediately. The bundle, which is on
the quintic, is defined via the short exact sequence, \beq 0
\rightarrow V \rightarrow \cO_X(2)^{\oplus 3}\oplus \cO_X(1)^{\oplus 4}
\rightarrow \cO_X(4) \oplus\cO_X(3)^{\oplus 2} \rightarrow 0~~.  \eeq We
find that $c_2(V)=-45$ which is not divisible by $25$ as required by
Table \ref{constraints_table}. Hence this bundle does not admit an
equivariant structure and will be of no use to us in model building.

The remaining four $E_6$ models and a single $SU(5)$ model survive our
preliminary checks. We shall see in the following sections that all of
these do admit equivariant structures and produce three-generation
models after quotienting $X$ by $G$. 

\subsection{The $E_6$ Models}
\subsubsection{$E_6$ GUTs and colored triplets}
In this section, we demonstrate that, as expected by the standard
arguments \cite{GSW,Breit:1985ud}, when breaking the $E_6$ GUTs to the
Standard Model using Wilson lines, we will always have colored triplet
Higgs. As a result, these $E_6$ models are less interesting than the
$SO(10)$ or $SU(5)$ models, and are of limited use without further
fine tuning to split the doublet-triplet.

A key feature of $E_6$ GUTs, as opposed to $SO(10)$ or $SU(5)$, is
that the fermions and Higgs multiplets all reside in the same
representation, namely the {\bf 27} (for all positive monad bundles,
$h^1(V^*)=0$ and hence the $\overline{{\bf 27}}$ anti-families all
vanish).  As usual, we will break $E_6$ with Wilson lines to obtain
the standard model symmetry (with extra $U(1)$ gauge factors). Let
${\cal S}_a$ denote the representations of the low-energy group ${\cal H}$ contained in
${\bf 27}$. Then, as explained in Ref.~\cite{GSW}, the number $n_a^{\pm}$, of massless positive and
negative chirality fermions transforming as ${\cal S}_a$ under the low-energy
gauge group satisfies: ${\rm Ind}({\cal S}_a) = n_a^+ - n_a^- = N_{gen} \; \forall
\;a$. Thus we expect massless colour-triplets to be present in the low-energy spectrum.

\subsubsection{An example $SU(3)$ bundle} \label{E6example} To confirm
the expectation above, we give as an example the single $E_6$
three-generation model available on $[\mathbb{P}^{5}\,|\, 3\,3\,]$, with
$\mathbb{Z}_3\times \mathbb{Z}_3$ Wilson lines. 
\beq\label{E6eg} 0\rightarrow V \rightarrow \cO_X(1)^{\oplus 6}
\stackrel{f}{\rightarrow} \cO_X(2)^{\oplus 3} \rightarrow 0 \eeq

The bundle satisfies $\textnormal{Ind}(V)=-27$ and hence, under the
$\mathbb{Z}_3\times \mathbb{Z}_3$ symmetry available on $X$, can produce a
three-generation model. Labeling the coordinates of $\mathbb{P}^5$,
as $\left(x_i,y_i\right)$, where $i=0,\ldots 2$, the group action is
defined by
\begin{equation}
\begin{array}{llll}
{\mathbb{Z}_3}^{(1)}&:& g_1:x_k \to x_{k+1},&g_1:y_k \to y_{k+1}~~\\
{\mathbb{Z}_3}^{(2)}&:& g_2:x_k \to \alpha^{-k}x_{k},&g_2:y_k \to
\alpha^{k} y_{k}~,
\end{array}
\end{equation}
where $\alpha^{3}=1$ is a primitive root of
unity. For the monad bundle in \eref{E6eg}, an equivariant structure
can be defined for $B=\cO(1)^{\oplus 6}$ and $C=\cO(2)^{\oplus 3}$, as
described in Section \ref{sections}, by the following actions on the
spaces $\Gamma(X,B)$ and $\Gamma(X,C)$: \beq
\Phi_{B,g_{i}}(s)=\phi_{B,g_{i}} \circ s \circ
g_{i}^{-1},~~~~~\Phi_{C,g_{i}}(s)=\phi_{C,g_{i}} \circ s \circ
g_{i}^{-1} \eeq Here \beq \phi_{B,g_1}=\left(
\begin{array}
[c]{cc}%
\gamma_1 &0 \\
0& \gamma_1 
\end{array}
\right),~
\phi_{B,g_2}=\left(
\begin{array}
[c]{cc}%
\gamma_2 &0 \\
0& \gamma_2 
\end{array}
\right),~
\phi_{C,g_1}=\left(\gamma_1
\right),~\phi_{C,g_2}=\left(\gamma_{2}^{2}\right)
\eeq
with
\beq
\gamma_1=\left(
\begin{array}
[c]{ccc}%
0 & 1&0 \\
0 & 0&1\\
1&0& 0 
\end{array}
\right),~~\gamma_2=\left(
\begin{array}
[c]{ccc}%
1 & 0&0 \\
0 & \alpha&0\\
0&0& \alpha^2 
\end{array}
\right) \;.  \eeq With these explicit matrices in hand, we can write
out $\Phi_{B,g_{i}},\Phi_{C,g_{i}}$ as matrices acting on bases of
monomials. Next, we can compute their characters, \beq
\chi^{i}_{\Phi}(g)=\textnormal{tr}(\Phi^{i}_g) \;,\eeq and use
\eref{chi_prod} to find the explicit decompositions,
\eref{thellamabadger}, of the representations in terms of irreducible
representations of $\mathbb{Z}_3 \times \mathbb{Z}_3$. We find that
$\Phi_B$ contains $h^0(X,B)/9$ copies, and $\Phi_C$ $h^0(X,C)/9$ copies,
of the regular representation. As a result, by \eref{h1} and
\eref{h1diff}, we find that, $H^1(X,V)$ carries $27$ copies of the
regular representation. Combining this with any Wilson line that could
break $E_6$ to the gauge group $SU(3)\times SU(2)\times U(1) \times
U(1) \times U(1)$ results in a low energy particle spectrum containing
exactly three of each of the standard model fields and three exotic colour
triplets, as predicted above.

Similarly, the other three $E_6$ bundles in our list produce models with the
standard model spectrum, plus colour triplets and two additional gauged
$U(1)$ symmetries. With this observation in hand, we turn to the final
model of the positive monad scan.

\subsection{The $SU(5)$ model} \label{su5section} 

From the entire positive monad data set, we find only a single $SU(5)$
model survives the three-generation test~\eqref{3fam}. The bundle,
\beq\label{su5} 0 \rightarrow V \rightarrow \cO_X(2)^{\oplus 5} \oplus
\cO_X(1)^{\oplus 5} \stackrel{f}{\rightarrow} \cO_X(3)^{\oplus 5}
\rightarrow 0 \eeq on the quintic, satisfies all of the conditions on
Chern classes given in Table \ref{constraints_table}, and admits an
equivariant structure with respect to the $\mathbb{Z}_5 \times
\mathbb{Z}_5$ symmetry of the quintic.

This bundle was first presented as a potential three-generation model in
Ref.~\cite{Kachru:1995em}\footnote{In addition, while the current paper was
  in preparation, this bundle was also studied in detail in
 Ref.~\cite{Braun:2009mb}.}  . Unfortunately, since the structure group of this bundle
is $SU(5)$, and we have only $\mathbb{Z}_5$ (or $\mathbb{Z}_5 \times \mathbb{Z}_5$) Wilson lines at our
disposal, it is not possible to break the gauge group down to that of
the standard model. As a result, this model is of limited use from a
phenomenological point of view.
 
On the quintic, the freely acting $\mathbb{Z}_5 \times \mathbb{Z}_5$
symmetry acts on the coordinates $x_k$, $k=0,\ldots 4$ and is
generated by \bea\label{z5}
{\mathbb{Z}_5}^{(1)}:~~~~~~~g_1:x_k \to x_{k+1}\\ \nonumber
{\mathbb{Z}_5}^{(2)}:~~~~~~~g_2:x_k \to \alpha^{k} x_{k} \eea where
$\alpha^5=1$.

By equivariant obstruction theory, we know that a single copy of the
line bundle $\cO_X(1)$ on the quintic does not admit a $\mathbb{Z}_5
\times \mathbb{Z}_5$-equivariant structure (see Appendix
\ref{obstruction}). Indeed, using \eref{sPhi} and \eref{z5} one can
immediately show that the two $\mathbb{Z}_5$ group actions on the
space of global sections $\Gamma(X,\cO_X(1))$ do not commute. Rather,
the matrices induced from \eref{z5} form a representation of the order
$125$ Heisenberg group. Fortunately, however, the sum
$\cO_X(1)^{\oplus 5}$ {\it does} admit an equivariant structure.

Similarly to the $SU(3)$ case of the previous
section, we can define the group action on $V$ in terms of equivariant
structures on $B=\cO_X(2)^{\oplus 5} \oplus \cO_X(1)^{\oplus 5}$ and
$C=\cO_X(3)^{\oplus 5}$. Following Section \ref{sections}, we can
define the equivariant structures on $B$ and $C$ via group actions on their sections.
Explicitly, we take the action on $\Gamma(X,B)$ and $\Gamma(X,C)$ to be
\beq\label{su5eq} \Phi_{B,g_{i}}(s)=\phi_{B,g_{i}} \circ s \circ
g_{i}^{-1},~~~~~\Phi_{C,g_{i}}=\phi_{C,g_{i}} \circ s \circ g_{i}^{-1}
\eeq where \beq \phi_{B,g_1}=\left(
\begin{array}
[c]{cc}%
\gamma_1 &0 \\
0& \gamma_1 
\end{array}
\right),~
\phi_{B,g_2}=\left(
\begin{array}
[c]{cc}%
\gamma_{2}^4 &0 \\
0& \gamma_{2}^{3} 
\end{array}
\right),~
\phi_{C,g_1}=\left(\gamma_1
\right),~\phi_{C,g_2}=\left(\gamma_{2}^{2}\right)
\eeq
and
\beq
\gamma_1=\left(
\begin{array}
[c]{ccccc}%
0 & 1&0&0&0 \\
0 & 0&1&0&0\\
0&0&0&1&0\\
0&0&0&0&1\\
1&0&0&0&0 
\end{array}
\right),~~\gamma_2=\left(
\begin{array}
[c]{ccccc}%
1 & 0&0&0&0 \\
0 & \alpha^4&0&0&0\\
0&0&\alpha^3 &0&0\\
0&0&0&\alpha^2&0\\
0&0&0&0&\alpha
\end{array}
\right) \;.
\eeq

With the equivariant structure in hand, we turn now to the particle
spectrum.  Recall that the particle spectrum of an $SU(5)$ heterotic
model is given by the cohomologies: $n_{\bf 10}=h^1(X,V)$,
$n_{\overline{\bf 5}}=h^1(X,\wedge^{2}V)$ and $n_{\bf 5}=h^1(X,\wedge^{2}
V^*)=h^2(X,\wedge^{2} V)$. Since $V$ in \eref{su5} is a positive
monad, $h^1(X,V^*)=0$ and hence $n_{{\bf \overline{10}}}=0$. Furthermore,
observing that $\textnormal{Ind}(V)=-75$ it follows that
$\textnormal{Ind}(\hat{V})=-3$ on $X/G$ and that the invariant
subspace of $H^{1}(X,V)$ has dimension $h^1(X/G,\hat{V})=3$. Note that
this also follows immediately from the form of \eref{su5eq} given
above and the arguments given in Sections \ref{sections} and
\ref{h1v}. The $SU(5)$ model on $X/G$ will, therefore, contain three
${\bf 10}$-multiplets on $X/G$.

As was shown in Refs.~\cite{Anderson:2007nc,Anderson:2008uw}, for a generic
choice of the morphism, $f$, this bundle has $h^2(X,\wedge^2 V)=0$
and, hence, cannot produce ${\bf 5}$-multiplets in its low energy
spectrum. However, at a special locus in bundle moduli space (that is, for a
special choice of the map $f$ in \eref{su5}), the cohomology can
change \cite{Anderson:2007nc,Donagi:2004qk}. Specifically, the
following choice of map \beq \tilde{f}=\left(
\begin{array}
[c]{cccccccccc}%
0 & x^{2}_{4}&0&0&x^{2}_{1}&0&x_2&0&x_4&x_3 \\
0 & x^{2}_{0}&0&x^{2}_{4}&0&x_3&0&x_2&x_1&0\\
x^{2}_{3}&0&0&x^{2}_{4}&0&0&x_3&x_0&0&x_1\\
x^{2}_{2}&0&x^{2}_{1}&0&0&x_1&x_4&0&x_0&0 \\
0&0&x^{2}_{3}&0&x^{2}_{0}&x_2&0&x_4&0&x_0 
\end{array}
\right) \eeq satisfies the intertwining condition,
\eref{intertwining}, and gives rise to $h^{2}(X,\wedge^2 V)=6$. The
invariant subspace of this under the group action is one-dimensional
and thus $h^{2}(X/G, \hat{V})=1$. Thus, we have a single ${\bf 5}$ on
$X/G$.

In general, for an $SU(n)$ bundle, the indices of $\hat{V}$ and
$\wedge^{2}\hat{V}$ are related via \beq {\rm Ind}(\hat{V})=(n-4){\rm
  Ind}(\wedge^{2}\hat{V}) \;.\eeq Since $\hat{V}$ is a rank $5$
bundle, giving rise to three generations on $X/G$, we know that on
$X/G$, \beq {\rm Ind}(\hat{V})=-3={\rm Ind}(\wedge^{2}\hat{V}) \eeq
and as a result, it is clear that $h^1(X,\wedge^2 \hat{V})=4$. Thus
the spectrum consists of three $\bf{10}$s, four $\bf {\overline{5}}$s,
and a single $\bf{5}$ multiplet.

\section{Extending the class: A look ahead}\label{the_model}
As we can see from the previous sections, the data set of positive monad, three-generation models turns out to be a surprisingly restricted one. None of the models listed in the previous section produce the exact symmetries and particle content of the standard model. However, the techniques we have developed are readily applicable to the broader class of monad bundles, and a systematic scan of general monads of the form \eref{V},\eref{BC} is already underway \cite{zeropaper}. As an example, we will demonstrate here that the data set of semi-positive monads (those that allow zero entries in the line bundles in \eref{BC}) will be much richer. Specifically, below we will describe a three-generation $SO(10)$ model, which leads exactly to the particle spectrum of the supersymmetric standard model with gauge group $SU(3) \times SU(2) \times U(1)_Y \times U(1)_{B-L}$. 

\subsection{A $SO(10)$ heterotic standard model from a monad}
In this section, we present a new ``$SO(10)$ heterotic standard model". In particular, we present a $SU(4)$ bundle which admits a $\mathbb{Z}_3 \times \mathbb{Z}_3$-equivariant structure on the ``bi-cubic" manifold, defined by a bi-degree $(3,3)$ polynomial in $\mathbb{P}^2 \times \mathbb{P}^2$. The low-energy spectrum is such that the bundle produces exactly $3$ generations of quarks and leptons on the manifold $X/G$. While this model contains no Higgs doublets generically, at a special locus in bundle moduli space, exactly one Higgs doublet pair is added to the spectrum. 

\subsubsection{The manifold}

Consider the ``bi-cubic" three-fold,
\begin{equation}
 X= \left[\begin{array}[c]{c}\mathbb{P}^2\\\mathbb{P}^2\end{array}\left|\begin{array}[c]{ccc}3 \\3
 \end{array}\right.  \right]^{2,83}\; .
\end{equation} 
If we denote the coordinates on $\mathbb{P}^2\times \mathbb{P}^{2}$ by $\{x_i,y_i\}$, where $i=0,1,2$, then a freely acting $\mathbb{Z}_3 \times \mathbb{Z}_3$ symmetry is generated by \cite{Candelas:2007ac},
\begin{equation}\label{z3z3}
\begin{array}{llll}
{\mathbb{Z}_3}^(1)&:&x_k \to x_{k+1},&y_k \to y_{k+1}\\
{\mathbb{Z}_3}^(2)&:&x_k \to \alpha^k x_{k},&y_k \to \alpha^{-k} y_{k}~.
\end{array}
\end{equation}
where $\alpha=\exp (2\pi i/3)$. 
As shown in Ref.~\cite{Candelas:2007ac}, the most general bi-degree $(3,3)$ polynomial invariant under the above symmetry is given by
\begin{eqnarray}\label{bicubic_poly}
p_{(3,3)}&=&A_{1}^{k,\pm} \sum_{j} x_{j}^{2}x_{j\pm 1}y^{2}_{j+k}y_{j+k\pm1}+A_{2}^{k}\sum_{j}x^{3}_{j}y^{3}_{j+k}+A_{3}x_1x_2x_3\sum_{j}y^{3}_j\nonumber\\
&&+A_{4}y_1y_2y_3\sum_{j}x^{3}_{j}+A_5x_1x_2x_3y_1y_2y_3
\end{eqnarray}
where $j,k=0,1,2$ and there are a total of $12$ free coefficients, denoted by $A$ with various indices. In the explicit computations carried out below, we shall take these coefficients to be generic (that is, random) integers.

Having chosen an invariant polynomial, we can quotient $X$ by the $\mathbb{Z}_{3}\times \mathbb{Z}_3$ symmetry, to produce a non-simply connected manifold, $\hat{X}=X/G$. The resulting manifold, $\hat{X}$,  has moduli $h^{1,1}(\hat{X})=2$ and $h^{2,1}(\hat{X})=11$. In the next subsection, we will consider a rank four vector bundle which also admits a $\mathbb{Z}_3 \times \mathbb{Z}_3$ equivariant structure and, hence, descends to a bundle $\hat{V}$ on $\hat{X}$. To begin, however, we consider how the divisors of $X$, and the line bundles associated to them behave under the symmetry.

Let the restrictions of the two hyperplane classes in $\mathbb{P}^2\times \mathbb{P}^2$, be denoted by $\{H_1,H_2\}$. If we take this set as the basis for the divisor classes of $X$, then it is straightforward to check that the invariant divisor classes invariant under the symmetry action \eref{z3z3}) are generated by $\{H_1+ H_2, 3H_1\}$ (see Table \ref{constraints_table}). Thus, we may choose a basis $\hat{H}_1,\hat{H}_2$, for the generators of the divisor classes of $\hat{X}$ that are related to the divisors of $X$ via the pullback map: $q^*(\hat{H}_1)=H_1+H_2$ and $q^*(\hat{H}_2)=3H_1$.  Furthermore, we note that in the basis $\{H_r\}$, the triple intersection numbers, $d_{rst}$ of $X$ are given by $d_{111}=d_{222}=0$ and otherwise, $d_{rst}=3$. With these preliminary observations in hand, we turn now to the description of the bundle.

\subsubsection{The bundle}
On this manifold, we consider the bundle defined by the monad sequence
\beq\label{good_bundle}
0 \to V \to \cO_X(1,0)^{\oplus 3} \oplus \cO_X(0,1)^{\oplus 3} \stackrel{f}{\rightarrow} \cO_X(1,1) \oplus \cO_X(2,2) \to 0
\eeq
If this bundle is to produce a physically interesting model, we require it to satisfy a number of important physical constraints --  namely it that it be a slope stable, holomorphic vector bundle \cite{duy} and that it is consistent with heterotic anomaly cancellation. We shall discuss the stability of $V$ in a subsequent section, and begin here by observing that from the formulas of Section \ref{thebuns}, it is easy to verify that this bundle satisfies the anomaly cancellation condition, \eref{anomaly}, 
\beq\label{c2good}
c_{2r}(V) = \frac12  d_{rst} \left(\sum_{a=1}^{r_C} c^s_a c^t_a- \sum_{i=1}^{r_B} b^s_i b^t_i \right)
=(18,18)\leq c_{2r}(TX)=(36,36)~.
\eeq
From the above, we see that $c_2(TX)-c_2(V)$ is an effective class, hence we are free to take the Hidden sector bundle to be trivial and to satisfy the anomaly cancellation condition with $M5$-branes.
Of course, the fact that $V$ satisfies the anomaly cancellation condition on $X$ is not enough. We must also verify that the associated bundle $\hat{V}$ is part of an anomaly free theory on $\hat{X}$. To this end, we note that a necessary condition for $\hat{V}$ to be anomaly free is
\beq
q^*(c_2(\hat{X})) -q^*(c_2(\hat{V}))=q^*([\hat{{\cal W}}])
\eeq
where $[\hat{{\cal W}}] \in H^4(\hat{X}, \mathbb{Z})$ is the class of an effective curve on $\hat{X}$. However, we are fortunate that in this example, the integral cohomologies $H^4(\hat{X}, \mathbb{Z})$ and $H^4(X, \mathbb{Z})$ are related via the pullback map, which is an injection
\beq
q^*: H^4(\hat{X}, \mathbb{Z}) \rightarrow H^4(X, \mathbb{Z})
\eeq
defined by the integer correspondence $(n,m) \rightarrow (9n,9(n-m))$ (see Table \ref{constraints_table}). As a result, the anomaly cancellation condition condition on the covering space and on the quotient $\hat{X}$ are equivalent so long as they are satisfied in the integral cohomology.\footnote{See Ref.~\cite{Braun:2007xh} for comments regarding finite torsion components of $H^4(X,\mathbb{Z})$ and discrete anomalies.}
Finally, it is worth noting that $c_2(V)_r$ in \eref{c2good} satisfies the necessary condition on the second Chern class of equivariant bundles derived in Table \ref{constraints_table}
Also, observe here that $c_2(V)$ in Eq.~\eref{c2good} satisfies the necessary condition on the second Chern class for equivariant bundles derived in Table \ref{constraints_table}.

In addition, from Eq.~\eref{c3}, we have
\begin{equation}\label{c3_good}
  c_3(V) = \frac13 d_{rst} \left(\sum_{i=1}^{r_B} b^r_i b^s_i b^t_i - \sum_{a=1}^{r_C} c^r_ac^s_a c^t_a \right)=-27,
\end{equation}
and thus, this model passes our initial constraints for a three-family model. 

\subsubsection{The ``Upstairs'' spectrum}

Before constructing the ``downstairs" bundle, $\hat{V}$, we must obtain the particle content of the ``upstairs" theory, that is, the number of ${\bf 16}$ and $\overline{\bf 16}$ multiplets, given by $H^1(X,V)$ and $H^1(X,V^*)$ respectively, and the number of ${\bf 10}$ multiplets\footnote{Recall that for rank $4$ holomorphic vector bundles with $c_1(V)=0$, the following isomorphism holds: $\wedge^2 V \approx \wedge^2 V^*$.} given by $H^1(X,\wedge^2 V)\approx H^1(X, \wedge^2 V^*)$. First, we observe that from Eq.~\eref{c3_good} that ${\rm Ind(V)}=-27$. Moreover, despite the fact that this is a semi-positive monad, there are no anti-generations, that is $H^i(X,V)=0$ for $i\neq 1$. To see this, consider the long exact sequence in cohomology associated to \eref{good_bundle}. 
\beq\label{good_co}
0\to H^0(X,V) \to H^0(X,B) \to H^0(X,C) \to H^1(X,V) \to \ldots
\eeq
From the results of Ref. \cite{Anderson:2008ex}, for the cohomology of line bundles on the bi-cubic, we find that 
\beq\label{line_coho}
H^i(X,\cO(k,0))=0~ {\rm for}~~ 0< k<3~~{\rm and}~~ i \neq 0~. 
\eeq
Using the above, we have $H^i(X,B)=0$ for $i >0$. Moreover, as mentioned in the previous sections, since all the line bundles in $C$ have strictly positive entries, $H^i(X,C)=0$ for $i\neq 0$ by the Kodaira vanishing theorem \cite{AG,AG2}. As a result, \eref{good_co} reduces to a four-term exact sequence. Next, using the techniques described in Ch. $8$ of Ref. \cite{Anderson:2008ex}, we find the map $\tilde{f}: H^0(X,B) \rightarrow H^0(X,C)$ to be injective. Hence, $H^0(X,V)=0$ (a necessary condition for $V$ to be a slope-stable bundle). Combining these results we have the exact sequence
\beq\label{littleh1}
0 \to H^0(X,B) \to H^0(X,C) \to H^1(X,V) \to 0~.
\eeq
and $H^i(X,V)=0$, for $i \neq 1$. As a result, the number of {\bf 16} multiplets is given by $H^1(X,V)=-27$ and there are no $\overline{{\bf 16}}$ multiplets.

Next, to compute the number of ${\bf 10}$ multiplets, consider the exterior power sequence, \eref{extp},
\beq\label{wedge_again}
0 \to \wedge^2 V \to \wedge^2 B \to B\otimes C \to S^{2} C \to 0
\eeq

For the bundle in \eref{good_bundle}, this leads to the following sequences: 
\begin{eqnarray}\label{wedge_explicit}
&0&\to \wedge^2 V \to \cO_X(2,0)^{\oplus 3}\oplus \cO_X(1,1)^{\oplus 9}\oplus \cO_X(0,2)^{\oplus 3} \to K \to 0 \\
&0& \to K \to \cO_X(3,2)^{\oplus 3}\oplus \cO_X(2,3)^{\oplus 3}\oplus \cO_X(2,1)^{\oplus 3} \oplus \cO_X(1,2)^{\oplus 3}\nonumber\\
&& \to \cO_X(2,2)\oplus \cO_X(3,3) \oplus \cO_X(4,4) \to 0
\eea
Using \eref{line_coho} and the ampleness of $C$ once again, the long exact sequences in cohomology associated to \eref{wedge_again} and \eref{wedge_explicit} yield
\beq
\ba
{l}
0\to H^0(\wedge^2 V) \to H^0(X, \wedge^2 B) \to H^0(X,K) \to H^1(X, \wedge^2 V) \to 0 \\
0\to H^1(X, K) \to H^2(X,\wedge^2 V) \to 0 \\
0 \to H^0(X,K) \to H^0(X, B\otimes C) \stackrel{\tilde{F}}{\rightarrow} H^0(X, S^2 C) \to H^1(X, K) \to 0
\ea
\eeq

From the above, we have
\beq
\ba
{lll}
H^1(X, \wedge^2 V)={\rm Coker}(\tilde{H})&,&\tilde{H}: H^0(X, \wedge^2 B) \to H^0(X,K) \\
H^2(X, \wedge^2 V)=H^1(X,K)={\rm Coker}(\tilde{F})&,&\tilde{F}: H^0(X, B\otimes C) \to H^0(X, S^2 C)
\ea
\eeq
As was proven for positive monads in Ref.~\cite{Anderson:2008uw,Anderson:2008ex}, and verified explicitly above, for a generic choice of map, $f$, in
\eref{good_bundle}, the induced map $\tilde{F}$ has maximal rank. As a result, its cokernel, $H^1(X,K)$ must vanish and so
\beq
H^2(X,\wedge^2 V)=0~~{\rm generically}~.
\eeq
Further, by Serre duality, $H^1(X, \wedge^2 V^*)^*=H^2(X, \wedge^2 V)$ and thus, we see that the total number of {\bf 10} multiplets vanishes. Since we are attempting to build a phenomenologically interesting model and the Higgs doublets reside in the {\bf 10} of $SO(10)$ this would seem like an unfortunate result. Fortunately, however, at special loci in moduli space of $V$, the spectrum
can become enhanced by additional numbers of {\bf 10} multiplets
\cite{Donagi:2004qk,Anderson:2007nc}. As we will demonstrate below, by careful choice of the map $f$ in \eref{good_bundle} it is possible to increase the number of {\bf 10} multiplets by $1,2$ or more.

Finally, it is worth noting that the number of bundle moduli, $H^1(X, V \otimes V^*)$, can be computed from  Eq. $(7.37)$ of Ref. \cite{Anderson:2008uw}:
\beq
\ba
{l}
h^1(X, V\otimes V^*)=h^0(X, B^* \otimes C)-h^0(X, B^*\otimes B)-h^0(X, C^*\otimes C) \\
~~~~~~~~~~~~~~~~~~~~~~~+h^0(X, C^* \otimes B)-h^1(X, C^* \otimes B)+h^1(X, B^*\otimes B)+1~.
\ea
\eeq
Using the above, we find that the number of singlets is given by $n_1=h^1(X, V\otimes V^*)=98$.

\subsubsection{The equivariant structure}
To define an equivariant structure on $V$, we begin by noting that under the group action defined in \eref{z3z3}, the line bundle $\cO_X(1,1)$ is manifestly equivariant and hence $C=\cO_X(1,1) \oplus \cO_X(2,2)$ admits an equivariant structure. As in Section \ref{su5section}, we shall see that one has to look more carefully in order to define an equivariant structure on the sum of line bundles $B=\cO_X(1,0)^{\oplus 3} \oplus \cO_X(0,1)^{\oplus 3}$ in \eref{good_bundle}.

First, consider the line bundle $\cL =O_X(1,0)$. As we shall see, this line bundle is equivariant with respect to each of the ${\mathbb{Z}_3}$ symmetries independently. Choosing the basis of sections of $\cL$ by $\{x_0,x_1,x_2\}$, we find that for ${\mathbb{Z}_3}^{(1)}$, the section-wise map is given by 

\beq
\Phi_{\cL,g_{1}}=\left(
\begin{array}
[c]{ccc}%
0 & 1&0\\
0&0 & 1\\
1&0 &0 
\end{array}
\right)
\eeq
and for the second ${\mathbb{Z}_{3}}^{(2)}$,
\beq
\Phi_{\cL,g_2}=\left(
\begin{array}
[c]{ccc}%
1 & 0&0 \\
0 & \alpha^2&0\\
0&0& \alpha 
\end{array}
\right)
\eeq
where $\alpha=\exp (2\pi i/3)$.

It is clear that the equivariance (cocycle) condition \eref{eqPhi} for $\cL$ is satisfied for each of $\Phi_{\cL,g_i}$ independently, since each of the matrices above generate a representation of $\mathbb{Z}_3$. Since $\cL$ is globally generated, by the arguments of Section \ref{glob_gen}, $\cL$ is equivariant with respect to $\mathbb{Z}_3$. Note that this agrees with the trivial check of equivariance mentioned in Section \ref{simple_tests}, since $\text{Ind}(\cL)=3$ which is divisible by $|\mathbb{Z}_3|$. 

However, from the above, we can immediately note that these section mappings, $\Phi_{\cL,g_i}$ cannot be used to define an equivariant structure with respect to the full, $\mathbb{Z}_3\times \mathbb{Z}_3$ symmetry, since the matrices $\Phi_{\cL,g_1}$ and $\Phi_{\cL,g_2}$ do not commute. Instead, they form a representation of the order $27$ Heisenberg group, $H_{27}=(\mathbb{Z}_3 \times \mathbb{Z}_3) \rtimes \mathbb{Z}_3$. 

Since the equivariant structure of a line bundle is unique up to a character, the above argument shows that $\cL$ cannot admit any $\mathbb{Z}_3\times \mathbb{Z}_3$ action. Furthermore, this agrees with the index check and what we would expect from obstruction theory (see Appendix \ref{obstruction}).

However, from the fact that the obstruction to equivariance, $H^2(\mathbb{Z}_3\times \mathbb{Z}_3, \mathbb{Z}_3)=\mathbb{Z}_3$, it is clear that we can ``fix" this obstruction, given sufficiently many copies of $\cO_X(1,0)$. Specifically, while $\cO_X(1,0)$ has no equivariant structure, three copies of the line bundle {\it can} admit an equivariant structure.

Consider $\cL^{\oplus 3}=\cO_X(1,0)^{\oplus 3}$ and the $9 \times 9$ matrices defined by
\bea\label{leq1}
\Phi_{\cL^{\oplus 3},g_{1}}=\phi_{\cL^{\oplus 3},g_1} \otimes \Phi_{\cL,g_1}&=&\left(
\begin{array}
[c]{ccc}%
0 & 1&0\\
0&0 & 1\\
1&0 &0 
\end{array}
\right) \otimes 
\left(
\begin{array}
[c]{ccc}%
0 & 1&0\\
0&0 & 1\\
1&0 &0 
\end{array}
\right) \\ \nonumber &=&\left(
\begin{array}
[c]{ccc}%
0 & \Phi_{\cL,g_1}&0\\
0&0 & \Phi_{\cL,g_1}\\
\Phi_{\cL,g_1}&0 &0 
\end{array}
\right) 
\eea
and
\bea\label{leq2}
\Phi_{\cL^{\oplus 3},g_2}=\phi_{\cL^{\oplus 3},g_2} \otimes \Phi_{\cL,g_2}&=&\left(
\begin{array}
[c]{ccc}%
1 & 0&0 \\
0 & \alpha&0\\
0&0& \alpha^2 
\end{array}
\right) \otimes
\left(
\begin{array}
[c]{ccc}%
1 & 0&0 \\
0 & \alpha^2&0\\
0&0& \alpha 
\end{array}
\right) \\ \nonumber &=&
\left(
\begin{array}
[c]{ccc}%
\Phi_{\cL,g_2} & 0&0 \\
0 & \alpha\Phi_{\cL,g_2}&0\\
0&0& \alpha^2\Phi_{\cL,g_2} 
\end{array}
\right)
\eea
where $\otimes$ refers to the matrix outer product. These matrices clearly commute and form a representation of $\mathbb{Z}_3\times \mathbb{Z}_3$. The nontrivial bundle morphisms given by $\phi_{\cL^{\oplus 3},g_i}$ rearrange the line bundles ${\cal O}_X(1,0)$ in the sum and introduce the characters, $\alpha^k$ necessary to form an equivariant structure. 

Using the above definitions, we define the full equivariant structures on $B$ and $C$ in \eref{good_bundle} as
\beq\label{good_phis}
\ba
{l}
\Phi_{B,g_{i}}=\Phi_{\cL_{1},g_{i}}\oplus \Phi_{\cL_{2},g_{i}} \\
\Phi_{C, g_{i}}=\phi_{\cO(1,1),g_{i}} \circ s \circ g_{i}^{-1} \oplus \phi_{\cO(2,2),g_{i}} \circ s \circ g_{i}^{-1}
\ea
\eeq
where $\Phi_{\cL_{1},g_{i}}$ is the equivariant action on the global sections of $\cO_X(1,0)^{\oplus 3}$ defined in \eref{leq1} and \eref{leq2} and $\Phi_{\cL_{2},g_{i}}$ is its direct analogue for $\cO_X(0,1)^{\oplus 3}$. 

With these definitions in hand, we turn now to the equivariance of $V$ itself. From the Lemma in Section \ref{glob_gen}, recall that the monad bundle $V$ will admit an equivariant structure if $B$ and $C$ are equivariant and the ``intertwining condition", $\phi_{C,g}\circ f = f\circ \phi_{B,g}$ is satisfied. Moreover, in the case where $B$ and $C$ are generated by their global sections, we can express the intertwining condition on global sections as in Eq.~\eref{section_intertwining}, by writing
\beq \label{intertwining2}
 \Phi_{C,g}\circ\tilde{f} =
\tilde{f}\circ \Phi_{B,g}\; . \eeq

Note the section-wise mapping, $\tilde{f}: \Gamma(X, B) \rightarrow \Gamma(X,C)$, induced from  \eref{good_bundle}, is a generic $2\times 6$ matrix of polynomials of the form 
\beq
\tilde{f}=\left(
\ba
{cccccc}
p^{1}_{(0,1)}&p^{2}_{(0,1)}&p^{3}_{(0,1)}&p^{4}_{(1,0)}&p^{5}_{(1,0)}&p^{6}_{(1,0)}\\
p^{7}_{(1,2)}&p^{8}_{(1,2)}&p^{9}_{(1,2)}&p^{10}_{(2,1)}&p^{11}_{(2,1)}&p^{12}_{(2,1)}\\
\ea
\right)
\eeq
where $p^{i}_{(m,n)}$ denotes a polynomial of bi-degree $(m,n)$. Substituting $\tilde{f}$ and the morphisms \eref{good_phis} into \eref{intertwining2}, we can solve exactly for the set of ``equivariant" maps. The most general map associated to an equivariant structure on $V$ in \eref{good_bundle} is 
\beq\label{gen_map}
\tilde{f}^{T}=\left(
\begin{array}{ll}
a y_0 & c_{1} x_0 y_0^2+c_{2} x_1 y_1
 Ê y_0+c_{3} x_2 y_2 y_0+c_{4} x_2
 Ê y_1^2+c_{5} x_1 y_2^2+c_{6} x_0 y_1 y_2 \\
a y_2 & c_{4} x_1 y_0^2+c_{6} x_2 y_1
 Ê y_0+c_{2} x_0 y_2 y_0+c_{5} x_0
 Ê y_1^2+c_{1} x_2 y_2^2+c_{3} x_1 y_1 y_2 \\
a y_1 & c_{5} x_2 y_0^2+c_{3} x_0 y_1
 Ê y_0+c_{6} x_1 y_2 y_0+c_{1} x_1
 Ê y_1^2+c_{4} x_0 y_2^2+c_{2} x_2 y_1 y_2 \\
b x_0 & d_{1} y_0 x_0^2+d_{2} x_1 y_1
 Ê x_0+d_{3} x_2 y_2 x_0+d_{4} x_1 x_2
 Ê y_0+d_{5} x_2^2 y_1+d_{6} x_1^2 y_2 \\
b x_2 & d_{6} y_1 x_0^2+d_{2} x_2 y_0
 Ê x_0+d_{4} x_1 y_2 x_0+d_{5} x_1^2
 Ê y_0+d_{3} x_1 x_2 y_1+d_{1} x_2^2 y_2 \\
b x_1 & d_{5} y_2 x_0^2+d_{3} x_1 y_0
 Ê x_0+d_{4} x_2 y_1 x_0+d_{6} x_2^2
 Ê y_0+d_{1} x_1^2 y_1+d_{2} x_1 x_2 y_2
\end{array}
\right)
\eeq
with $14$ free parameters given by $a,b,c_j,d_j$, $j=1,\ldots 6$.

\vspace{.2in}

Returning to the observations made at the end of the previous section, we now note that by specializing this general equivariant map $\tilde{f}$ in \eref{gen_map} still further, the number of {\bf 10} multiplets can ``jump", giving rise to the possibility of a model with Higgs doublets on $\hat{X}$. We have analysed this possibility using the computer algebraic geometry packages Macaulay2 and Singular~\cite{macaulay2,sing}. For a monad map 
\beq\label{good_map} \tilde{f}^T=\left( \ba{cc}
-2y_0& -x_2y_1^2 + 2x_0y_1y_2 - x_1y_2^2\\ 
-2y_2& x_1y_0^2 + 2x_2y_0y_1 - x_0y_1^2\\
 -2y_1& -x_2y_0^2 + 2x_1y_0y_2 - x_0y_2^2\\ 
 -x_0& -2x_1x_2y_0 + x_0x_1y_1 + x_2^2y_1 + 2x_1^2y_2 - 2x_0x_2y_2\\
  -x_2& x_1^2y_0 + x_0x_2y_0 + 2x_0^2y_1 - 2x_1x_2y_1 - 2x_0x_1y_2\\
   -x_1& -2x_0x_1y_0 + 2x_2^2y_0 - 2x_0x_2y_1 + x_0^2y_2 + x_1x_2y_2
 \ea
  \right) \;,
 \eeq
we find that indeed $h^1(X,\wedge^2V)=h^1(X,\wedge^2 V^*)=1$, and, hence,
that there is a single ${\bf 10}$ multiplet in the low energy spectrum. At this locus in
moduli space, we will consider the descent of this equivariant bundle
to $\hat{V}$ on the non-simply connected manifold, $X/G$.

\subsubsection{The equivariant action on cohomology}
Given the equivariant structures defined in \eref{good_phis}, we can induce the action of the group on the cohomology of $V$, i.e. $H^1(X,V)$ and $H^1(X,\wedge^2 V)$. We will employ the results of Section \ref{h1v} and specifically Eqs.~\eref{h1}, \eref{h1diff} and \eref{thirdso10}. To begin, note that from \eref{littleh1} we can write 
\beq\label{h1again} 
H^1(V)\cong\frac{\Gamma(X,C)}{\tilde{f}(\Gamma(X,B))} \;. 
\eeq
where $\tilde{f}$ is given in \eref{good_map}. To determine the group action on $H^1(X,V)$ we will find the action of $G$ on $\Gamma(X,C)$, $\Gamma(X,B)$ and its decomposition in terms of irreducible representations.

Computing the traces of the matrices in \eref{good_phis} we can use standard character theory as described in Section \ref{chars} to compute the $\mathbb{Z}_3 \times \mathbb{Z}_3$ representation content of $H^1(V)$. The traces $\chi_{B,g_{i}},\chi_{C,g_{i}}$ of $\Phi_{B,g_{i}}$ and $\Phi_{C,g_{i}}$ respectively, completely determine the decomposition of these matrices into irreducible representations, \eref{thellamabadger}. Denoting the irreducible $\mathbb{Z}_3 \times \mathbb{Z}_3$ representations by $R_p$, let
\begin{equation}
\begin{array}{lll}
R_{H^0(X,B)}&=&\bigoplus_{p}n^{p}_{B}R_{p} \\
R_{H^0(X,C)}&=&\bigoplus_{p}n^{p}_{C}R_{p}~. 
\end{array}
\end{equation}
Then the multiplicity of the irreducible representations may be uniquely determined by the traces and the inner product on characters \eref{char_sub} and \eref{chi_prod}
\beq\label{ns}
\ba
{c}
n^{p}_{B}=(\chi_{p},\chi_{B,g_{i}}) \\
n^{p}_{C}=(\chi_{p},\chi_{C,g_{i}}) 
\ea
\eeq

Explicitly computing the traces of \eref{good_phis}, we find that the group actions on $\Gamma(X, C)$ and $\Gamma(X,C)$ are traceless except for the identity element. Expressed as a vector of length $|G|$, the characters are $\chi_{\Gamma(X,C)}=(45,0,\ldots,0)$ and $\chi_{\Gamma(X,B)}=(18,0,\ldots,0)$ ($n^{p}=3$ for all $R_{p}$) hence by \eref{ns}, we see that the multiplicities are given by $n^{p}_{B}=5$ and $n^{p}_{C}=2$ for all $p$. That is $\Gamma(X,B)$ carries two copies of the regular representation and $\Gamma(X, C)$ five copies. Finally, from Eq.~\eref{h1diff}, we recall that the traces associated to the action of $G$ on $H^1(X,V)$ are given by
\begin{equation}
 \chi_{H^{1}(X,V)}=\chi_{\Gamma(X,C)}-\chi_{\Gamma(X,B)}\; .
\end{equation} 
and hence, we see that the multiplicity of each irreducible representation of $\mathbb{Z}_3\times \mathbb{Z}_3$ in $H^1(V)$ is exactly three. That is, $H^1(V)$ carries three copies of the regular representation of $G$. To make this more explicit, we can write this out as a formal sum over the nine irreducible representations (labeled by their characters/roots of unity) as
\beq\label{h1decomp}
\Phi_{H^{1}(X,V)}=3 \oplus 3\alpha_1 \oplus 3\alpha_{1}^{2}\oplus 3\alpha_2\oplus 3\alpha_{2}^{2}\oplus 3\alpha_1\alpha_2\oplus 3\alpha_{1}^2\alpha_2\oplus 3\alpha_{1}\alpha_{2}^{2}\oplus  3\alpha_{1}^2\alpha_{2}^{2}~.
\eeq
Here $\alpha_i$ is the character associated to $\mathbb{Z}_{3}^{(i)}$, $i=1,2$. The invariant part of $H^1(X,V)$ under the action \eref{h1decomp} will descend to the quotient space, $\hat{X}$. That, is $H^1(\hat{X},\hat{V})=H^1(X,V)^{inv}$. By inspection of \eref{h1decomp}, it is clear that there exactly $3$ singlets under this action, and thus, $h^{1}(\hat{X},\hat{V})=3$ as expected by index arguments.

Next, we repeat this analysis for $H^1(X,\wedge^2 V)$. For the map \eref{good_map}, chosen at the end of the last section, we can explicitly compute the cokernel of the map, $\tilde{F}: H^0(X,B\otimes C)\rightarrow H^0(X,S^2 C)$. Using
\beq\label{h1wedge}
h^1(X, \wedge^2 V)={\rm dim}({\rm(Coker}(\tilde{F}))=1
\eeq
we see that the representation content of $H^1(X, \wedge^2 V)$ is just a single character (irreducible representation) of $\mathbb{Z}_3\times \mathbb{Z}_3$. That is $\Phi_{H^1(X,\wedge^2 V)}=\alpha_{1}^{i}\alpha_{2}^{j}$ for some $i,j=0,1,2$. Since we are free to re-define the equivariant structure on $\wedge^2 V$ by an overall character, without loss of generality, we take
\beq\label{onwedge2}
\Phi_{H^1(X,\wedge^2 V)}=\alpha_{1}~.
\eeq

However, we now note that if we take a character $\alpha$ to act on $\wedge^2 V$, the {\it dual} representation, $\alpha^{-1}$ acts on $\wedge^2 V^*$. Thus, although the vector spaces $H^1(X,\wedge^2 V)$ and $H^1(X,\wedge^2 V^*)$ are isomorphic, they transform as dual representations under the equivariant $G$-action and
\beq\label{onwedge2dual}
\Phi_{H^1(X,\wedge^2 V^*)}=\alpha_{1}^{2}~.
\eeq

Having determined the transformation properties of the cohomology of $V$, we turn now to the definition of the Wilson lines necessary to break the visible $SO(10)$ symmetry down to $SU(3)\times SU(2)\times U(1)_{Y} \times U(1)_{B-L}$.

\subsubsection{Wilson Lines}
With Wilson line breaking alone it is clear that we can at best break $SO(10)$ to the Standard Model gauge group with an extra $U(1)$ factor. In this section, we will choose the Wilson lines, so that this additional Abelian symmetry is $U(1)_{B-L}$. We will take the same choice of Wilson lines~\footnote{On phenomenological grounds it is worth noting that if $U(1)_{B-L}$ is broken just above the electroweak scale, it can be useful for preventing nucleon decay \cite{Ambroso:2009jd,Ambroso:2009sc}.} as was made in Refs.~\cite{Braun:2004xv,Braun:2005zv}. We choose the Wilson line corresponding to $\mathbb{Z}_{3}^{(1)}$ to act on the {\bf 16} of $Spin(10)$ as
\beq
\mathbb{Z}_3^{(1)}=\left(
\ba
{ccc}
\alpha_1^2 \mathbbm{1}_{10} & &\\
&\mathbbm{1}_{5} &\\
& &\alpha_1
\ea \right)~.
\eeq
This embedding breaks $Spin(10)$ to $SU(5) \times U(1)_1$. Next, we choose $\mathbb{Z}_{3}^{(2)}$ to embed into the structure group as
\beq
\mathbb{Z}_3^{(2)}=\left(
\ba
{llllll}
\alpha_2 \mathbbm{1}_{6} & &&&&\\
&1&&&&\\
& &\alpha_2^2\mathbbm{1}_{3}&&&\\
&&&\mathbbm{1}_{2}&&\\
&&&&\alpha_2^2\mathbbm{1}_3&\\
&&&&&1
\ea \right)
\eeq
which in turn breaks $Spin(10)$ to $SU(3)\times SU(2)\times SU(2) \times U(1)_2$. 

Considering the combined structure of the two $\mathbb{Z}_3$ symmetries above, we find that the commutant in $Spin(10)$ of our total, $\mathbb{Z}_3 \times \mathbb{Z}_3$ Wilson line is simply $SU(3)\times SU(2) \times U(1)_1 \times U(1)_2$. Setting,
\bea
Y_H=\frac{1}{2}Y_1 + \frac{5}{6}Y_2\\
Y_{B-L}=\frac{1}{3}Y_2
\eea
we recover exactly the standard model symmetry $SU(3)\times SU(2) \times U(1)_Y$ with an additional $U(1)_{B-L}$. 

In order to finally determine the particle spectrum of our four dimensional effective theory, we must determine the action of these Wilson lines on the {\bf 10} and {\bf 16} representations of $SO(10)$. Also, we can decompose the action on the {\bf 10} and {\bf 16} representations as a formal sum of characters as in \eref{h1decomp}. Similarly to Refs.~\cite{Braun:2004xv,Braun:2005zv}, we find the following transformation properties for each of the $(SU(3),S(2))_{(U(1)_Y,U(1)_{B-L})}$ representations
\bea\label{16badger}
{\bf 16}=\alpha_{1}^{2}\alpha_2({\bf 3},{\bf 2})_{(1,1)}\oplus\alpha_{1}^{2}({\bf 1},{\bf 1})_{(6,3)}\oplus \alpha_{1}^{2}\alpha_{2}^2({\bf \overline{3}},{\bf 1})_{(-4,-1)}\nonumber\\
+\alpha_{2}^{2}({\bf \overline{3}},{\bf 1})_{(2,-1)}\oplus({\bf 1},{\bf 2})_{(-3,-3)}\oplus\alpha_{1}({\bf 1},{\bf 1})_{(0,3)}
\eea
and
\beq\label{wilson_10}
{\bf 10}=\alpha_{1}({\bf 1},{\bf 2})_{(3,0)}\oplus\alpha_{1}\alpha_2({\bf 3},{\bf 1})_{(-2,-2)}\oplus\alpha_{1}^{2}({\bf 1},{\bf 2})_{(-3,0)}\oplus\alpha_{1}^{2}\alpha_{2}^{2}({\bf 3},{\bf 1})_{(2,2)}
\eeq

We can now combine the Wilson line actions given in Eqs.~\eref{16badger} and \eref{wilson_10} with the equivariant action of the symmetry on the cohomology $H^1(X,V)$ in \eref{h1decomp} and $H^1(X,\wedge^2 V)$ and $H^1(X,\wedge^2 V^*)$ in \eref{onwedge2} and \eref{onwedge2dual}, respectively. As discussed in Section \ref{chars} , the singlets under this combined action, form the final low energy particle spectrum.

\subsubsection{The ``Downstairs" Spectrum}
Combining the results of the last two sections we can obtain the final particle spectrum on $X/G$. We shall find the invariant elements of cohomology under the combined action of the Wilson line and the equivariant action, as in Eq.~\eref{wilson_inv}.

First, from the decomposition of the {\bf 16} of $SO(10)$ in Eq.~\eqref{16badger} and the representation content of $H^1(X,V)$ in Eq.~\eqref{h1decomp} we see that
\beq
\ba
{l}
[{{\bf 16}} \otimes H^1(X,V)]_{\rm inv}=\\
([\alpha_{1}^{2}\alpha_2({\bf 3},{\bf 2})_{(1,1)}\oplus\alpha_{1}^{2}({\bf 1},{\bf 1})_{(6,3)}\oplus \alpha_{1}^{2}\alpha_{2}^2({\bf \overline{3}},{\bf 1})_{(-4,-1)}
\oplus\alpha_{2}^{2}({\bf \overline{3}},{\bf 1})_{(2,-1)}\oplus({\bf 1},{\bf 2})_{(-3,-3)}\oplus\alpha_{1}({\bf 1},{\bf 1})_{(0,3)} \\
\otimes (3 \oplus 3\alpha_1 \oplus 3\alpha_{1}^{2}\oplus 3\alpha_2\oplus 3\alpha_{2}^{2}\oplus 3\alpha_1\alpha_2\oplus 3\alpha_{1}^2\alpha_2\oplus 3\alpha_{1}\alpha_{2}^{2}\oplus  3\alpha_{1}^2\alpha_{2}^{2})]_{\rm inv}\\
\stackrel{{\rm invariant}}{\to} 3({\bf 3},{\bf 2})_{(1,1)}\oplus 3({\bf 1},{\bf 1})_{(6,3)}\oplus 3({\bf \overline{3}},{\bf 1})_{(-4,-1)}\oplus3({\bf \overline{3}},{\bf 1})_{(2,-1)}\oplus 3({\bf 1},{\bf 2})_{(-3,-3)}\oplus 3({\bf 1},{\bf 1})_{(0,3)}~.
\ea
\eeq 
Therefore, exactly three families of each standard model quark and lepton and three right-handed neutrinos survive the Wilson line projection.

Next, to find the downstairs spectrum from the {\bf 10} of $SO(10)$ we combine \eref{onwedge2} and \eref{wilson_10},
\beq
\ba
{l}
[W_{{\bf 10}}\otimes H^1(X, \wedge^2 V)]_{\rm inv}=\\
=[(\alpha_{1}({\bf 1},{\bf 2})_{(3,0)}\oplus\alpha_{1}\alpha_2({\bf 3},{\bf 1})_{(-2,-2)}\oplus\alpha_{1}^{2}({\bf 1},{\bf 2})_{(-3,0)}\oplus\alpha_{1}^{2}\alpha_{2}^{2}({\bf 3},{\bf 1})_{(2,2)})\otimes(\alpha_1)]\\
\stackrel{{\rm invariant}}{\rightarrow}
({\bf 1},{\bf 2})_{(-3,0)}
\ea
\eeq
and finally combining \eref{onwedge2dual} and \eref{wilson_10}:
\beq
\ba
{l}
[W_{{\bf 10}}\otimes H^1(X, \wedge^2 V^*)]_{\rm inv}\\
=[(\alpha_{1}({\bf 1},{\bf 2})_{(3,0)}\oplus\alpha_{1}\alpha_2({\bf 3},{\bf 1})_{(-2,-2)}\oplus\alpha_{1}^{2}({\bf 1},{\bf 2})_{(-3,0)}\oplus\alpha_{1}^{2}\alpha_{2}^{2}({\bf 3},{\bf 1})_{(2,2)})\otimes(\alpha_{1}^{2})]\\
\stackrel{{\rm invariant}}{\rightarrow}
({\bf 1},{\bf 2})_{(3,0)}
\ea
\eeq

From the above expressions it is clear that though there is a single {\bf 10} of $SO(10)$ on $X$, the overall characters induced on $\wedge^2 V$ and $\wedge^2 V^*$, combined with the Wilson line action, project out different invariant components of $H^1(X,\wedge^2 V)$ and $H^1(X, \wedge^2 V^*)$, thereby preserving the chiral asymmetry of the spectrum on $\hat{X}$. The overall character on $\wedge^2 V$ is chosen so that the surviving elements of the ${\bf 10}$ multiplet are precisely a single Higgs doublet pair. That is, exactly one Higgs up/down pair survives the combined projections and all exotic color triplets are projected out. As a result, we obtain no exotic particles and exactly the MSSM spectrum.

In Table~\ref{spec} we list the complete matter spectrum of our model together with its associated cohomological origin.
\begin{table}[h]
\begin{center}
\begin{tabular}{|l|l|l|l|}\hline
${\rm Cohomology}$&${\rm Representation}$&${\rm Multiplicity}$&${\rm Name}$\\\hline\hline

$[\alpha_{1}^{2}\alpha_2\otimes H^1(X,V)]^{inv}$ & $({\bf 3,2})_{1,1}$ & $3$ & left-handed quark\\\hline

$[\alpha_{1}^{2}\otimes H^1(X,V)]^{inv}$  & $({\bf 1,1})_{6,3}$ &$3$&left-handed anti-lepton\\\hline

$[\alpha_{1}^{2}\alpha_{2}^{2}\otimes H^1(X,V)]^{inv}$  &$({\bf \overline{3},1})_{-4,-1}$&$3$&left-handed  anti-up\\ \hline

$[\alpha_{2}^{2}\otimes H^1(X,V)]^{inv}$  &$({\bf \overline{3},1})_{2,-1}$&$3$&left-handed anti-down\\ \hline

$[H^1(X,V)]^{inv}$  &$({\bf 1,2})_{-3,-3}$&$3$& left-handed lepton\\ \hline

$[\alpha_1\otimes H^1(X,V)]^{inv}$  &$({\bf 1,1})_{0,3}$&$3$&left-handed anti-neutrino\\ \hline

$[\alpha_1\otimes H^1(X,\wedge^2V)]^{inv}$  &$({\bf 1,2})_{3,0}$&$1$&up Higgs\\ \hline

$[\alpha_{1}^{2}\otimes H^1(X,\wedge^2V)]^{inv}$  &$({\bf 1,2})_{-3,0}$&$1$& down Higgs\\\hline
\end{tabular}
\caption{The complete low energy particle spectrum of our model. Note that there are no exotic fields.}
\label{spec}
\end{center}
\end{table}

\subsubsection{Stability}
To have a supersymmetric heterotic vacuum, we require that the gauge connection associated to $V$, \eref{good_bundle} satisfies the Hermitian-Yang-Mills equations. This means $V$ needs to be a ``slope stable" vector bundle.

Recall that a holomorphic bundle is called stable if for all torsion-free sub-sheaves $\cF \in V$, with ${\rm rk}(\cF)<{\rm rk}(V)$, satisfy
\beq\label{slope}
\mu(\cF) < \mu(V)
\eeq
where the slope of a sheaf $\cF$ is defined by
\beq
\mu(\cF)=\frac{1}{rk(\cF)} \int_{X}c_1(\cF)\wedge J \wedge J
\eeq
for a given choice of polarization (that is, K\"ahler form), $J$.

Using the techniques described in Refs.~\cite{Anderson:2008ex,Anderson:2009nt,stabpaper} it is possible to algorithmically scan all possible sub-sheaves of the the monad bundle $V$ to verify that \eref{slope} is satisfied. We will not reproduce this lengthy calculation here, but simply refer the reader to the explicit tools described in \cite{Anderson:2008ex,Anderson:2009nt,stabpaper}. 

We check the stability of $V$ ``upstairs" on $X$ and find it to be slope stable throughout its entire two-dimensional K\"ahler cone (that is, for all possible choices of polarization). For the ``downstairs" bundle $\hat{V}$ on $X/G$, we observe that since $V=q^*\hat{V}$, $\hat{V}$ will be stable if $V$ is stable for all {\it equivariant sub-sheaves} ${\cal \tilde{F}} \in V$. That is, the regions of stability in the K\"ahler cone can only get bigger in passing from $V$ on $X$ to $\hat{V}$ on $X/G$ (destabilizing, non-equivariant sub-sheaves could disappear upon quotienting). Hence, since $V$ is stable for all polarizations on $X$, it follows that $\hat{V}$ is also stable for all choices of polarization $\hat{J}$.

\vspace{.2in}

To summarise, the above example provides a stable holomorphic bundle on the bi-cubic CICY, which satisfies the anomaly cancellation condition and produces a low energy theory with the exact particle spectrum of the MSSM: three families of quarks and leptons, a single Higgs doublet pair and three right-handed neutrinos. In view of this example, we expect that the class of general two-term monad bundles over complete intersection manifolds to be phenomenologically much more promising than the positive monads alone. A systematic scan for standard-model- bundles within this class has already begun~\cite{zeropaper}.

\section{Conclusions and future work}\label{conc}
In the spirit of exploring the space of vacuum solutions of the heterotic string, an ``algorithmic'' programme had been launched by constructing large data-sets of $SU(n)$ bundles with $n=3,4,5$ \cite{Anderson:2007nc,Anderson:2008uw,Anderson:2009ge}. These bundles break the $E_8$ gauge theory down to $E_6$, $SO(10)$ and $SU(5)$ GUT theories, and with discrete Wilson line turned on, can break the latter further down to the standard model gauge group (with possible extra $U(1)$ factors). A traditional method of manufacturing bundles is the so-called {\it monad construction} and this method has been employed within the heterotic string literature over the past two decades \cite{Distler:1987ee,Kachru:1995em,Douglas:2004yv,Blumenhagen:2006wj,Anderson:2007nc}. However, one of the largest data sets of such bundles was only recently obtained and systematically studied in Ref.~\cite{Anderson:2008uw}; these are the positive monad bundles on favourable complete intersection Calabi-Yau threefolds (CICYs). Within the context of $N = 1$ supersymmetric compactifications of the $E_8\times E_8$ heterotic string, the class of such bundles was shown in Ref.~\cite{Anderson:2008uw} to be finite and in fact to consist of $7118$ bundles arising on just $36$ manifolds (all the remaining $4500$ or so favorable CICYs do not allow positive monads which satisfy anomaly cancellation).

In this paper, we have explored further the phenomenology of these bundles by producing a systematic scan for three-generation models with the gauge symmetry and particle spectrum of the MSSM. The breaking of the GUT groups is accomplished by the introduction of Wilson lines.

Since all the CICYs are by themselves simply connected, to accomplish symmetry breaking we need to find freely acting discrete symmetries of these manifolds. With such symmetries in hand, it is possible to produce a smooth quotient, $X/G$, with non-trivial fundamental group $G$, suitable for the introduction of a $G$-Wilson line. The classification of such discrete groups $G$ on the CICY data-set was recently completed \cite{Braun:2009qy}, however, for the present work we find that simple arguments using twisted Euler indices can readily restrict the possible groups allowed.

In order to produce a heterotic model over the ``downstairs" manifold we need to decide which of the positive monads consistently descend to bundles on the quotient manifold. Given a specific CICY $X$ with freely-acting symmetry $G$, we have investigated systematically which bundles $V$ on $X$ descend to a bundle $\hat{V}$ on the quotient $X/G$, that is which bundles $V$ allow for a $G$-equivariant structure.

The results of the scan over the positive monads on CICYs turned out to be highly restrictive.
We found that of the $E_6$ models, only $87$ pass the initial three-generation test. Of these, $81$ fail to admit $G$-equivariant structures for a group of the appropriate order to produce three-family models. The remaining six $E_6$ models all suffer from the standard doublet-triplet splitting problem and thus, contain exotic particles. We find no three-generation $SO(10)$ models and only a single $SU(5)$ example. This $SU(5)$ model on the quintic is equivariant with respect to a $\mathbb{Z}_5 \times \mathbb{Z}_5$ symmetry. Upon quotienting this geometry by $G=\mathbb{Z}_5 \times \mathbb{Z}_5$ we produce an $SU(5)$ model with a particle spectrum consisting of three families in ${\bf 10}\oplus {\overline{\bf5}}$ and one pair of Higgs multiplets in a ${\bf 5}\oplus\overline{\bf 5}$ multiplet. Such a particle content could reduce to exactly the MSSM spectrum with the introduction of Wilson lines. Unfortunately, it is impossible to break the $SU(5)$ GUT group to the standard model gauge group by means of a $\mathbb{Z}_5 \times \mathbb{Z}_5$ Wilson line alone and thus this remaining possibility is found to be phenomenologically disfavoured.

With the results above in hand, it is clear that we must move beyond the positive monad data set in our search for phenomenologically viable models. We have initiated this study here, by demonstrating that it is possible to produce heterotic standard models within the broader class of two-term monads.

An illustrative example of a stable semi-positive monad of rank four on the bi-cubic hypersurface in $\mathbb{P}^2 \times \mathbb{P}^2$, giving rise to an "upstairs" GUT model with $SO(10)$ gauge group, was studied in detail.
We have explicitly constructed an $\IZ_3 \times \IZ_3$ equivariant structure for this bundle, and have determine the associated bundle $\hat{V}$ on the quotient manifold. With the introduction of Wilson lines, we obtain a supersymmetric low-energy theory with $SU(3) \times SU(2) \times U(1)_{Y}\times U(1)_{B-L}$ gauge group, exactly three generations of quarks and leptons and three right-handed neutrinos. No anti-families or exotic matter of any kind is present. By carefully choosing a $\IZ_3 \times \IZ_3$ Wilson line and the monad map (that is, for a special locus in bundle moduli space), the color triplets are projected out, leaving us with only a single pair of Higgs doublets. Hence, the matter spectrum is precisely that of the MSSM.

This work is an important step forward towards our ongoing goal of systematically producing a large data set of phenomenologically viable models. While we have demonstrated that the positive monad data set is a very restrictive one, we have also developed the necessary tools to analyze the broader class of non-positive monads. As the new heterotic standard model presented in this work demonstrates, such an analysis should produce interesting new models. A systematic study of this problem is currently underway \cite{zeropaper}.

\section*{Acknowledgments}
The authors would like to thank B.~Ovrut, T.~Pantev, R.~Donagi, and
B.~Szendroi for useful discussions. J.~Gray is supported by STFC and
would like to thank the University of Pennsylvania for hospitality
while part of this work was completed. L.A. is supported in part by
the DOE under contract No.  DE-AC02-76-ER-03071and by NSF RTG Grant
DMS-0636606. Y.-H.~H~is indebted to the UK STFC for an Advanced
Fellowship as well as the FitzJames Fellowship of Merton College,
Oxford.
\appendix
\section{Equivariance}
\subsection{Obstructions to Equivariance}\label{obstruction}
The question of whether or not an automorphism $G$ of a manifold $X$ lifts to an automorphism of a bundle $V \stackrel{\pi}{\rightarrow} X$ is a long-standing one and is referred to as equivariant obstruction theory\footnote{The lifting of an automorphism $G$ on $X$ to a linear, holomorphic action on the fibers of $V$ is also referred to in the literature as ``G-linearization" \cite{mumford}.}. In order to discuss this however it is convenient to re-phrase slightly the definitions of equivariance introduced in Section \ref{equivariance}. In the main text the relevant definitions (see Eqs.~\eref{invalt} and \eref{cocycle}) are as follows. A bundle is called $G$-invariant if there exist morphisms, $\phi_g$ such that the diagrams
\begin{equation}
\begin{array}{lllll}
&V&\stackrel{\phi_g}{\longrightarrow}&V&\\
\pi &\downarrow&&\downarrow&\pi\\
&X&\stackrel{g}{\longrightarrow}&X&
\end{array}
\label{first_def}
\end{equation}
commute for all $g \in G$. That is, $\phi_g$ is a map from $V$ to itself that covers the action of $g \in G$ on the base. Further, if the morphism $\phi_g$ satisfies the co-cycle condition,
\beq\label{cocycle2}
\phi_g \circ \phi_h =\phi_{gh}
\eeq
for all $g,h \in G$, then the bundle is called equivariant.

To discuss obstruction theory, we will find it useful to reformulate the above definitions in terms of morphisms $\rho$ which cover the identity on $X$. Let $g: X \rightarrow X$ be an automorphism of $X$ and $\rho_g: V \rightarrow g^* V$ be a map from V to the pullback bundle associated to $g$. This morphism covers the identity on the $X$ so that the diagram
\begin{equation}
\begin{array}{rrlll}
&V&\stackrel{\rho_g}{\longrightarrow}&g^*V&\\
&\pi&\searrow&\downarrow&\pi'\\
&&&X&
\end{array}
\label{invariance2}
\end{equation}
commutes. Above, $\pi'$ is the standard projection map of the pull-back bundle \cite{AG,AG2}. If $\rho_g$ is an isomorphism, the bundle is invariant and commutativity of \eref{invariance2} is equivalent to commutativity of \eref{first_def}. Further, formulated in terms of $\rho_g$, the cocycle condition, \eref{cocycle2} can be written as
\beq
g^*(\rho_h)\circ\rho_g=\rho_{h\cdot g}~. \label{eqvar}
\eeq
If an isomorphism $\rho_g$ satisfies \eref{eqvar} for all $g \in G$, then the bundle is equivariant. Written in this different, but equivalent, language, we can discuss obstructions to equivariance. The following discussion will follow closely the review given in Ref.~\cite{Donagi:2003tb}.

For any $G$-invariant bundle, there is a natural group ${\cal G}(V)$ associated to $V$. An element of ${\cal G}(V)$ is a pair,
\beq
(g,\rho),~~\text{where}~~ g \in G,~~\rho: V\rightarrow g^* V
\eeq
with $\rho$ a vector bundle isomorphism. The group multiplication rule is defined by
\beq
(g,\rho) \cdot (h,\psi)=(gh,h^*(\rho)\circ \psi)~.
\eeq
for any $(g,\rho),(h,\psi) \in {\cal G}(V)$. The group ${\cal G}(V)$ is known as the ``Theta" group~\cite{mumford} of an invariant bundle, $V$.

The Theta group fits into the following short exact sequence of groups:
\beq\label{obs_seq}
1\rightarrow GL(V) \rightarrow {\cal G}(V) \rightarrow G \rightarrow 1
\eeq
where $GL(V)$ is the group of vector bundle isomorphisms of $V$. We observe here that if there exists a group homomorphism, $G\rightarrow {\cal G}(V)$, then there exists a lifting of the group action $G$ to $V$ and the bundle is equivariant. Phrased in the language of group theory \cite{groupbook}, if the sequence \eref{obs_seq} is split, then $V$ is equivariant.

In the case that $V$ is simple (which will hold for the stable bundles considered in this work), $GL(V)=\mathbb{C}^*$. Using the fact that $G$ is a finite group, there is a sub-extension \cite{Donagi:2003tb} of \eref{obs_seq} which takes the simple form of a central extension:
\beq\label{obs_fin}
1\rightarrow \mu_d \rightarrow {\cal G}^{fin}(V) \rightarrow G \rightarrow 1
\eeq
where $\mu_d \in \mathbb{C}^*$ is the subgroup of $d$-th roots of unity and $d$ is just the least common multiple of the orders of all elements. The obstruction to the splitting of this sequence is given  by a cohomology class in $H^{2}(G,\mu_d)$, see, for example, Ref.~\cite{groupbook}.

In order to construct the equivariant structures for the monad bundles used in this work, it is convenient to understand the obstructions to equivariance for line bundles over $X$. For example, if on the quintic, $[\mathbb{P}^{4}\,|\,5]$, with its freely acting $\mathbb{Z}_{5}\times \mathbb{Z}_{5}$, we consider the line bundle $\cO_X(1)$, we find that the sequence \eref{obs_fin} is given by
\beq
1\rightarrow \mathbb{Z}_5 \rightarrow {\cal G}^{fin}(V) \rightarrow \mathbb{Z}_{5}\times \mathbb{Z}_{5}  \rightarrow 1~.
\eeq
As mentioned in Section \ref{su5section}, in this case, ${\cal G}^{fin}(V)=H_{125}=(\mathbb{Z}_{5}\times \mathbb{Z}_5) \ltimes \mathbb{Z}_5$, the order $125$ Heisenberg group. In this case, we can see that $H^{2}(\mathbb{Z}_{5}\times \mathbb{Z}_5,\mathbb{Z}_5)=\mathbb{Z}_5$. As a result, the single line bundle, $\cO_X(1)$ cannot be equivariant. However, the $\mathbb{Z}_5$ obstruction above indicates that it is possible for five copies of the line bundle to have an equivariant structure. This can be seen explicitly by the construction in Section \ref{su5section}. Using the explicit matrices, \eref{su5eq}, the direct sum, $\cO_X(1)^{\oplus 5}$ does indeed admit an equivariant structure. Similar arguments are used to determine the obstructions and equivariant sums of line bundles in Sections \ref{E6example} and \ref{the_model}.

\subsection{Quaternionic equivariant structures and non-Abelian Wilson lines}\label{quat_eq}

In the course of this work, equivariant structures corresponding to
non-Abelian discrete symmetries and Wilson lines were developed. While
the $81$ $SU(3)$ positive monad bundles defined over the tetra-quadric
manifold (see Appendix \ref{posmonad}) failed to produce
three-generation models, the techniques developed will likely be
useful in future constructions and are worth noting here for their
novelty.

The ``tetra-quadric" manifold, 
\beq
X=\left[\begin{array}{c}\mathbb{P}^1\\\mathbb{P}^1\\\mathbb{P}^1\\\mathbb{P}^1\end{array}\right.\left|\begin{array}{c}2\\2\\2\\2\end{array}\right]\;
\eeq
 has a fixed-point free action of the quaternions, $\mathbb{H}$, see Ref.~\cite{Candelas:2008wb}. To simply describe the action of this non-Abelian group of order $8$, let the coordinates on $\mathbb{P}^1\times\mathbb{P}^1\times\mathbb{P}^1\times\mathbb{P}^1$ be labeled by $(x_{\sigma},x_{-\sigma})$ where $\sigma \in \mathbb{H}_{+}=(1,i,j,k)$. Then the quaternionic symmetry acts via
\beq\label{quat}
\tau:(x_{\sigma},x_{-\sigma}) \rightarrow (x_{\tau\sigma},s_{-\tau\sigma})~~~\tau \in \mathbb{H}
\eeq
We can choose the polynomial of multi-degree $(2,2,2,2)$ as $\prod_{\tau \in \mathbb{H}}x_{\tau}$ so that this symmetry is an automorphism of $X$.

Note that this symmetry not only re-arranges the coordinates of $X$, but also re-arranges the ``ambient" $\mathbb{P}^{1}$'s themselves. Thus, unlike the other examples in this work, the Picard group, ${\rm Pic}(X)$ itself experiences a non-trivial automorphism. For example, if we denote by $H_i$ the four divisors of $X$ associated to the $\mathbb{P}^1$'s, then under the group action $i \in \mathbb{H}$, the divisor $H_1$ is interchanged with $H_2$ and $H_3$ with $H_4$.

We can now ask, whether it is possible to lift this group action to monad bundles defined over $X$? To illustrate this, we will consider the first bundle in the list in Appendix \ref{posmonad}. As we will see, the bundle
\beq\label{quat_ex}
0\rightarrow V \rightarrow \cO_X(1,1,1,1)^{\oplus 7} \stackrel{f}{\rightarrow} \cO_X(4,1,1,1)\oplus\cO_X(1,4,1,1)\oplus \cO_X(1,1,4,1)\oplus \cO_X(1,1,1,4) \rightarrow 0
\eeq
does admit an $\mathbb{H}$-equivariant structure (though unfortunately not the $\mathbb{H} \times \mathbb{Z}_2$ action necessary to produce a three-generation model). Considering \eref{quat_ex}, our first observation is that the line bundle $\cO_X(1,1,1,1)$ is clearly invariant under the group action (since under all group elements, the associated divisor class $\sum_i H_i$ is left invariant) and as we shall see, this line bundle is actually equivariant. However, unlike in the other examples in this paper, the component line bundles of $C$ are not even invariant.

To begin, note that unlike the other discrete group actions discussed in this work, the quaternionic symmetry in \eref{quat} actually interchanges the {\it divisor classes}, $H_i$ of $X$, rather than simply rotation the divisors within a given class. That is, under the group element $i \in \mathbb{H}$, for example, $\cO_X(4,1,1,1) \rightarrow \cO_X(1,4,1,1)$. Thus, the line bundle $\cO_X(4,1,1,1)$ is clearly not invariant (much less equivariant) under $\mathbb{H}$ since ${\rm Hom}(\cO_X(4,1,1,1),\cO_X(1,4,1,1))=0$. However, for sums of such line bundles the situation is different. Consider the same group element and the sum of line bundles 
\beq
{\cal U}=\cO_X(4,1,1,1)_X\oplus \cO_X(1,4,1,1)~.
\eeq
Since the two divisor classes associated to the line bundles in ${\cal U}$ are interchanged under $ \pm i \in \mathbb{H}$ we can define for example, a $\mathbb{Z}_4^{(i)}=\{1,-1,i,-i\}$ equivariant structure on ${\cal U}$ via
\beq
\Phi_{{\cal U},\rho}(s)=\phi_{\cal{U},\rho} \circ s \circ \rho^{-1},
\eeq
where $\rho \in \mathbb{Z}_{4}^{(i)}$ above and the bundle morphism, $\phi_{\cal{U},\rho}$ is given by
\beq
\left(
\ba
{cc}
1 &0\\
0 & 1
\ea
\right)~.
\eeq
Then, $\Phi_{{\cal U},\rho}$ satisfies the cocycle condition \eref{cocycle}, and ${\cal U}$ is equivariant. This is of particular importance since although there are no non-trivial maps between $\cO(4,1,1,1)$ and $\cO(1,4,1,1)$ {\it which cover the identity on} $X$, maps such as $\phi_{\cal{U},\rho}$ exist which {\it cover} $\rho$ {\it over the base}. We will use this observation to construct full $\mathbb{H}$-equivariant structures below.

Returning now to the bundle, $V$, in \eref{quat_ex}, we will induce an equivariant structure on $V$ by first defining equivariant structures on $B$ and $C$ via an action on their spaces of global sections.
An equivariant structure can be defined as described in Section \ref{sections} (see Eq.~\eref{sPhi}) for $B=\cO_X(1,1,1,1)^{\oplus 7}$ and $C=\cO_X(4,1,1,1)\oplus\cO_X(1,4,1,1)\oplus \cO_X(1,1,4,1)\oplus \cO_X(1,1,1,4)$ by the following actions on $\Gamma(X,B)$ and $\Gamma(X,C)$
\beq
\Phi_{B,\tau}(s)=\phi_{B,\tau} \circ s \circ \tau^{-1},~~~~~\Phi_{C,\tau}=\phi_{C,\tau} \circ s \circ \tau^{-1}
\eeq
where
\beq
\phi_{B,\tau}=\mathbbm{1}_{7\times 7}~~{\rm and}~~\phi_{C,\tau}=\mathbbm{1}_{4\times 4}
\eeq
Despite the fact that $\phi_{C,\tau}=\mathbbm{1}_{4\times 4}$ is the identity, the composition $\circ s \circ \tau^{-1}$ above, causes the elements of $C$ to be interchanged. To demonstrate these non-trivial interchanges of line bundle components in $C$ we write out the $160\times 160$ section-wise morphisms $\Phi_C,\tau$ below schematically. Each of the four line bundles has ${\rm dim}(\Gamma(\cO(4,1,1,1)))=40$ independent sections and $\tau$ acts non-trivially on each space of sections individually, while also interchanging them. For simplicity we will denote this action on each $40\times 40$ block by $x$. Covering $\tau$ on $X$ and acting on $C=\cO(4,1,1,1)\oplus\cO(1,4,1,1)\oplus \cO(1,1,4,1)\oplus \cO(1,1,1,4)$ the section mappings are

\bea
\label{quatmorphs}
\Phi_{C,\pm 1}&=&\left(
\begin{array}
[c]{cccc}%
x &0&0&0 \\
0&x&0&0\\
0&0&x&0\\
0&0&0&x
\end{array}
\right),~
\Phi_{C,\pm i}=\left(
\begin{array}
[c]{cccc}%
0 &x&0&0 \\
x&0&0&0\\
0&0&0&x\\
0&0&x&0
\end{array}
\right), \\ \nonumber
\Phi_{C,\pm j}&=&\left(
\begin{array}
[c]{cccc}%
0 &0&x&0 \\
0&0&0&x\\
x&0&0&0\\
0&x&0&0
\end{array}
\right),~
\Phi_{C,\pm k}=\left(
\begin{array}
[c]{cccc}%
0 &0&0&x \\
0&0&x&0\\
0&x&0&0\\
x&0&0&0
\end{array}
\right),~
\eea
Writing out the explicit matrices $\Phi_B$ and $\Phi_C$ acting on a basis of monomials in $\Gamma(X,B)$ and $\Gamma(X,C)$, we find that not only do they form a representation of the quaternions, (that is, they solve the cocycle condition \eref{cocycle}), but there also exists a monad map $f$ satisfying the intertwining condition \eref{intertwining}, and hence $V$ admits an equivariant structure.

\section{Positive monads passing the three-family test}\label{posmonad}
In Section~\ref{setup} we introduced the positive monads over the
favourable CICYs and reviewed their classification which leads to a
data set of 7118 bundles. In Section~\ref{initconst} we discussed the
simple ``three-family'' constraint~\eqref{3fam}. It requires the
existence of a possible freely-acting symmetry which, after being
divided out, leads to three families of matter ``downstairs''.  It
turns out that only $91$ positive monad bundles on five CICYs satisfy
this constraint. These bundles are explicitly presented in the
subsequent tables, ordered by the base CICY on which they arise.  The
sums of line bundles $B$ and $C$ which, from Eqs.~\eqref{V},
\eqref{BC}, define the monad bundle, are denoted by matrices $B\sim
({b^r}_i)$ and $C\sim ({c^r}_a)$ with each column representing a line
bundle. We also provide ${\rm rk}(V)$, the rank of $V$, the components $(c_{2r}(V)$ of the second
Chern class $c_2(V)=c_{2r}(V)\nu^r$, the order
$|G|$ of the symmetry group and the Euler characteristics ${\rm ind}(B)$,
${\rm ind}(C)$ of $B$ and $C$. 

\begin{table}[h]
\begin{center}
\begin{tabular}{|l|l|l|l|l|l|}\hline
$B$&$C$&\footnotesize{rk$(V)$}&$c_{2r}(V)$&\footnotesize{(${\rm ind}(B)$,${\rm ind}(C)$)}&\footnotesize{$|G|$}\\\hline\hline
$\tiny{\left(
\begin{array}{cccccccccc}
 2 & 2 & 2 & 2 & 2 & 1 & 1 & 1 & 1 & 1
\end{array}
\right)}$&$\tiny{\left(
\begin{array}{ccccc}
 3 & 3 & 3 & 3 & 3
\end{array}
\right)}$&5&(50)&(100,175)&25\\\hline

$\tiny{\left(
\begin{array}{ccccccc}
 2 & 2 & 2 & 1 & 1 & 1 & 1
\end{array}
\right)}$&$\tiny{\left(
\begin{array}{ccc}
 4 & 3 & 3
\end{array}
\right)}$&4&(45)&(65,140)&25\\\hline

$\tiny{\left(
\begin{array}{ccccc}
 3 & 2 & 1 & 1 & 1
\end{array}
\right)}$&$\tiny{\left(
\begin{array}{cc}
 4 & 4
\end{array}
\right)}$&3&(40)&(65,140)&25\\\hline

$\tiny{\left(
\begin{array}{cccccc}
 1 & 1 & 1 & 1 & 1 & 1
\end{array}
\right)}$&$\tiny{\left(
\begin{array}{ccc}
 2 & 2 & 2
\end{array}
\right)}$&3&(15)&(30,45)&5\\\hline

$\tiny{\left(
\begin{array}{ccccccccc}
 2 & 2 & 2 & 2 & 2 & 2 & 2 & 2 & 2
\end{array}
\right)}$&$\tiny{\left(
\begin{array}{cccccc}
 3 & 3 & 3 & 3 & 3 & 3
\end{array}
\right)}$&3&(45)&(135,210)&25\\\hline
\end{tabular}
\caption{Results for the quintic, $[\mathbb{P}^4|5]$. The bundles require groups of order $|G|=5$ and  $|G|=25$, given by the two well-known freely acting ${\mathbb{Z}_5}$ symmetries and by $G=\mathbb{Z}_5\times\mathbb{Z}_5$. }
\label{tablequintic}
\end{center}
\end{table}

\vspace{-0.9cm}

\begin{table}[h]
\begin{center}
\begin{tabular}{|l|l|l|l|l|l|}\hline
$B$&$C$&\footnotesize{rk$(V)$}&$(c_{2r}(V))$&\footnotesize{(${\rm ind}(B)$,${\rm ind}(C)$)}&\footnotesize{$|G |$}\\\hline\hline

$\tiny{\left(
\begin{array}{ccccccc}
 1 & 1 & 1 & 1 & 1 & 1 & 1
\end{array}
\right)}$&$\tiny{\left(
\begin{array}{ccc}
 3 & 2 & 2
\end{array}
\right)}$&4&(45)&(42,96)&18\\\hline

$\tiny{\left(
\begin{array}{cccccccc}
 1 & 1 & 1 & 1 & 1 & 1 & 1 & 1
\end{array}
\right)}$&$\tiny{\left(
\begin{array}{cccc}
 2 & 2 & 2 & 2
\end{array}
\right)}$&4&(36)&(48,84)&12\\\hline

$\tiny{\left(
\begin{array}{cccccc}
 1 & 1 & 1 & 1 & 1 & 1
\end{array}
\right)}$&$\tiny{\left(
\begin{array}{ccc}
 2 & 2 & 2
\end{array}
\right)}$&3&(27)&(36,63)&9\\\hline

\end{tabular}
\caption{Results for $[\,\mathbb{P}^5\, |\, 3\, 3\,]$. Freely-acting
  symmetries with orders $|G|=12$ and $18$ do not exist for this
  manifold, so the first two examples cannot lead to viable
  models. The freely-acting symmetry of order $|G|=9$ is given by
  $G=\mathbb{Z}_3\times\mathbb{Z}_3$.}
\label{table33}
\end{center}
\end{table}

\vspace{-0.9cm}

\begin{table}[h]
\begin{center}
\begin{tabular}{|l|l|l|l|l|l|}\hline
$B$&$C$&\footnotesize{rk$(V)$}&$(c_{2r}(V))$&\footnotesize{(${\rm ind}(B)$,${\rm ind}(C)$)}&\footnotesize{$|G |$}\\\hline\hline

$\tiny{\left(
\begin{array}{cccccc}
 1 & 1 & 1 & 1 & 1 & 1
\end{array}
\right)}$&$\tiny{\left(
\begin{array}{ccc}
 2 & 2 & 2
\end{array}
\right)}$&3&(48)&(48,96)&16\\\hline

\end{tabular}
\caption{Results for $[\,\mathbb{P}^7\,|\,2\,2\,2\,2\,]$ (\# 7861). The possible symmetry groups $G$ are the order 16 sub-groups of $\mathbb{Z}_8\times\mathbb{Z}_4$ or $\mathbb{H}\times\mathbb{Z}_2$.}
\label{table2222}
\end{center}
\end{table}


\begin{table}[!h]
\begin{center}
\begin{tabular}{|l|l|l|l|l|l|}\hline
$B$&$C$&\footnotesize{rk$(V)$}&$(c_{2r}(V))$&\footnotesize{(${\rm ind}(B)$,${\rm ind}(C)$)}&\footnotesize{$|G |$}\\\hline\hline

$\tiny{\left(
 \begin{array}{ccccccccc}
  1 & 1 & 1 & 1 & 1 & 1 & 1 & 1 & 1 \\
 1 & 1 & 1 & 1 & 1 & 1 & 1 & 1 & 1 
\end{array}
\right)}$&${\tiny \left(
\begin{array}{cccccc}
 2 & 2 & 2 & 1 & 1 & 1 \\
 1 & 1 & 1 & 2 & 2 & 2
\end{array}
\right)}$&3&(18,18)&(81,108)&9\\\hline

\end{tabular}
\caption{Results for the bi-cubic in
  $\mathbb{P}^2\times\mathbb{P}^2$. The symmetry of order $|G|=9$ is
  $G=\mathbb{Z}_3\times\mathbb{Z}_3$.}
\label{tablebicubic}
\end{center}
\end{table}

\clearpage 

\begin{table}[!h]
\begin{center}
\begin{tabular}{|l|l|l|}\hline
$B$&$C$&\footnotesize{(${\rm ind}(B)$,${\rm ind}(C)$)}\\\hline\hline

$\tiny{\left(
\begin{array}{ccccccc}
 1 & 1 & 1 & 1 & 1 & 1 & 1 \\
 1 & 1 & 1 & 1 & 1 & 1 & 1 \\
 1 & 1 & 1 & 1 & 1 & 1 & 1 \\
 1 & 1 & 1 & 1 & 1 & 1 & 1
\end{array}
\right)}$&$\tiny{\left(
\begin{array}{cccc}
 4 & 1 & 1 & 1 \\
 1 & 4 & 1 & 1 \\
 1 & 1 & 4 & 1 \\
 1 & 1 & 1 & 4
\end{array}
\right)}$&(112,160)\\\hline

$\tiny{\left(
\begin{array}{cccccccc}
 1 & 1 & 1 & 1 & 1 & 1 & 1 & 1 \\
 1 & 1 & 1 & 1 & 1 & 1 & 1 & 1 \\
 1 & 1 & 1 & 1 & 1 & 1 & 1 & 1 \\
 1 & 1 & 1 & 1 & 1 & 1 & 1 & 1
\end{array}
\right)}$&$\tiny{\left(
\begin{array}{ccccc}
 3 & 2 & 1 & 1 & 1 \\
 1 & 1 & 4 & 1 & 1 \\
 1 & 1 & 1 & 4 & 1 \\
 1 & 1 & 1 & 1 & 4
\end{array}
\right)}$&(128,176)\\\hline

$\tiny{\left(
\begin{array}{cccccccc}
 1 & 1 & 1 & 1 & 1 & 1 & 1 & 1 \\
 1 & 1 & 1 & 1 & 1 & 1 & 1 & 1 \\
 1 & 1 & 1 & 1 & 1 & 1 & 1 & 1 \\
 1 & 1 & 1 & 1 & 1 & 1 & 1 & 1
\end{array}
\right)}$&$\tiny{\left(
\begin{array}{ccccc}
 4 & 1 & 1 & 1 & 1 \\
 1 & 3 & 2 & 1 & 1 \\
 1 & 1 & 1 & 4 & 1 \\
 1 & 1 & 1 & 1 & 4
\end{array}
\right)}$&(128,176)\\\hline

$\tiny{\left(
\begin{array}{cccccccc}
 1 & 1 & 1 & 1 & 1 & 1 & 1 & 1 \\
 1 & 1 & 1 & 1 & 1 & 1 & 1 & 1 \\
 1 & 1 & 1 & 1 & 1 & 1 & 1 & 1 \\
 1 & 1 & 1 & 1 & 1 & 1 & 1 & 1
\end{array}
\right)}$&$\tiny{\left(
\begin{array}{ccccc}
 4 & 1 & 1 & 1 & 1 \\
 1 & 4 & 1 & 1 & 1 \\
 1 & 1 & 3 & 2 & 1 \\
 1 & 1 & 1 & 1 & 4
\end{array}
\right)}$&(128,176)\\\hline

$\tiny{\left(
\begin{array}{cccccccc}
 1 & 1 & 1 & 1 & 1 & 1 & 1 & 1 \\
 1 & 1 & 1 & 1 & 1 & 1 & 1 & 1 \\
 1 & 1 & 1 & 1 & 1 & 1 & 1 & 1 \\
 1 & 1 & 1 & 1 & 1 & 1 & 1 & 1
\end{array}
\right)}$&$\tiny{\left(
\begin{array}{ccccc}
 4 & 1 & 1 & 1 & 1 \\
 1 & 4 & 1 & 1 & 1 \\
 1 & 1 & 4 & 1 & 1 \\
 1 & 1 & 1 & 3 & 2
\end{array}
\right)}$&(128,176)\\\hline

$\tiny{\left(
\begin{array}{ccccccccc}
 1 & 1 & 1 & 1 & 1 & 1 & 1 & 1 & 1 \\
 1 & 1 & 1 & 1 & 1 & 1 & 1 & 1 & 1 \\
 1 & 1 & 1 & 1 & 1 & 1 & 1 & 1 & 1 \\
 1 & 1 & 1 & 1 & 1 & 1 & 1 & 1 & 1
\end{array}
\right)}$&$\tiny{\left(
\begin{array}{cccccc}
 2 & 2 & 2 & 1 & 1 & 1 \\
 1 & 1 & 1 & 4 & 1 & 1 \\
 1 & 1 & 1 & 1 & 4 & 1 \\
 1 & 1 & 1 & 1 & 1 & 4
\end{array}
\right)}$&(144,192)\\\hline

$\tiny{\left(
\begin{array}{ccccccccc}
 1 & 1 & 1 & 1 & 1 & 1 & 1 & 1 & 1 \\
 1 & 1 & 1 & 1 & 1 & 1 & 1 & 1 & 1 \\
 1 & 1 & 1 & 1 & 1 & 1 & 1 & 1 & 1 \\
 1 & 1 & 1 & 1 & 1 & 1 & 1 & 1 & 1
\end{array}
\right)}$&$\tiny{\left(
\begin{array}{cccccc}
 3 & 2 & 1 & 1 & 1 & 1 \\
 1 & 1 & 3 & 2 & 1 & 1 \\
 1 & 1 & 1 & 1 & 4 & 1 \\
 1 & 1 & 1 & 1 & 1 & 4
\end{array}
\right)}$&(144,192)\\\hline

$\tiny{\left(
\begin{array}{ccccccccc}
 1 & 1 & 1 & 1 & 1 & 1 & 1 & 1 & 1 \\
 1 & 1 & 1 & 1 & 1 & 1 & 1 & 1 & 1 \\
 1 & 1 & 1 & 1 & 1 & 1 & 1 & 1 & 1 \\
 1 & 1 & 1 & 1 & 1 & 1 & 1 & 1 & 1
\end{array}
\right)}$&$\tiny{\left(
\begin{array}{cccccc}
 3 & 2 & 1 & 1 & 1 & 1 \\
 1 & 1 & 4 & 1 & 1 & 1 \\
 1 & 1 & 1 & 3 & 2 & 1 \\
 1 & 1 & 1 & 1 & 1 & 4
\end{array}
\right)}$&(144,192)\\\hline

$\tiny{\left(
\begin{array}{ccccccccc}
 1 & 1 & 1 & 1 & 1 & 1 & 1 & 1 & 1 \\
 1 & 1 & 1 & 1 & 1 & 1 & 1 & 1 & 1 \\
 1 & 1 & 1 & 1 & 1 & 1 & 1 & 1 & 1 \\
 1 & 1 & 1 & 1 & 1 & 1 & 1 & 1 & 1
\end{array}
\right)}$&$\tiny{\left(
\begin{array}{cccccc}
 3 & 2 & 1 & 1 & 1 & 1 \\
 1 & 1 & 4 & 1 & 1 & 1 \\
 1 & 1 & 1 & 4 & 1 & 1 \\
 1 & 1 & 1 & 1 & 3 & 2
\end{array}
\right)}$&(144,192)\\\hline

$\tiny{\left(
\begin{array}{ccccccccc}
 1 & 1 & 1 & 1 & 1 & 1 & 1 & 1 & 1 \\
 1 & 1 & 1 & 1 & 1 & 1 & 1 & 1 & 1 \\
 1 & 1 & 1 & 1 & 1 & 1 & 1 & 1 & 1 \\
 1 & 1 & 1 & 1 & 1 & 1 & 1 & 1 & 1
\end{array}
\right)}$&$\tiny{\left(
\begin{array}{cccccc}
 4 & 1 & 1 & 1 & 1 & 1 \\
 1 & 2 & 2 & 2 & 1 & 1 \\
 1 & 1 & 1 & 1 & 4 & 1 \\
 1 & 1 & 1 & 1 & 1 & 4
\end{array}
\right)}$&(144,192)\\\hline

$\tiny{\left(
\begin{array}{ccccccccc}
 1 & 1 & 1 & 1 & 1 & 1 & 1 & 1 & 1 \\
 1 & 1 & 1 & 1 & 1 & 1 & 1 & 1 & 1 \\
 1 & 1 & 1 & 1 & 1 & 1 & 1 & 1 & 1 \\
 1 & 1 & 1 & 1 & 1 & 1 & 1 & 1 & 1
\end{array}
\right)}$&$\tiny{\left(
\begin{array}{cccccc}
 4 & 1 & 1 & 1 & 1 & 1 \\
 1 & 3 & 2 & 1 & 1 & 1 \\
 1 & 1 & 1 & 3 & 2 & 1 \\
 1 & 1 & 1 & 1 & 1 & 4
\end{array}
\right)}$&(144,192)\\\hline

$\tiny{\left(
\begin{array}{ccccccccc}
 1 & 1 & 1 & 1 & 1 & 1 & 1 & 1 & 1 \\
 1 & 1 & 1 & 1 & 1 & 1 & 1 & 1 & 1 \\
 1 & 1 & 1 & 1 & 1 & 1 & 1 & 1 & 1 \\
 1 & 1 & 1 & 1 & 1 & 1 & 1 & 1 & 1
\end{array}
\right)}$&$\tiny{\left(
\begin{array}{cccccc}
 4 & 1 & 1 & 1 & 1 & 1 \\
 1 & 3 & 2 & 1 & 1 & 1 \\
 1 & 1 & 1 & 4 & 1 & 1 \\
 1 & 1 & 1 & 1 & 3 & 2
\end{array}
\right)}$&(144,192)\\\hline

$\tiny{\left(
\begin{array}{ccccccccc}
 1 & 1 & 1 & 1 & 1 & 1 & 1 & 1 & 1 \\
 1 & 1 & 1 & 1 & 1 & 1 & 1 & 1 & 1 \\
 1 & 1 & 1 & 1 & 1 & 1 & 1 & 1 & 1 \\
 1 & 1 & 1 & 1 & 1 & 1 & 1 & 1 & 1
\end{array}
\right)}$&$\tiny{\left(
\begin{array}{cccccc}
 4 & 1 & 1 & 1 & 1 & 1 \\
 1 & 4 & 1 & 1 & 1 & 1 \\
 1 & 1 & 2 & 2 & 2 & 1 \\
 1 & 1 & 1 & 1 & 1 & 4
\end{array}
\right)}$&(144,192)\\\hline

$\tiny{\left(
\begin{array}{ccccccccc}
 1 & 1 & 1 & 1 & 1 & 1 & 1 & 1 & 1 \\
 1 & 1 & 1 & 1 & 1 & 1 & 1 & 1 & 1 \\
 1 & 1 & 1 & 1 & 1 & 1 & 1 & 1 & 1 \\
 1 & 1 & 1 & 1 & 1 & 1 & 1 & 1 & 1
\end{array}
\right)}$&$\tiny{\left(
\begin{array}{cccccc}
 4 & 1 & 1 & 1 & 1 & 1 \\
 1 & 4 & 1 & 1 & 1 & 1 \\
 1 & 1 & 3 & 2 & 1 & 1 \\
 1 & 1 & 1 & 1 & 3 & 2
\end{array}
\right)}$&(144,192)\\\hline

$\tiny{\left(
\begin{array}{ccccccccc}
 1 & 1 & 1 & 1 & 1 & 1 & 1 & 1 & 1 \\
 1 & 1 & 1 & 1 & 1 & 1 & 1 & 1 & 1 \\
 1 & 1 & 1 & 1 & 1 & 1 & 1 & 1 & 1 \\
 1 & 1 & 1 & 1 & 1 & 1 & 1 & 1 & 1
\end{array}
\right)}$&$\tiny{\left(
\begin{array}{cccccc}
 4 & 1 & 1 & 1 & 1 & 1 \\
 1 & 4 & 1 & 1 & 1 & 1 \\
 1 & 1 & 4 & 1 & 1 & 1 \\
 1 & 1 & 1 & 2 & 2 & 2
\end{array}
\right)}$&(144,192)\\\hline

$\tiny{\left(
\begin{array}{cccccccccc}
 1 & 1 & 1 & 1 & 1 & 1 & 1 & 1 & 1 & 1 \\
 1 & 1 & 1 & 1 & 1 & 1 & 1 & 1 & 1 & 1 \\
 1 & 1 & 1 & 1 & 1 & 1 & 1 & 1 & 1 & 1 \\
 1 & 1 & 1 & 1 & 1 & 1 & 1 & 1 & 1 & 1
\end{array}
\right)}$&$\tiny{\left(
\begin{array}{ccccccc}
 2 & 2 & 2 & 1 & 1 & 1 & 1 \\
 1 & 1 & 1 & 3 & 2 & 1 & 1 \\
 1 & 1 & 1 & 1 & 1 & 4 & 1 \\
 1 & 1 & 1 & 1 & 1 & 1 & 4
\end{array}
\right)}$&(160,208)\\\hline

$\tiny{\left(
\begin{array}{cccccccccc}
 1 & 1 & 1 & 1 & 1 & 1 & 1 & 1 & 1 & 1 \\
 1 & 1 & 1 & 1 & 1 & 1 & 1 & 1 & 1 & 1 \\
 1 & 1 & 1 & 1 & 1 & 1 & 1 & 1 & 1 & 1 \\
 1 & 1 & 1 & 1 & 1 & 1 & 1 & 1 & 1 & 1
\end{array}
\right)}$&$\tiny{\left(
\begin{array}{ccccccc}
 2 & 2 & 2 & 1 & 1 & 1 & 1 \\
 1 & 1 & 1 & 4 & 1 & 1 & 1 \\
 1 & 1 & 1 & 1 & 3 & 2 & 1 \\
 1 & 1 & 1 & 1 & 1 & 1 & 4
\end{array}
\right)}$&(160,208)\\\hline

$\tiny{\left(
\begin{array}{cccccccccc}
 1 & 1 & 1 & 1 & 1 & 1 & 1 & 1 & 1 & 1 \\
 1 & 1 & 1 & 1 & 1 & 1 & 1 & 1 & 1 & 1 \\
 1 & 1 & 1 & 1 & 1 & 1 & 1 & 1 & 1 & 1 \\
 1 & 1 & 1 & 1 & 1 & 1 & 1 & 1 & 1 & 1
\end{array}
\right)}$&$\tiny{\left(
\begin{array}{ccccccc}
 2 & 2 & 2 & 1 & 1 & 1 & 1 \\
 1 & 1 & 1 & 4 & 1 & 1 & 1 \\
 1 & 1 & 1 & 1 & 4 & 1 & 1 \\
 1 & 1 & 1 & 1 & 1 & 3 & 2
\end{array}
\right)}$&(160,208)\\\hline

\end{tabular}
\caption{Results for the tetra-quadric in
  $\mathbb{P}^1\times\mathbb{P}^1\times\mathbb{P}^1\times\mathbb{P}^1$. There
  are $81$ bundles in total, all of them with rank $3$. The symmetry
  order is $|G|=16$, the relevant symmetry is
  $G=\mathbb{H}\times\mathbb{Z}_2$ and $(c_{2r}(V))=(18,18,18,18)$ in all cases.}
  \label{tabletetraquadric}
\end{center}
\end{table}

\clearpage

\begin{table}[!h]
\begin{center}
\begin{tabular}{|l|l|l|}\hline
$B$&$C$&\footnotesize{(${\rm ind}(B)$,${\rm ind}(C)$)}\\\hline\hline

$\tiny{\left(
\begin{array}{cccccccccc}
 1 & 1 & 1 & 1 & 1 & 1 & 1 & 1 & 1 & 1 \\
 1 & 1 & 1 & 1 & 1 & 1 & 1 & 1 & 1 & 1 \\
 1 & 1 & 1 & 1 & 1 & 1 & 1 & 1 & 1 & 1 \\
 1 & 1 & 1 & 1 & 1 & 1 & 1 & 1 & 1 & 1
\end{array}
\right)}$&$\tiny{\left(
\begin{array}{ccccccc}
 3 & 2 & 1 & 1 & 1 & 1 & 1 \\
 1 & 1 & 2 & 2 & 2 & 1 & 1 \\
 1 & 1 & 1 & 1 & 1 & 4 & 1 \\
 1 & 1 & 1 & 1 & 1 & 1 & 4
\end{array}
\right)}$&(160,208) \\\hline

$\tiny{\left(
\begin{array}{cccccccccc}
 1 & 1 & 1 & 1 & 1 & 1 & 1 & 1 & 1 & 1 \\
 1 & 1 & 1 & 1 & 1 & 1 & 1 & 1 & 1 & 1 \\
 1 & 1 & 1 & 1 & 1 & 1 & 1 & 1 & 1 & 1 \\
 1 & 1 & 1 & 1 & 1 & 1 & 1 & 1 & 1 & 1
\end{array}
\right)}$&$\tiny{\left(
\begin{array}{ccccccc}
 3 & 2 & 1 & 1 & 1 & 1 & 1 \\
 1 & 1 & 3 & 2 & 1 & 1 & 1 \\
 1 & 1 & 1 & 1 & 3 & 2 & 1 \\
 1 & 1 & 1 & 1 & 1 & 1 & 4
\end{array}
\right)}$&(160,208) \\\hline

$\tiny{\left(
\begin{array}{cccccccccc}
 1 & 1 & 1 & 1 & 1 & 1 & 1 & 1 & 1 & 1 \\
 1 & 1 & 1 & 1 & 1 & 1 & 1 & 1 & 1 & 1 \\
 1 & 1 & 1 & 1 & 1 & 1 & 1 & 1 & 1 & 1 \\
 1 & 1 & 1 & 1 & 1 & 1 & 1 & 1 & 1 & 1
\end{array}
\right)}$&$\tiny{\left(
\begin{array}{ccccccc}
 3 & 2 & 1 & 1 & 1 & 1 & 1 \\
 1 & 1 & 3 & 2 & 1 & 1 & 1 \\
 1 & 1 & 1 & 1 & 4 & 1 & 1 \\
 1 & 1 & 1 & 1 & 1 & 3 & 2
\end{array}
\right)}$&(160,208) \\\hline

$\tiny{\left(
\begin{array}{cccccccccc}
 1 & 1 & 1 & 1 & 1 & 1 & 1 & 1 & 1 & 1 \\
 1 & 1 & 1 & 1 & 1 & 1 & 1 & 1 & 1 & 1 \\
 1 & 1 & 1 & 1 & 1 & 1 & 1 & 1 & 1 & 1 \\
 1 & 1 & 1 & 1 & 1 & 1 & 1 & 1 & 1 & 1
\end{array}
\right)}$&$\tiny{\left(
\begin{array}{ccccccc}
 3 & 2 & 1 & 1 & 1 & 1 & 1 \\
 1 & 1 & 4 & 1 & 1 & 1 & 1 \\
 1 & 1 & 1 & 2 & 2 & 2 & 1 \\
 1 & 1 & 1 & 1 & 1 & 1 & 4
\end{array}
\right)}$&(160,208) \\\hline

$\tiny{\left(
\begin{array}{cccccccccc}
 1 & 1 & 1 & 1 & 1 & 1 & 1 & 1 & 1 & 1 \\
 1 & 1 & 1 & 1 & 1 & 1 & 1 & 1 & 1 & 1 \\
 1 & 1 & 1 & 1 & 1 & 1 & 1 & 1 & 1 & 1 \\
 1 & 1 & 1 & 1 & 1 & 1 & 1 & 1 & 1 & 1
\end{array}
\right)}$&$\tiny{\left(
\begin{array}{ccccccc}
 3 & 2 & 1 & 1 & 1 & 1 & 1 \\
 1 & 1 & 4 & 1 & 1 & 1 & 1 \\
 1 & 1 & 1 & 3 & 2 & 1 & 1 \\
 1 & 1 & 1 & 1 & 1 & 3 & 2
\end{array}
\right)}$&(160,208) \\\hline

$\tiny{\left(
\begin{array}{cccccccccc}
 1 & 1 & 1 & 1 & 1 & 1 & 1 & 1 & 1 & 1 \\
 1 & 1 & 1 & 1 & 1 & 1 & 1 & 1 & 1 & 1 \\
 1 & 1 & 1 & 1 & 1 & 1 & 1 & 1 & 1 & 1 \\
 1 & 1 & 1 & 1 & 1 & 1 & 1 & 1 & 1 & 1
\end{array}
\right)}$&$\tiny{\left(
\begin{array}{ccccccc}
 3 & 2 & 1 & 1 & 1 & 1 & 1 \\
 1 & 1 & 4 & 1 & 1 & 1 & 1 \\
 1 & 1 & 1 & 4 & 1 & 1 & 1 \\
 1 & 1 & 1 & 1 & 2 & 2 & 2
\end{array}
\right)}$&(160,208) \\\hline

$\tiny{\left(
\begin{array}{cccccccccc}
 1 & 1 & 1 & 1 & 1 & 1 & 1 & 1 & 1 & 1 \\
 1 & 1 & 1 & 1 & 1 & 1 & 1 & 1 & 1 & 1 \\
 1 & 1 & 1 & 1 & 1 & 1 & 1 & 1 & 1 & 1 \\
 1 & 1 & 1 & 1 & 1 & 1 & 1 & 1 & 1 & 1
\end{array}
\right)}$&$\tiny{\left(
\begin{array}{ccccccc}
 4 & 1 & 1 & 1 & 1 & 1 & 1 \\
 1 & 2 & 2 & 2 & 1 & 1 & 1 \\
 1 & 1 & 1 & 1 & 3 & 2 & 1 \\
 1 & 1 & 1 & 1 & 1 & 1 & 4
\end{array}
\right)}$&(160,208) \\\hline

$\tiny{\left(
\begin{array}{cccccccccc}
 1 & 1 & 1 & 1 & 1 & 1 & 1 & 1 & 1 & 1 \\
 1 & 1 & 1 & 1 & 1 & 1 & 1 & 1 & 1 & 1 \\
 1 & 1 & 1 & 1 & 1 & 1 & 1 & 1 & 1 & 1 \\
 1 & 1 & 1 & 1 & 1 & 1 & 1 & 1 & 1 & 1
\end{array}
\right)}$&$\tiny{\left(
\begin{array}{ccccccc}
 4 & 1 & 1 & 1 & 1 & 1 & 1 \\
 1 & 2 & 2 & 2 & 1 & 1 & 1 \\
 1 & 1 & 1 & 1 & 4 & 1 & 1 \\
 1 & 1 & 1 & 1 & 1 & 3 & 2
\end{array}
\right)}$&(160,208) \\\hline

$\tiny{\left(
\begin{array}{cccccccccc}
 1 & 1 & 1 & 1 & 1 & 1 & 1 & 1 & 1 & 1 \\
 1 & 1 & 1 & 1 & 1 & 1 & 1 & 1 & 1 & 1 \\
 1 & 1 & 1 & 1 & 1 & 1 & 1 & 1 & 1 & 1 \\
 1 & 1 & 1 & 1 & 1 & 1 & 1 & 1 & 1 & 1
\end{array}
\right)}$&$\tiny{\left(
\begin{array}{ccccccc}
 4 & 1 & 1 & 1 & 1 & 1 & 1 \\
 1 & 3 & 2 & 1 & 1 & 1 & 1 \\
 1 & 1 & 1 & 2 & 2 & 2 & 1 \\
 1 & 1 & 1 & 1 & 1 & 1 & 4
\end{array}
\right)}$&(160,208) \\\hline

$\tiny{\left(
\begin{array}{cccccccccc}
 1 & 1 & 1 & 1 & 1 & 1 & 1 & 1 & 1 & 1 \\
 1 & 1 & 1 & 1 & 1 & 1 & 1 & 1 & 1 & 1 \\
 1 & 1 & 1 & 1 & 1 & 1 & 1 & 1 & 1 & 1 \\
 1 & 1 & 1 & 1 & 1 & 1 & 1 & 1 & 1 & 1
\end{array}
\right)}$&$\tiny{\left(
\begin{array}{ccccccc}
 4 & 1 & 1 & 1 & 1 & 1 & 1 \\
 1 & 3 & 2 & 1 & 1 & 1 & 1 \\
 1 & 1 & 1 & 3 & 2 & 1 & 1 \\
 1 & 1 & 1 & 1 & 1 & 3 & 2
\end{array}
\right)}$&(160,208) \\\hline

$\tiny{\left(
\begin{array}{cccccccccc}
 1 & 1 & 1 & 1 & 1 & 1 & 1 & 1 & 1 & 1 \\
 1 & 1 & 1 & 1 & 1 & 1 & 1 & 1 & 1 & 1 \\
 1 & 1 & 1 & 1 & 1 & 1 & 1 & 1 & 1 & 1 \\
 1 & 1 & 1 & 1 & 1 & 1 & 1 & 1 & 1 & 1
\end{array}
\right)}$&$\tiny{\left(
\begin{array}{ccccccc}
 4 & 1 & 1 & 1 & 1 & 1 & 1 \\
 1 & 3 & 2 & 1 & 1 & 1 & 1 \\
 1 & 1 & 1 & 4 & 1 & 1 & 1 \\
 1 & 1 & 1 & 1 & 2 & 2 & 2
\end{array}
\right)}$&(160,208) \\\hline

$\tiny{\left(
\begin{array}{cccccccccc}
 1 & 1 & 1 & 1 & 1 & 1 & 1 & 1 & 1 & 1 \\
 1 & 1 & 1 & 1 & 1 & 1 & 1 & 1 & 1 & 1 \\
 1 & 1 & 1 & 1 & 1 & 1 & 1 & 1 & 1 & 1 \\
 1 & 1 & 1 & 1 & 1 & 1 & 1 & 1 & 1 & 1
\end{array}
\right)}$&$\tiny{\left(
\begin{array}{ccccccc}
 4 & 1 & 1 & 1 & 1 & 1 & 1 \\
 1 & 4 & 1 & 1 & 1 & 1 & 1 \\
 1 & 1 & 2 & 2 & 2 & 1 & 1 \\
 1 & 1 & 1 & 1 & 1 & 3 & 2
\end{array}
\right)}$&(160,208) \\\hline

$\tiny{\left(
\begin{array}{cccccccccc}
 1 & 1 & 1 & 1 & 1 & 1 & 1 & 1 & 1 & 1 \\
 1 & 1 & 1 & 1 & 1 & 1 & 1 & 1 & 1 & 1 \\
 1 & 1 & 1 & 1 & 1 & 1 & 1 & 1 & 1 & 1 \\
 1 & 1 & 1 & 1 & 1 & 1 & 1 & 1 & 1 & 1
\end{array}
\right)}$&$\tiny{\left(
\begin{array}{ccccccc}
 4 & 1 & 1 & 1 & 1 & 1 & 1 \\
 1 & 4 & 1 & 1 & 1 & 1 & 1 \\
 1 & 1 & 3 & 2 & 1 & 1 & 1 \\
 1 & 1 & 1 & 1 & 2 & 2 & 2
\end{array}
\right)}$&(160,208) \\\hline

$\tiny{\left(
\begin{array}{ccccccccccc}
 1 & 1 & 1 & 1 & 1 & 1 & 1 & 1 & 1 & 1 & 1 \\
 1 & 1 & 1 & 1 & 1 & 1 & 1 & 1 & 1 & 1 & 1 \\
 1 & 1 & 1 & 1 & 1 & 1 & 1 & 1 & 1 & 1 & 1 \\
 1 & 1 & 1 & 1 & 1 & 1 & 1 & 1 & 1 & 1 & 1
\end{array}
\right)}$&$\tiny{\left(
\begin{array}{cccccccc}
 2 & 2 & 2 & 1 & 1 & 1 & 1 & 1 \\
 1 & 1 & 1 & 2 & 2 & 2 & 1 & 1 \\
 1 & 1 & 1 & 1 & 1 & 1 & 4 & 1 \\
 1 & 1 & 1 & 1 & 1 & 1 & 1 & 4
\end{array}
\right)}$&(176,224) \\\hline

$\tiny{\left(
\begin{array}{ccccccccccc}
 1 & 1 & 1 & 1 & 1 & 1 & 1 & 1 & 1 & 1 & 1 \\
 1 & 1 & 1 & 1 & 1 & 1 & 1 & 1 & 1 & 1 & 1 \\
 1 & 1 & 1 & 1 & 1 & 1 & 1 & 1 & 1 & 1 & 1 \\
 1 & 1 & 1 & 1 & 1 & 1 & 1 & 1 & 1 & 1 & 1
\end{array}
\right)}$&$\tiny{\left(
\begin{array}{cccccccc}
 2 & 2 & 2 & 1 & 1 & 1 & 1 & 1 \\
 1 & 1 & 1 & 3 & 2 & 1 & 1 & 1 \\
 1 & 1 & 1 & 1 & 1 & 3 & 2 & 1 \\
 1 & 1 & 1 & 1 & 1 & 1 & 1 & 4
\end{array}
\right)}$&(176,224) \\\hline

$\tiny{\left(
\begin{array}{ccccccccccc}
 1 & 1 & 1 & 1 & 1 & 1 & 1 & 1 & 1 & 1 & 1 \\
 1 & 1 & 1 & 1 & 1 & 1 & 1 & 1 & 1 & 1 & 1 \\
 1 & 1 & 1 & 1 & 1 & 1 & 1 & 1 & 1 & 1 & 1 \\
 1 & 1 & 1 & 1 & 1 & 1 & 1 & 1 & 1 & 1 & 1
\end{array}
\right)}$&$\tiny{\left(
\begin{array}{cccccccc}
 2 & 2 & 2 & 1 & 1 & 1 & 1 & 1 \\
 1 & 1 & 1 & 3 & 2 & 1 & 1 & 1 \\
 1 & 1 & 1 & 1 & 1 & 4 & 1 & 1 \\
 1 & 1 & 1 & 1 & 1 & 1 & 3 & 2
\end{array}
\right)}$&(176,224) \\\hline

$\tiny{\left(
\begin{array}{ccccccccccc}
 1 & 1 & 1 & 1 & 1 & 1 & 1 & 1 & 1 & 1 & 1 \\
 1 & 1 & 1 & 1 & 1 & 1 & 1 & 1 & 1 & 1 & 1 \\
 1 & 1 & 1 & 1 & 1 & 1 & 1 & 1 & 1 & 1 & 1 \\
 1 & 1 & 1 & 1 & 1 & 1 & 1 & 1 & 1 & 1 & 1
\end{array}
\right)}$&$\tiny{\left(
\begin{array}{cccccccc}
 2 & 2 & 2 & 1 & 1 & 1 & 1 & 1 \\
 1 & 1 & 1 & 4 & 1 & 1 & 1 & 1 \\
 1 & 1 & 1 & 1 & 2 & 2 & 2 & 1 \\
 1 & 1 & 1 & 1 & 1 & 1 & 1 & 4
\end{array}
\right)}$&(176,224) \\\hline

$\tiny{\left(
\begin{array}{ccccccccccc}
 1 & 1 & 1 & 1 & 1 & 1 & 1 & 1 & 1 & 1 & 1 \\
 1 & 1 & 1 & 1 & 1 & 1 & 1 & 1 & 1 & 1 & 1 \\
 1 & 1 & 1 & 1 & 1 & 1 & 1 & 1 & 1 & 1 & 1 \\
 1 & 1 & 1 & 1 & 1 & 1 & 1 & 1 & 1 & 1 & 1
\end{array}
\right)}$&$\tiny{\left(
\begin{array}{cccccccc}
 2 & 2 & 2 & 1 & 1 & 1 & 1 & 1 \\
 1 & 1 & 1 & 4 & 1 & 1 & 1 & 1 \\
 1 & 1 & 1 & 1 & 3 & 2 & 1 & 1 \\
 1 & 1 & 1 & 1 & 1 & 1 & 3 & 2
\end{array}
\right)}$&(176,224) \\\hline

\end{tabular}
\caption{Results for the tetra-quadric in $\mathbb{P}^1\times\mathbb{P}^1\times\mathbb{P}^1\times\mathbb{P}^1$, continued.}
\end{center}
\end{table}

\clearpage

\begin{table}[!h]
\begin{center}
\begin{tabular}{|l|l|l|}\hline
$B$&$C$&\footnotesize{(${\rm ind}(B)$,${\rm ind}(C)$)}\\\hline\hline

$\tiny{\left(
\begin{array}{ccccccccccc}
 1 & 1 & 1 & 1 & 1 & 1 & 1 & 1 & 1 & 1 & 1 \\
 1 & 1 & 1 & 1 & 1 & 1 & 1 & 1 & 1 & 1 & 1 \\
 1 & 1 & 1 & 1 & 1 & 1 & 1 & 1 & 1 & 1 & 1 \\
 1 & 1 & 1 & 1 & 1 & 1 & 1 & 1 & 1 & 1 & 1
\end{array}
\right)}$&$\tiny{\left(
\begin{array}{cccccccc}
 2 & 2 & 2 & 1 & 1 & 1 & 1 & 1 \\
 1 & 1 & 1 & 4 & 1 & 1 & 1 & 1 \\
 1 & 1 & 1 & 1 & 4 & 1 & 1 & 1 \\
 1 & 1 & 1 & 1 & 1 & 2 & 2 & 2
\end{array}
\right)}$&(176,224) \\\hline

$\tiny{\left(
\begin{array}{ccccccccccc}
 1 & 1 & 1 & 1 & 1 & 1 & 1 & 1 & 1 & 1 & 1 \\
 1 & 1 & 1 & 1 & 1 & 1 & 1 & 1 & 1 & 1 & 1 \\
 1 & 1 & 1 & 1 & 1 & 1 & 1 & 1 & 1 & 1 & 1 \\
 1 & 1 & 1 & 1 & 1 & 1 & 1 & 1 & 1 & 1 & 1
\end{array}
\right)}$&$\tiny{\left(
\begin{array}{cccccccc}
 3 & 2 & 1 & 1 & 1 & 1 & 1 & 1 \\
 1 & 1 & 2 & 2 & 2 & 1 & 1 & 1 \\
 1 & 1 & 1 & 1 & 1 & 3 & 2 & 1 \\
 1 & 1 & 1 & 1 & 1 & 1 & 1 & 4
\end{array}
\right)}$&(176,224) \\\hline

$\tiny{\left(
\begin{array}{ccccccccccc}
 1 & 1 & 1 & 1 & 1 & 1 & 1 & 1 & 1 & 1 & 1 \\
 1 & 1 & 1 & 1 & 1 & 1 & 1 & 1 & 1 & 1 & 1 \\
 1 & 1 & 1 & 1 & 1 & 1 & 1 & 1 & 1 & 1 & 1 \\
 1 & 1 & 1 & 1 & 1 & 1 & 1 & 1 & 1 & 1 & 1
\end{array}
\right)}$&$\tiny{\left(
\begin{array}{cccccccc}
 3 & 2 & 1 & 1 & 1 & 1 & 1 & 1 \\
 1 & 1 & 2 & 2 & 2 & 1 & 1 & 1 \\
 1 & 1 & 1 & 1 & 1 & 4 & 1 & 1 \\
 1 & 1 & 1 & 1 & 1 & 1 & 3 & 2
\end{array}
\right)}$&(176,224) \\\hline

$\tiny{\left(
\begin{array}{ccccccccccc}
 1 & 1 & 1 & 1 & 1 & 1 & 1 & 1 & 1 & 1 & 1 \\
 1 & 1 & 1 & 1 & 1 & 1 & 1 & 1 & 1 & 1 & 1 \\
 1 & 1 & 1 & 1 & 1 & 1 & 1 & 1 & 1 & 1 & 1 \\
 1 & 1 & 1 & 1 & 1 & 1 & 1 & 1 & 1 & 1 & 1
\end{array}
\right)}$&$\tiny{\left(
\begin{array}{cccccccc}
 3 & 2 & 1 & 1 & 1 & 1 & 1 & 1 \\
 1 & 1 & 3 & 2 & 1 & 1 & 1 & 1 \\
 1 & 1 & 1 & 1 & 2 & 2 & 2 & 1 \\
 1 & 1 & 1 & 1 & 1 & 1 & 1 & 4
\end{array}
\right)}$&(176,224) \\\hline

$\tiny{\left(
\begin{array}{ccccccccccc}
 1 & 1 & 1 & 1 & 1 & 1 & 1 & 1 & 1 & 1 & 1 \\
 1 & 1 & 1 & 1 & 1 & 1 & 1 & 1 & 1 & 1 & 1 \\
 1 & 1 & 1 & 1 & 1 & 1 & 1 & 1 & 1 & 1 & 1 \\
 1 & 1 & 1 & 1 & 1 & 1 & 1 & 1 & 1 & 1 & 1
\end{array}
\right)}$&$\tiny{\left(
\begin{array}{cccccccc}
 3 & 2 & 1 & 1 & 1 & 1 & 1 & 1 \\
 1 & 1 & 3 & 2 & 1 & 1 & 1 & 1 \\
 1 & 1 & 1 & 1 & 3 & 2 & 1 & 1 \\
 1 & 1 & 1 & 1 & 1 & 1 & 3 & 2
\end{array}
\right)}$&(176,224) \\\hline

$\tiny{\left(
\begin{array}{ccccccccccc}
 1 & 1 & 1 & 1 & 1 & 1 & 1 & 1 & 1 & 1 & 1 \\
 1 & 1 & 1 & 1 & 1 & 1 & 1 & 1 & 1 & 1 & 1 \\
 1 & 1 & 1 & 1 & 1 & 1 & 1 & 1 & 1 & 1 & 1 \\
 1 & 1 & 1 & 1 & 1 & 1 & 1 & 1 & 1 & 1 & 1
\end{array}
\right)}$&$\tiny{\left(
\begin{array}{cccccccc}
 3 & 2 & 1 & 1 & 1 & 1 & 1 & 1 \\
 1 & 1 & 3 & 2 & 1 & 1 & 1 & 1 \\
 1 & 1 & 1 & 1 & 4 & 1 & 1 & 1 \\
 1 & 1 & 1 & 1 & 1 & 2 & 2 & 2
\end{array}
\right)}$&(176,224) \\\hline

$\tiny{\left(
\begin{array}{ccccccccccc}
 1 & 1 & 1 & 1 & 1 & 1 & 1 & 1 & 1 & 1 & 1 \\
 1 & 1 & 1 & 1 & 1 & 1 & 1 & 1 & 1 & 1 & 1 \\
 1 & 1 & 1 & 1 & 1 & 1 & 1 & 1 & 1 & 1 & 1 \\
 1 & 1 & 1 & 1 & 1 & 1 & 1 & 1 & 1 & 1 & 1
\end{array}
\right)}$&$\tiny{\left(
\begin{array}{cccccccc}
 3 & 2 & 1 & 1 & 1 & 1 & 1 & 1 \\
 1 & 1 & 4 & 1 & 1 & 1 & 1 & 1 \\
 1 & 1 & 1 & 2 & 2 & 2 & 1 & 1 \\
 1 & 1 & 1 & 1 & 1 & 1 & 3 & 2
\end{array}
\right)}$&(176,224) \\\hline

$\tiny{\left(
\begin{array}{ccccccccccc}
 1 & 1 & 1 & 1 & 1 & 1 & 1 & 1 & 1 & 1 & 1 \\
 1 & 1 & 1 & 1 & 1 & 1 & 1 & 1 & 1 & 1 & 1 \\
 1 & 1 & 1 & 1 & 1 & 1 & 1 & 1 & 1 & 1 & 1 \\
 1 & 1 & 1 & 1 & 1 & 1 & 1 & 1 & 1 & 1 & 1
\end{array}
\right)}$&$\tiny{\left(
\begin{array}{cccccccc}
 3 & 2 & 1 & 1 & 1 & 1 & 1 & 1 \\
 1 & 1 & 4 & 1 & 1 & 1 & 1 & 1 \\
 1 & 1 & 1 & 3 & 2 & 1 & 1 & 1 \\
 1 & 1 & 1 & 1 & 1 & 2 & 2 & 2
\end{array}
\right)}$&(176,224) \\\hline

$\tiny{\left(
\begin{array}{ccccccccccc}
 1 & 1 & 1 & 1 & 1 & 1 & 1 & 1 & 1 & 1 & 1 \\
 1 & 1 & 1 & 1 & 1 & 1 & 1 & 1 & 1 & 1 & 1 \\
 1 & 1 & 1 & 1 & 1 & 1 & 1 & 1 & 1 & 1 & 1 \\
 1 & 1 & 1 & 1 & 1 & 1 & 1 & 1 & 1 & 1 & 1
\end{array}
\right)}$&$\tiny{\left(
\begin{array}{cccccccc}
 4 & 1 & 1 & 1 & 1 & 1 & 1 & 1 \\
 1 & 2 & 2 & 2 & 1 & 1 & 1 & 1 \\
 1 & 1 & 1 & 1 & 2 & 2 & 2 & 1 \\
 1 & 1 & 1 & 1 & 1 & 1 & 1 & 4
\end{array}
\right)}$&(176,224) \\\hline

$\tiny{\left(
\begin{array}{ccccccccccc}
 1 & 1 & 1 & 1 & 1 & 1 & 1 & 1 & 1 & 1 & 1 \\
 1 & 1 & 1 & 1 & 1 & 1 & 1 & 1 & 1 & 1 & 1 \\
 1 & 1 & 1 & 1 & 1 & 1 & 1 & 1 & 1 & 1 & 1 \\
 1 & 1 & 1 & 1 & 1 & 1 & 1 & 1 & 1 & 1 & 1
\end{array}
\right)}$&$\tiny{\left(
\begin{array}{cccccccc}
 4 & 1 & 1 & 1 & 1 & 1 & 1 & 1 \\
 1 & 2 & 2 & 2 & 1 & 1 & 1 & 1 \\
 1 & 1 & 1 & 1 & 3 & 2 & 1 & 1 \\
 1 & 1 & 1 & 1 & 1 & 1 & 3 & 2
\end{array}
\right)}$&(176,224) \\\hline

$\tiny{\left(
\begin{array}{ccccccccccc}
 1 & 1 & 1 & 1 & 1 & 1 & 1 & 1 & 1 & 1 & 1 \\
 1 & 1 & 1 & 1 & 1 & 1 & 1 & 1 & 1 & 1 & 1 \\
 1 & 1 & 1 & 1 & 1 & 1 & 1 & 1 & 1 & 1 & 1 \\
 1 & 1 & 1 & 1 & 1 & 1 & 1 & 1 & 1 & 1 & 1
\end{array}
\right)}$&$\tiny{\left(
\begin{array}{cccccccc}
 4 & 1 & 1 & 1 & 1 & 1 & 1 & 1 \\
 1 & 2 & 2 & 2 & 1 & 1 & 1 & 1 \\
 1 & 1 & 1 & 1 & 4 & 1 & 1 & 1 \\
 1 & 1 & 1 & 1 & 1 & 2 & 2 & 2
\end{array}
\right)}$&(176,224) \\\hline

$\tiny{\left(
\begin{array}{ccccccccccc}
 1 & 1 & 1 & 1 & 1 & 1 & 1 & 1 & 1 & 1 & 1 \\
 1 & 1 & 1 & 1 & 1 & 1 & 1 & 1 & 1 & 1 & 1 \\
 1 & 1 & 1 & 1 & 1 & 1 & 1 & 1 & 1 & 1 & 1 \\
 1 & 1 & 1 & 1 & 1 & 1 & 1 & 1 & 1 & 1 & 1
\end{array}
\right)}$&$\tiny{\left(
\begin{array}{cccccccc}
 4 & 1 & 1 & 1 & 1 & 1 & 1 & 1 \\
 1 & 3 & 2 & 1 & 1 & 1 & 1 & 1 \\
 1 & 1 & 1 & 2 & 2 & 2 & 1 & 1 \\
 1 & 1 & 1 & 1 & 1 & 1 & 3 & 2
\end{array}
\right)}$&(176,224) \\\hline

$\tiny{\left(
\begin{array}{ccccccccccc}
 1 & 1 & 1 & 1 & 1 & 1 & 1 & 1 & 1 & 1 & 1 \\
 1 & 1 & 1 & 1 & 1 & 1 & 1 & 1 & 1 & 1 & 1 \\
 1 & 1 & 1 & 1 & 1 & 1 & 1 & 1 & 1 & 1 & 1 \\
 1 & 1 & 1 & 1 & 1 & 1 & 1 & 1 & 1 & 1 & 1
\end{array}
\right)}$&$\tiny{\left(
\begin{array}{cccccccc}
 4 & 1 & 1 & 1 & 1 & 1 & 1 & 1 \\
 1 & 3 & 2 & 1 & 1 & 1 & 1 & 1 \\
 1 & 1 & 1 & 3 & 2 & 1 & 1 & 1 \\
 1 & 1 & 1 & 1 & 1 & 2 & 2 & 2
\end{array}
\right)}$&(176,224) \\\hline

$\tiny{\left(
\begin{array}{ccccccccccc}
 1 & 1 & 1 & 1 & 1 & 1 & 1 & 1 & 1 & 1 & 1 \\
 1 & 1 & 1 & 1 & 1 & 1 & 1 & 1 & 1 & 1 & 1 \\
 1 & 1 & 1 & 1 & 1 & 1 & 1 & 1 & 1 & 1 & 1 \\
 1 & 1 & 1 & 1 & 1 & 1 & 1 & 1 & 1 & 1 & 1
\end{array}
\right)}$&$\tiny{\left(
\begin{array}{cccccccc}
 4 & 1 & 1 & 1 & 1 & 1 & 1 & 1 \\
 1 & 4 & 1 & 1 & 1 & 1 & 1 & 1 \\
 1 & 1 & 2 & 2 & 2 & 1 & 1 & 1 \\
 1 & 1 & 1 & 1 & 1 & 2 & 2 & 2
\end{array}
\right)}$&(176,224) \\\hline

$\tiny{\left(
\begin{array}{cccccccccccc}
 1 & 1 & 1 & 1 & 1 & 1 & 1 & 1 & 1 & 1 & 1 & 1 \\
 1 & 1 & 1 & 1 & 1 & 1 & 1 & 1 & 1 & 1 & 1 & 1 \\
 1 & 1 & 1 & 1 & 1 & 1 & 1 & 1 & 1 & 1 & 1 & 1 \\
 1 & 1 & 1 & 1 & 1 & 1 & 1 & 1 & 1 & 1 & 1 & 1
\end{array}
\right)}$&$\tiny{\left(
\begin{array}{ccccccccc}
 2 & 2 & 2 & 1 & 1 & 1 & 1 & 1 & 1 \\
 1 & 1 & 1 & 2 & 2 & 2 & 1 & 1 & 1 \\
 1 & 1 & 1 & 1 & 1 & 1 & 3 & 2 & 1 \\
 1 & 1 & 1 & 1 & 1 & 1 & 1 & 1 & 4
\end{array}
\right)}$&(192,240) \\\hline

$\tiny{\left(
\begin{array}{cccccccccccc}
 1 & 1 & 1 & 1 & 1 & 1 & 1 & 1 & 1 & 1 & 1 & 1 \\
 1 & 1 & 1 & 1 & 1 & 1 & 1 & 1 & 1 & 1 & 1 & 1 \\
 1 & 1 & 1 & 1 & 1 & 1 & 1 & 1 & 1 & 1 & 1 & 1 \\
 1 & 1 & 1 & 1 & 1 & 1 & 1 & 1 & 1 & 1 & 1 & 1
\end{array}
\right)}$&$\tiny{\left(
\begin{array}{ccccccccc}
 2 & 2 & 2 & 1 & 1 & 1 & 1 & 1 & 1 \\
 1 & 1 & 1 & 2 & 2 & 2 & 1 & 1 & 1 \\
 1 & 1 & 1 & 1 & 1 & 1 & 4 & 1 & 1 \\
 1 & 1 & 1 & 1 & 1 & 1 & 1 & 3 & 2
\end{array}
\right)}$&(192,240) \\\hline

$\tiny{\left(
\begin{array}{cccccccccccc}
 1 & 1 & 1 & 1 & 1 & 1 & 1 & 1 & 1 & 1 & 1 & 1 \\
 1 & 1 & 1 & 1 & 1 & 1 & 1 & 1 & 1 & 1 & 1 & 1 \\
 1 & 1 & 1 & 1 & 1 & 1 & 1 & 1 & 1 & 1 & 1 & 1 \\
 1 & 1 & 1 & 1 & 1 & 1 & 1 & 1 & 1 & 1 & 1 & 1
\end{array}
\right)}$&$\tiny{\left(
\begin{array}{ccccccccc}
 2 & 2 & 2 & 1 & 1 & 1 & 1 & 1 & 1 \\
 1 & 1 & 1 & 3 & 2 & 1 & 1 & 1 & 1 \\
 1 & 1 & 1 & 1 & 1 & 2 & 2 & 2 & 1 \\
 1 & 1 & 1 & 1 & 1 & 1 & 1 & 1 & 4
\end{array}
\right)}$&(192,240) \\\hline

$\tiny{\left(
\begin{array}{cccccccccccc}
 1 & 1 & 1 & 1 & 1 & 1 & 1 & 1 & 1 & 1 & 1 & 1 \\
 1 & 1 & 1 & 1 & 1 & 1 & 1 & 1 & 1 & 1 & 1 & 1 \\
 1 & 1 & 1 & 1 & 1 & 1 & 1 & 1 & 1 & 1 & 1 & 1 \\
 1 & 1 & 1 & 1 & 1 & 1 & 1 & 1 & 1 & 1 & 1 & 1
\end{array}
\right)}$&$\tiny{\left(
\begin{array}{ccccccccc}
 2 & 2 & 2 & 1 & 1 & 1 & 1 & 1 & 1 \\
 1 & 1 & 1 & 3 & 2 & 1 & 1 & 1 & 1 \\
 1 & 1 & 1 & 1 & 1 & 3 & 2 & 1 & 1 \\
 1 & 1 & 1 & 1 & 1 & 1 & 1 & 3 & 2
\end{array}
\right)}$&(192,240) \\\hline

\end{tabular}
\caption{Results for the tetra-quadric in $\mathbb{P}^1\times\mathbb{P}^1\times\mathbb{P}^1\times\mathbb{P}^1$, continued.}
\end{center}
\end{table}

\clearpage

\begin{table}[!h]
\begin{center}
\begin{tabular}{|l|l|l|}\hline
$B$&$C$&\footnotesize{(${\rm ind}(B)$,${\rm ind}(C)$)}\\\hline\hline

$\tiny{\left(
\begin{array}{cccccccccccc}
 1 & 1 & 1 & 1 & 1 & 1 & 1 & 1 & 1 & 1 & 1 & 1 \\
 1 & 1 & 1 & 1 & 1 & 1 & 1 & 1 & 1 & 1 & 1 & 1 \\
 1 & 1 & 1 & 1 & 1 & 1 & 1 & 1 & 1 & 1 & 1 & 1 \\
 1 & 1 & 1 & 1 & 1 & 1 & 1 & 1 & 1 & 1 & 1 & 1
\end{array}
\right)}$&$\tiny{\left(
\begin{array}{ccccccccc}
 2 & 2 & 2 & 1 & 1 & 1 & 1 & 1 & 1 \\
 1 & 1 & 1 & 3 & 2 & 1 & 1 & 1 & 1 \\
 1 & 1 & 1 & 1 & 1 & 4 & 1 & 1 & 1 \\
 1 & 1 & 1 & 1 & 1 & 1 & 2 & 2 & 2
\end{array}
\right)}$&(192,240) \\\hline

$\tiny{\left(
\begin{array}{cccccccccccc}
 1 & 1 & 1 & 1 & 1 & 1 & 1 & 1 & 1 & 1 & 1 & 1 \\
 1 & 1 & 1 & 1 & 1 & 1 & 1 & 1 & 1 & 1 & 1 & 1 \\
 1 & 1 & 1 & 1 & 1 & 1 & 1 & 1 & 1 & 1 & 1 & 1 \\
 1 & 1 & 1 & 1 & 1 & 1 & 1 & 1 & 1 & 1 & 1 & 1
\end{array}
\right)}$&$\tiny{\left(
\begin{array}{ccccccccc}
 2 & 2 & 2 & 1 & 1 & 1 & 1 & 1 & 1 \\
 1 & 1 & 1 & 4 & 1 & 1 & 1 & 1 & 1 \\
 1 & 1 & 1 & 1 & 2 & 2 & 2 & 1 & 1 \\
 1 & 1 & 1 & 1 & 1 & 1 & 1 & 3 & 2
\end{array}
\right)}$&(192,240) \\\hline

$\tiny{\left(
\begin{array}{cccccccccccc}
 1 & 1 & 1 & 1 & 1 & 1 & 1 & 1 & 1 & 1 & 1 & 1 \\
 1 & 1 & 1 & 1 & 1 & 1 & 1 & 1 & 1 & 1 & 1 & 1 \\
 1 & 1 & 1 & 1 & 1 & 1 & 1 & 1 & 1 & 1 & 1 & 1 \\
 1 & 1 & 1 & 1 & 1 & 1 & 1 & 1 & 1 & 1 & 1 & 1
\end{array}
\right)}$&$\tiny{\left(
\begin{array}{ccccccccc}
 2 & 2 & 2 & 1 & 1 & 1 & 1 & 1 & 1 \\
 1 & 1 & 1 & 4 & 1 & 1 & 1 & 1 & 1 \\
 1 & 1 & 1 & 1 & 3 & 2 & 1 & 1 & 1 \\
 1 & 1 & 1 & 1 & 1 & 1 & 2 & 2 & 2
\end{array}
\right)}$&(192,240) \\\hline

$\tiny{\left(
\begin{array}{cccccccccccc}
 1 & 1 & 1 & 1 & 1 & 1 & 1 & 1 & 1 & 1 & 1 & 1 \\
 1 & 1 & 1 & 1 & 1 & 1 & 1 & 1 & 1 & 1 & 1 & 1 \\
 1 & 1 & 1 & 1 & 1 & 1 & 1 & 1 & 1 & 1 & 1 & 1 \\
 1 & 1 & 1 & 1 & 1 & 1 & 1 & 1 & 1 & 1 & 1 & 1
\end{array}
\right)}$&$\tiny{\left(
\begin{array}{ccccccccc}
 3 & 2 & 1 & 1 & 1 & 1 & 1 & 1 & 1 \\
 1 & 1 & 2 & 2 & 2 & 1 & 1 & 1 & 1 \\
 1 & 1 & 1 & 1 & 1 & 2 & 2 & 2 & 1 \\
 1 & 1 & 1 & 1 & 1 & 1 & 1 & 1 & 4
\end{array}
\right)}$&(192,240) \\\hline

$\tiny{\left(
\begin{array}{cccccccccccc}
 1 & 1 & 1 & 1 & 1 & 1 & 1 & 1 & 1 & 1 & 1 & 1 \\
 1 & 1 & 1 & 1 & 1 & 1 & 1 & 1 & 1 & 1 & 1 & 1 \\
 1 & 1 & 1 & 1 & 1 & 1 & 1 & 1 & 1 & 1 & 1 & 1 \\
 1 & 1 & 1 & 1 & 1 & 1 & 1 & 1 & 1 & 1 & 1 & 1
\end{array}
\right)}$&$\tiny{\left(
\begin{array}{ccccccccc}
 3 & 2 & 1 & 1 & 1 & 1 & 1 & 1 & 1 \\
 1 & 1 & 2 & 2 & 2 & 1 & 1 & 1 & 1 \\
 1 & 1 & 1 & 1 & 1 & 3 & 2 & 1 & 1 \\
 1 & 1 & 1 & 1 & 1 & 1 & 1 & 3 & 2
\end{array}
\right)}$&(192,240) \\\hline

$\tiny{\left(
\begin{array}{cccccccccccc}
 1 & 1 & 1 & 1 & 1 & 1 & 1 & 1 & 1 & 1 & 1 & 1 \\
 1 & 1 & 1 & 1 & 1 & 1 & 1 & 1 & 1 & 1 & 1 & 1 \\
 1 & 1 & 1 & 1 & 1 & 1 & 1 & 1 & 1 & 1 & 1 & 1 \\
 1 & 1 & 1 & 1 & 1 & 1 & 1 & 1 & 1 & 1 & 1 & 1
\end{array}
\right)}$&$\tiny{\left(
\begin{array}{ccccccccc}
 3 & 2 & 1 & 1 & 1 & 1 & 1 & 1 & 1 \\
 1 & 1 & 2 & 2 & 2 & 1 & 1 & 1 & 1 \\
 1 & 1 & 1 & 1 & 1 & 4 & 1 & 1 & 1 \\
 1 & 1 & 1 & 1 & 1 & 1 & 2 & 2 & 2
\end{array}
\right)}$&(192,240) \\\hline

$\tiny{\left(
\begin{array}{cccccccccccc}
 1 & 1 & 1 & 1 & 1 & 1 & 1 & 1 & 1 & 1 & 1 & 1 \\
 1 & 1 & 1 & 1 & 1 & 1 & 1 & 1 & 1 & 1 & 1 & 1 \\
 1 & 1 & 1 & 1 & 1 & 1 & 1 & 1 & 1 & 1 & 1 & 1 \\
 1 & 1 & 1 & 1 & 1 & 1 & 1 & 1 & 1 & 1 & 1 & 1
\end{array}
\right)}$&$\tiny{\left(
\begin{array}{ccccccccc}
 3 & 2 & 1 & 1 & 1 & 1 & 1 & 1 & 1 \\
 1 & 1 & 3 & 2 & 1 & 1 & 1 & 1 & 1 \\
 1 & 1 & 1 & 1 & 2 & 2 & 2 & 1 & 1 \\
 1 & 1 & 1 & 1 & 1 & 1 & 1 & 3 & 2
\end{array}
\right)}$&(192,240) \\\hline

$\tiny{\left(
\begin{array}{cccccccccccc}
 1 & 1 & 1 & 1 & 1 & 1 & 1 & 1 & 1 & 1 & 1 & 1 \\
 1 & 1 & 1 & 1 & 1 & 1 & 1 & 1 & 1 & 1 & 1 & 1 \\
 1 & 1 & 1 & 1 & 1 & 1 & 1 & 1 & 1 & 1 & 1 & 1 \\
 1 & 1 & 1 & 1 & 1 & 1 & 1 & 1 & 1 & 1 & 1 & 1
\end{array}
\right)}$&$\tiny{\left(
\begin{array}{ccccccccc}
 3 & 2 & 1 & 1 & 1 & 1 & 1 & 1 & 1 \\
 1 & 1 & 3 & 2 & 1 & 1 & 1 & 1 & 1 \\
 1 & 1 & 1 & 1 & 3 & 2 & 1 & 1 & 1 \\
 1 & 1 & 1 & 1 & 1 & 1 & 2 & 2 & 2
\end{array}
\right)}$&(192,240) \\\hline

$\tiny{\left(
\begin{array}{cccccccccccc}
 1 & 1 & 1 & 1 & 1 & 1 & 1 & 1 & 1 & 1 & 1 & 1 \\
 1 & 1 & 1 & 1 & 1 & 1 & 1 & 1 & 1 & 1 & 1 & 1 \\
 1 & 1 & 1 & 1 & 1 & 1 & 1 & 1 & 1 & 1 & 1 & 1 \\
 1 & 1 & 1 & 1 & 1 & 1 & 1 & 1 & 1 & 1 & 1 & 1
\end{array}
\right)}$&$\tiny{\left(
\begin{array}{ccccccccc}
 3 & 2 & 1 & 1 & 1 & 1 & 1 & 1 & 1 \\
 1 & 1 & 4 & 1 & 1 & 1 & 1 & 1 & 1 \\
 1 & 1 & 1 & 2 & 2 & 2 & 1 & 1 & 1 \\
 1 & 1 & 1 & 1 & 1 & 1 & 2 & 2 & 2
\end{array}
\right)}$&(192,240) \\\hline

$\tiny{\left(
\begin{array}{cccccccccccc}
 1 & 1 & 1 & 1 & 1 & 1 & 1 & 1 & 1 & 1 & 1 & 1 \\
 1 & 1 & 1 & 1 & 1 & 1 & 1 & 1 & 1 & 1 & 1 & 1 \\
 1 & 1 & 1 & 1 & 1 & 1 & 1 & 1 & 1 & 1 & 1 & 1 \\
 1 & 1 & 1 & 1 & 1 & 1 & 1 & 1 & 1 & 1 & 1 & 1
\end{array}
\right)}$&$\tiny{\left(
\begin{array}{ccccccccc}
 4 & 1 & 1 & 1 & 1 & 1 & 1 & 1 & 1 \\
 1 & 2 & 2 & 2 & 1 & 1 & 1 & 1 & 1 \\
 1 & 1 & 1 & 1 & 2 & 2 & 2 & 1 & 1 \\
 1 & 1 & 1 & 1 & 1 & 1 & 1 & 3 & 2
\end{array}
\right)}$&(192,240) \\\hline

$\tiny{\left(
\begin{array}{cccccccccccc}
 1 & 1 & 1 & 1 & 1 & 1 & 1 & 1 & 1 & 1 & 1 & 1 \\
 1 & 1 & 1 & 1 & 1 & 1 & 1 & 1 & 1 & 1 & 1 & 1 \\
 1 & 1 & 1 & 1 & 1 & 1 & 1 & 1 & 1 & 1 & 1 & 1 \\
 1 & 1 & 1 & 1 & 1 & 1 & 1 & 1 & 1 & 1 & 1 & 1
\end{array}
\right)}$&$\tiny{\left(
\begin{array}{ccccccccc}
 4 & 1 & 1 & 1 & 1 & 1 & 1 & 1 & 1 \\
 1 & 2 & 2 & 2 & 1 & 1 & 1 & 1 & 1 \\
 1 & 1 & 1 & 1 & 3 & 2 & 1 & 1 & 1 \\
 1 & 1 & 1 & 1 & 1 & 1 & 2 & 2 & 2
\end{array}
\right)}$&(192,240) \\\hline

$\tiny{\left(
\begin{array}{cccccccccccc}
 1 & 1 & 1 & 1 & 1 & 1 & 1 & 1 & 1 & 1 & 1 & 1 \\
 1 & 1 & 1 & 1 & 1 & 1 & 1 & 1 & 1 & 1 & 1 & 1 \\
 1 & 1 & 1 & 1 & 1 & 1 & 1 & 1 & 1 & 1 & 1 & 1 \\
 1 & 1 & 1 & 1 & 1 & 1 & 1 & 1 & 1 & 1 & 1 & 1
\end{array}
\right)}$&$\tiny{\left(
\begin{array}{ccccccccc}
 4 & 1 & 1 & 1 & 1 & 1 & 1 & 1 & 1 \\
 1 & 3 & 2 & 1 & 1 & 1 & 1 & 1 & 1 \\
 1 & 1 & 1 & 2 & 2 & 2 & 1 & 1 & 1 \\
 1 & 1 & 1 & 1 & 1 & 1 & 2 & 2 & 2
\end{array}
\right)}$&(192,240) \\\hline

$\tiny{\left(
\begin{array}{ccccccccccccc}
 1 & 1 & 1 & 1 & 1 & 1 & 1 & 1 & 1 & 1 & 1 & 1 & 1 \\
 1 & 1 & 1 & 1 & 1 & 1 & 1 & 1 & 1 & 1 & 1 & 1 & 1 \\
 1 & 1 & 1 & 1 & 1 & 1 & 1 & 1 & 1 & 1 & 1 & 1 & 1 \\
 1 & 1 & 1 & 1 & 1 & 1 & 1 & 1 & 1 & 1 & 1 & 1 & 1
\end{array}
\right)}$&$\tiny{\left(
\begin{array}{cccccccccc}
 2 & 2 & 2 & 1 & 1 & 1 & 1 & 1 & 1 & 1 \\
 1 & 1 & 1 & 2 & 2 & 2 & 1 & 1 & 1 & 1 \\
 1 & 1 & 1 & 1 & 1 & 1 & 2 & 2 & 2 & 1 \\
 1 & 1 & 1 & 1 & 1 & 1 & 1 & 1 & 1 & 4
\end{array}
\right)}$&(208,256) \\\hline

$\tiny{\left(
\begin{array}{ccccccccccccc}
 1 & 1 & 1 & 1 & 1 & 1 & 1 & 1 & 1 & 1 & 1 & 1 & 1 \\
 1 & 1 & 1 & 1 & 1 & 1 & 1 & 1 & 1 & 1 & 1 & 1 & 1 \\
 1 & 1 & 1 & 1 & 1 & 1 & 1 & 1 & 1 & 1 & 1 & 1 & 1 \\
 1 & 1 & 1 & 1 & 1 & 1 & 1 & 1 & 1 & 1 & 1 & 1 & 1
\end{array}
\right)}$&$\tiny{\left(
\begin{array}{cccccccccc}
 2 & 2 & 2 & 1 & 1 & 1 & 1 & 1 & 1 & 1 \\
 1 & 1 & 1 & 2 & 2 & 2 & 1 & 1 & 1 & 1 \\
 1 & 1 & 1 & 1 & 1 & 1 & 3 & 2 & 1 & 1 \\
 1 & 1 & 1 & 1 & 1 & 1 & 1 & 1 & 3 & 2
\end{array}
\right)}$&(208,256) \\\hline

$\tiny{\left(
\begin{array}{ccccccccccccc}
 1 & 1 & 1 & 1 & 1 & 1 & 1 & 1 & 1 & 1 & 1 & 1 & 1 \\
 1 & 1 & 1 & 1 & 1 & 1 & 1 & 1 & 1 & 1 & 1 & 1 & 1 \\
 1 & 1 & 1 & 1 & 1 & 1 & 1 & 1 & 1 & 1 & 1 & 1 & 1 \\
 1 & 1 & 1 & 1 & 1 & 1 & 1 & 1 & 1 & 1 & 1 & 1 & 1
\end{array}
\right)}$&$\tiny{\left(
\begin{array}{cccccccccc}
 2 & 2 & 2 & 1 & 1 & 1 & 1 & 1 & 1 & 1 \\
 1 & 1 & 1 & 2 & 2 & 2 & 1 & 1 & 1 & 1 \\
 1 & 1 & 1 & 1 & 1 & 1 & 4 & 1 & 1 & 1 \\
 1 & 1 & 1 & 1 & 1 & 1 & 1 & 2 & 2 & 2
\end{array}
\right)}$&(208,256) \\\hline

$\tiny{\left(
\begin{array}{ccccccccccccc}
 1 & 1 & 1 & 1 & 1 & 1 & 1 & 1 & 1 & 1 & 1 & 1 & 1 \\
 1 & 1 & 1 & 1 & 1 & 1 & 1 & 1 & 1 & 1 & 1 & 1 & 1 \\
 1 & 1 & 1 & 1 & 1 & 1 & 1 & 1 & 1 & 1 & 1 & 1 & 1 \\
 1 & 1 & 1 & 1 & 1 & 1 & 1 & 1 & 1 & 1 & 1 & 1 & 1
\end{array}
\right)}$&$\tiny{\left(
\begin{array}{cccccccccc}
 2 & 2 & 2 & 1 & 1 & 1 & 1 & 1 & 1 & 1 \\
 1 & 1 & 1 & 3 & 2 & 1 & 1 & 1 & 1 & 1 \\
 1 & 1 & 1 & 1 & 1 & 2 & 2 & 2 & 1 & 1 \\
 1 & 1 & 1 & 1 & 1 & 1 & 1 & 1 & 3 & 2
\end{array}
\right)}$&(208,256) \\\hline

$\tiny{\left(
\begin{array}{ccccccccccccc}
 1 & 1 & 1 & 1 & 1 & 1 & 1 & 1 & 1 & 1 & 1 & 1 & 1 \\
 1 & 1 & 1 & 1 & 1 & 1 & 1 & 1 & 1 & 1 & 1 & 1 & 1 \\
 1 & 1 & 1 & 1 & 1 & 1 & 1 & 1 & 1 & 1 & 1 & 1 & 1 \\
 1 & 1 & 1 & 1 & 1 & 1 & 1 & 1 & 1 & 1 & 1 & 1 & 1
\end{array}
\right)}$&$\tiny{\left(
\begin{array}{cccccccccc}
 2 & 2 & 2 & 1 & 1 & 1 & 1 & 1 & 1 & 1 \\
 1 & 1 & 1 & 3 & 2 & 1 & 1 & 1 & 1 & 1 \\
 1 & 1 & 1 & 1 & 1 & 3 & 2 & 1 & 1 & 1 \\
 1 & 1 & 1 & 1 & 1 & 1 & 1 & 2 & 2 & 2
\end{array}
\right)}$&(208,256) \\\hline

$\tiny{\left(
\begin{array}{ccccccccccccc}
 1 & 1 & 1 & 1 & 1 & 1 & 1 & 1 & 1 & 1 & 1 & 1 & 1 \\
 1 & 1 & 1 & 1 & 1 & 1 & 1 & 1 & 1 & 1 & 1 & 1 & 1 \\
 1 & 1 & 1 & 1 & 1 & 1 & 1 & 1 & 1 & 1 & 1 & 1 & 1 \\
 1 & 1 & 1 & 1 & 1 & 1 & 1 & 1 & 1 & 1 & 1 & 1 & 1
\end{array}
\right)}$&$\tiny{\left(
\begin{array}{cccccccccc}
 2 & 2 & 2 & 1 & 1 & 1 & 1 & 1 & 1 & 1 \\
 1 & 1 & 1 & 4 & 1 & 1 & 1 & 1 & 1 & 1 \\
 1 & 1 & 1 & 1 & 2 & 2 & 2 & 1 & 1 & 1 \\
 1 & 1 & 1 & 1 & 1 & 1 & 1 & 2 & 2 & 2
\end{array}
\right)}$&(208,256) \\\hline

\end{tabular}
\caption{Results for the tetra-quadric in $\mathbb{P}^1\times\mathbb{P}^1\times\mathbb{P}^1\times\mathbb{P}^1$, continued.}
\end{center}
\end{table}

\clearpage

\begin{table}[!h]
\begin{center}
\begin{tabular}{|l|l|l|}\hline
$B$&$C$&\footnotesize{(${\rm ind}(B)$,${\rm ind}(C)$)}\\\hline\hline

$\tiny{\left(
\begin{array}{ccccccccccccc}
 1 & 1 & 1 & 1 & 1 & 1 & 1 & 1 & 1 & 1 & 1 & 1 & 1 \\
 1 & 1 & 1 & 1 & 1 & 1 & 1 & 1 & 1 & 1 & 1 & 1 & 1 \\
 1 & 1 & 1 & 1 & 1 & 1 & 1 & 1 & 1 & 1 & 1 & 1 & 1 \\
 1 & 1 & 1 & 1 & 1 & 1 & 1 & 1 & 1 & 1 & 1 & 1 & 1
\end{array}
\right)}$&$\tiny{\left(
\begin{array}{cccccccccc}
 3 & 2 & 1 & 1 & 1 & 1 & 1 & 1 & 1 & 1 \\
 1 & 1 & 2 & 2 & 2 & 1 & 1 & 1 & 1 & 1 \\
 1 & 1 & 1 & 1 & 1 & 2 & 2 & 2 & 1 & 1 \\
 1 & 1 & 1 & 1 & 1 & 1 & 1 & 1 & 3 & 2
\end{array}
\right)}$&(208,256) \\\hline

$\tiny{\left(
\begin{array}{ccccccccccccc}
 1 & 1 & 1 & 1 & 1 & 1 & 1 & 1 & 1 & 1 & 1 & 1 & 1 \\
 1 & 1 & 1 & 1 & 1 & 1 & 1 & 1 & 1 & 1 & 1 & 1 & 1 \\
 1 & 1 & 1 & 1 & 1 & 1 & 1 & 1 & 1 & 1 & 1 & 1 & 1 \\
 1 & 1 & 1 & 1 & 1 & 1 & 1 & 1 & 1 & 1 & 1 & 1 & 1
\end{array}
\right)}$&$\tiny{\left(
\begin{array}{cccccccccc}
 3 & 2 & 1 & 1 & 1 & 1 & 1 & 1 & 1 & 1 \\
 1 & 1 & 2 & 2 & 2 & 1 & 1 & 1 & 1 & 1 \\
 1 & 1 & 1 & 1 & 1 & 3 & 2 & 1 & 1 & 1 \\
 1 & 1 & 1 & 1 & 1 & 1 & 1 & 2 & 2 & 2
\end{array}
\right)}$&(208,256) \\\hline

$\tiny{\left(
\begin{array}{ccccccccccccc}
 1 & 1 & 1 & 1 & 1 & 1 & 1 & 1 & 1 & 1 & 1 & 1 & 1 \\
 1 & 1 & 1 & 1 & 1 & 1 & 1 & 1 & 1 & 1 & 1 & 1 & 1 \\
 1 & 1 & 1 & 1 & 1 & 1 & 1 & 1 & 1 & 1 & 1 & 1 & 1 \\
 1 & 1 & 1 & 1 & 1 & 1 & 1 & 1 & 1 & 1 & 1 & 1 & 1
\end{array}
\right)}$&$\tiny{\left(
\begin{array}{cccccccccc}
 3 & 2 & 1 & 1 & 1 & 1 & 1 & 1 & 1 & 1 \\
 1 & 1 & 3 & 2 & 1 & 1 & 1 & 1 & 1 & 1 \\
 1 & 1 & 1 & 1 & 2 & 2 & 2 & 1 & 1 & 1 \\
 1 & 1 & 1 & 1 & 1 & 1 & 1 & 2 & 2 & 2
\end{array}
\right)}$&(208,256) \\\hline

$\tiny{\left(
\begin{array}{ccccccccccccc}
 1 & 1 & 1 & 1 & 1 & 1 & 1 & 1 & 1 & 1 & 1 & 1 & 1 \\
 1 & 1 & 1 & 1 & 1 & 1 & 1 & 1 & 1 & 1 & 1 & 1 & 1 \\
 1 & 1 & 1 & 1 & 1 & 1 & 1 & 1 & 1 & 1 & 1 & 1 & 1 \\
 1 & 1 & 1 & 1 & 1 & 1 & 1 & 1 & 1 & 1 & 1 & 1 & 1
\end{array}
\right)}$&$\tiny{\left(
\begin{array}{cccccccccc}
 4 & 1 & 1 & 1 & 1 & 1 & 1 & 1 & 1 & 1 \\
 1 & 2 & 2 & 2 & 1 & 1 & 1 & 1 & 1 & 1 \\
 1 & 1 & 1 & 1 & 2 & 2 & 2 & 1 & 1 & 1 \\
 1 & 1 & 1 & 1 & 1 & 1 & 1 & 2 & 2 & 2
\end{array}
\right)}$&(208,256) \\\hline

$\tiny{\left(
\begin{array}{cccccccccccccc}
 1 & 1 & 1 & 1 & 1 & 1 & 1 & 1 & 1 & 1 & 1 & 1 & 1 & 1 \\
 1 & 1 & 1 & 1 & 1 & 1 & 1 & 1 & 1 & 1 & 1 & 1 & 1 & 1 \\
 1 & 1 & 1 & 1 & 1 & 1 & 1 & 1 & 1 & 1 & 1 & 1 & 1 & 1 \\
 1 & 1 & 1 & 1 & 1 & 1 & 1 & 1 & 1 & 1 & 1 & 1 & 1 & 1
\end{array}
\right)}$&$\tiny{\left(
\begin{array}{ccccccccccc}
 2 & 2 & 2 & 1 & 1 & 1 & 1 & 1 & 1 & 1 & 1 \\
 1 & 1 & 1 & 2 & 2 & 2 & 1 & 1 & 1 & 1 & 1 \\
 1 & 1 & 1 & 1 & 1 & 1 & 2 & 2 & 2 & 1 & 1 \\
 1 & 1 & 1 & 1 & 1 & 1 & 1 & 1 & 1 & 3 & 2
\end{array}
\right)}$&(224,272) \\\hline

$\tiny{\left(
\begin{array}{cccccccccccccc}
 1 & 1 & 1 & 1 & 1 & 1 & 1 & 1 & 1 & 1 & 1 & 1 & 1 & 1 \\
 1 & 1 & 1 & 1 & 1 & 1 & 1 & 1 & 1 & 1 & 1 & 1 & 1 & 1 \\
 1 & 1 & 1 & 1 & 1 & 1 & 1 & 1 & 1 & 1 & 1 & 1 & 1 & 1 \\
 1 & 1 & 1 & 1 & 1 & 1 & 1 & 1 & 1 & 1 & 1 & 1 & 1 & 1
\end{array}
\right)}$&$\tiny{\left(
\begin{array}{ccccccccccc}
 2 & 2 & 2 & 1 & 1 & 1 & 1 & 1 & 1 & 1 & 1 \\
 1 & 1 & 1 & 2 & 2 & 2 & 1 & 1 & 1 & 1 & 1 \\
 1 & 1 & 1 & 1 & 1 & 1 & 3 & 2 & 1 & 1 & 1 \\
 1 & 1 & 1 & 1 & 1 & 1 & 1 & 1 & 2 & 2 & 2
\end{array}
\right)}$&(224,272) \\\hline

$\tiny{\left(
\begin{array}{cccccccccccccc}
 1 & 1 & 1 & 1 & 1 & 1 & 1 & 1 & 1 & 1 & 1 & 1 & 1 & 1 \\
 1 & 1 & 1 & 1 & 1 & 1 & 1 & 1 & 1 & 1 & 1 & 1 & 1 & 1 \\
 1 & 1 & 1 & 1 & 1 & 1 & 1 & 1 & 1 & 1 & 1 & 1 & 1 & 1 \\
 1 & 1 & 1 & 1 & 1 & 1 & 1 & 1 & 1 & 1 & 1 & 1 & 1 & 1
\end{array}
\right)}$&$\tiny{\left(
\begin{array}{ccccccccccc}
 2 & 2 & 2 & 1 & 1 & 1 & 1 & 1 & 1 & 1 & 1 \\
 1 & 1 & 1 & 3 & 2 & 1 & 1 & 1 & 1 & 1 & 1 \\
 1 & 1 & 1 & 1 & 1 & 2 & 2 & 2 & 1 & 1 & 1 \\
 1 & 1 & 1 & 1 & 1 & 1 & 1 & 1 & 2 & 2 & 2
\end{array}
\right)}$&(224,272) \\\hline

$\tiny{\left(
\begin{array}{cccccccccccccc}
 1 & 1 & 1 & 1 & 1 & 1 & 1 & 1 & 1 & 1 & 1 & 1 & 1 & 1 \\
 1 & 1 & 1 & 1 & 1 & 1 & 1 & 1 & 1 & 1 & 1 & 1 & 1 & 1 \\
 1 & 1 & 1 & 1 & 1 & 1 & 1 & 1 & 1 & 1 & 1 & 1 & 1 & 1 \\
 1 & 1 & 1 & 1 & 1 & 1 & 1 & 1 & 1 & 1 & 1 & 1 & 1 & 1
\end{array}
\right)}$&$\tiny{\left(
\begin{array}{ccccccccccc}
 3 & 2 & 1 & 1 & 1 & 1 & 1 & 1 & 1 & 1 & 1 \\
 1 & 1 & 2 & 2 & 2 & 1 & 1 & 1 & 1 & 1 & 1 \\
 1 & 1 & 1 & 1 & 1 & 2 & 2 & 2 & 1 & 1 & 1 \\
 1 & 1 & 1 & 1 & 1 & 1 & 1 & 1 & 2 & 2 & 2
\end{array}
\right)}$&(224,272) \\\hline

$\tiny{\left(
\begin{array}{ccccccccccccccc}
 1 & 1 & 1 & 1 & 1 & 1 & 1 & 1 & 1 & 1 & 1 & 1 & 1 & 1 & 1 \\
 1 & 1 & 1 & 1 & 1 & 1 & 1 & 1 & 1 & 1 & 1 & 1 & 1 & 1 & 1 \\
 1 & 1 & 1 & 1 & 1 & 1 & 1 & 1 & 1 & 1 & 1 & 1 & 1 & 1 & 1 \\
 1 & 1 & 1 & 1 & 1 & 1 & 1 & 1 & 1 & 1 & 1 & 1 & 1 & 1 & 1
\end{array}
\right)}$&$\tiny{\left(
\begin{array}{cccccccccccc}
 2 & 2 & 2 & 1 & 1 & 1 & 1 & 1 & 1 & 1 & 1 & 1 \\
 1 & 1 & 1 & 2 & 2 & 2 & 1 & 1 & 1 & 1 & 1 & 1 \\
 1 & 1 & 1 & 1 & 1 & 1 & 2 & 2 & 2 & 1 & 1 & 1 \\
 1 & 1 & 1 & 1 & 1 & 1 & 1 & 1 & 1 & 2 & 2 & 2
\end{array}
\right)}$&(240,288) \\\hline

\end{tabular}
\caption{Results for the tetra-quadric in
  $\mathbb{P}^1\times\mathbb{P}^1\times\mathbb{P}^1\times\mathbb{P}^1$,
  continued.}
\end{center}
\end{table}

\newpage



\end{document}